\documentclass[preprint,longabstract]{aastex}
\usepackage{comment}
\newcommand{\so}{$\sigma$~Orionis }
\usepackage{bm,graphicx}
\usepackage{epstopdf}
\usepackage{natbib}
\usepackage{amsmath, amsthm, amssymb}
\usepackage{lscape}
\usepackage{color}

\slugcomment{Accepted in ApJ}

\shorttitle{New isolated planetary mass objects and the stellar and substellar mass function of the  $\sigma$~Orionis cluster}
\shortauthors{Pe\~na Ram\'irez et al.}

\begin{document}

\title{New isolated planetary mass objects and the stellar and substellar mass function of the $\bm \sigma$~Orionis cluster}

\author{K$.$ Pe\~na Ram\'irez\altaffilmark{1}, V.\,J.\,S$.$ B\'ejar\altaffilmark{1}}
\affil{Instituto de Astrof\'isica de Canarias, C/$.$ V\'\i a L\'actea s/n, E-38205 La Laguna, Tenerife, Spain}
\email{karla@iac.es, vbejar@iac.es}
\author{M$.$\,R$.$ Zapatero Osorio}
\affil{Centro de Astrobiolog\'ia (CSIC-INTA), Crta$.$ Ajalvir km 4, E-28850 Torrej\'on de Ardoz, Madrid, Spain}
\email{mosorio@cab.inta-csic.es}
\author{M$.$\,G$.$ Petr-Gotzens}
\affil{European Southern Observatory, Karl-Schwarzschild-Str$.$ 2, 85748, Garching bei M\"unchen, Germany}
\email{mpetr@eso.org}
\and
\author{E.\,L$.$ Mart\'in}
\affil{Centro de Astrobiolog\'ia (CSIC-INTA), Crta$.$ Ajalvir km 4, E-28850 Torrej\'on de Ardoz, Madrid, Spain}
\email{ege@cab.inta-csic.es}

\altaffiltext{1}{Departamento de Astrof\'isica, Universidad de La Laguna, E-38205 La Laguna, Tenerife, Spain.}

\begin{abstract}
We report on our analysis of the VISTA Orion $ZYJHK_s$ photometric data (completeness magnitudes of $Z$\,=\,22.6 and $J$\,=\,21.0\,mag) focusing on a circular area of 2798.4 arcmin$^2$ around the young \so star cluster ($\sim$3\,Myr, $\sim$352\,pc, and solar metallicity). The combination of the VISTA photometry with optical, {\sl WISE} and {\sl Spitzer} data allows us to identify a total of 210 \so member candidates with masses in the interval 0.25--0.004\,$M_{\odot}$, 23 of which are new planetary-mass object findings. These discoveries double the number of cluster planetary-mass candidates known so far. One object has colors compatible with a T spectral type. The \so cluster harbors about as many brown dwarfs (69, 0.072--0.012\,$M_{\odot}$) and planetary-mass objects (37, 0.012--0.004\,$M_{\odot}$) as very low-mass stars (104, 0.25--0.072\,$M_{\odot}$). Based on {\sl Spitzer} data, we derive a disk frequency of $\sim$40\%~for very low-mass stars, brown dwarfs, and planetary mass objects in $\sigma$~
Orionis. The radial density distributions of these three mass intervals are alike: all are spatially concentrated within an effective radius of 12\arcmin~(1.2\,pc) around the multiple star $\sigma$ Ori, and no obvious segregation between disk-bearing and diskless objects is observed. Using the VISTA data and the Mayrit catalog, we derive the cluster mass spectrum  ($\Delta N / \Delta M \sim M^{-\alpha}$) from $\sim$19 to 0.006\,$M_{\odot}$ (VISTA $ZJ$ completeness), which is reasonably described by two power-law expressions with indices of $\alpha = 1.7 \pm 0.2$ for $M > 0.35$\,$M_{\odot}$, and $\alpha = 0.6 \pm 0.2$ for $M < 0.35$\,$M_{\odot}$. The \so mass spectrum smoothly extends into the planetary-mass regime down to 0.004\,$M_{\odot}$. Our findings of T-type sources ($< 0.004$\,$M_{\odot}$) in the VISTA \so exploration appear to be smaller than what is predicted by the extrapolation of the cluster mass spectrum down to the survey $J$-band completeness.
\end{abstract}


\section{Introduction}
The stellar and substellar mass function (\citealp{salpeter55}; \citealp[see][]{bastian10} for a recent review) is the most general outcome of the process of star formation in a given region. Young stellar clusters (ages less than about 10\,Myr) offer the opportunity to study the mass function soon after the main burst of star formation activity has ceased. Clusters provide a snapshot of a roughly coeval population that has emerged from the same molecular cloud. Several of these regions stand out for their properties of proximity, age, low internal extinction, and rich number of members. 

The \so cluster ($\sim$3\,Myr, \citealt{zapatero02,sherry08}; $\sim$300--450\,pc\footnote{Here, we will adopt the Hipparcos distance, 352\,pc, measured by \citet{perryman97}.}, \citealt{brown94,perryman97,mayne08,sherry08,caballero08a}; $E(B-V)$\,=\,0.05\,mag, \citealt{lee68}; and solar metallicity, \citealt{hernandez08}) is known since the early studies of \citet{garrison67} and \citet{lynga81,lynga83}. X-ray observations of the region around the $\sigma$ Ori multiple star revealed a very young stellar population \citep{wolk96,walter97}. The \so cluster is acknowledged as one suitable ground where observational efforts seek to establish to what extent there is a continuity of the star formation process from the OB-type stars (the star $\sigma$ Ori itself is a multiple, massive system in the main sequence) down to free-floating planetary mass objects of a few times the mass of Jupiter\footnote{1\,$M_\odot$\,=\,1047\,$M_{\rm Jup}$.} (e.g., see the Mayrit catalog by \citealt{caballero08b} and references therein). 

The stellar mass function of the \so cluster in the mass range 1--24\,$M_\odot$ was discussed in \citet{caballero07a}. This author found it consistent with the Salpeter mass function \citep{salpeter55}. Regarding the substellar domain, the first \so brown dwarf discoveries were reported about a decade ago \citep{bejar99,zapatero00}. Since then, combined deep optical and near-infrared photometric surveys have led to the detection of hundreds of cluster low-mass stars, brown dwarfs, and planetary-mass candidates (with a mass below the deuterium burning-mass limit at 13\, $M_{\rm Jup}$) down to about 5\, $M_{\rm Jup}$; a considerable fraction of them ($\sim$70\,\%) have follow-up spectroscopy and mid-infrared flux excesses confirming their likely membership in the cluster; (e.g., \citealt{martin01,barrado01,barrado02,zapatero02sori70,zapatero02,jeffries06,hernandez07,zapatero07,caballero07,caballero08d,sacco07,sacco08,scholz08,luhman08,rigliaco09}). The first derivation of the \so low-mass function (0.2--0.013\,$M_\odot$) was presented in \citet{bejar01} covering an area of 0.24 deg$^2$.

Only two surveys \citep{lodieu09,bejar11} have homogeneously explored a large area in the \so~cluster (0.8--1\,deg$^2$) down to the brown dwarf--planet borderline. These authors found that the cluster mass spectrum in the mass interval from low-mass stars to about 13\, $M_{\rm Jup}$ can be fit by a power-law expression $\Delta N / \Delta M \sim M^{-\alpha}$, where $\alpha$\,=\,0.5\,$\pm$\,0.2 \citep{lodieu09} and 0.7\,$\pm$\,0.3 \citep{bejar11} [Salpeter's $\alpha$\,=\,2.35]. 

Power law functions have been found appropriate to describe the mass function of several open clusters over the mass range  0.6--0.04\,$M_\odot$,  like Upper Scorpius ($\sim$5\,Myr, \citealt{lodieu07b}), IC\,4665 ($\sim$25\,Myr, \citealt{dewit06,lodieu11}), the Pleiades ($\sim$120\,Myr, e.g., \citealt{zapatero97,bouvier98,martin00,bihain06,lodieu07}), Blanco I ($\sim$130\,Myr, \citealt{moraux07}), the Orion Nebula Cluster ($\sim$1\,Myr, \citealt{muench02} and references therein) and, the field (\citealt{bastian10} and references therein). The extension of the \so mass function toward a smaller mass ($\sim$6\, $M_{\rm Jup}$) in tiny regions of the cluster yielded $\alpha$ values in the range 0.4--0.6 \citep{gonzalez06,caballero07}. \citet{bihain09} suggested that there may exist a turnover in the substellar mass spectrum of \so below 6\, $M_{\rm Jup}$. 

Here, we present VISTA (Visible and Infrared Survey Telescope for Astronomy) observations of \so using broadband filters $ZYJHK_s$ aiming at defining the substellar cluster sequence and its global properties in a consistent manner. One noteworthy characteristic of this survey is its homogeneity in mass coverage from 0.25 to 0.006\,$M_\odot$ ($\sim$\,262--6\, $M_{\rm Jup}$, completeness) all across an important fraction of the cluster area and uncovering $\geq$75\% of cluster members. VISTA and complementary photometry, and the identification of cluster member candidates are shown in Sections \ref{data} and \ref{search}. Mass estimates based on theoretical isochrones are presented in Section \ref{models}. In Section \ref{cont} we discuss the object contamination in the VISTA survey. By combining VISTA data with photometry from the literature, we study the presence of mid-infrared flux excesses in Section \ref{disksec}. Section \ref{spatial} deals with the spatial distribution of very low-mass stars, brown dwarfs and free-floating planets in $\sigma$~Orionis. Finally, in Section \ref{fmsec} we discuss the cluster mass function from the O-type stars through the free-floating planets. Conclusions are presented in Section \ref{final}.

\section{Observations}\label{data}
\subsection{Near-infrared photometry: VISTA}\label{limcomp}
Our survey of the \so cluster is based on the VISTA Orion survey data \citep{petr11}. This survey was a dedicated imaging exploration of 30~deg$^2$ in the Orion Belt region, including $\sigma$~Orionis, carried out as part of the VISTA science verification program. VISTA \citep{vista_1,vista_2} is a near-infrared survey facility equipped with a wide--field, near--infrared
camera (VIRCAM) and mounted on a 4.2-m telescope at ESO's Paranal Observatory. VIRCAM comprises sixteen 2048$\times$2048 pixel infrared detectors (Raytheon VIRGO HgCdTe) with a mean pixel size of 0\farcs339 and with signiÞcant interchip gaps. Therefore, the area covered by one VIRCAM exposure (or ``pawprint") is 0.6\,deg$^2$ in a field of view of 1.3$\times$1.0\,deg$^2$. For a uniform sky coverage, six pawprints offset by a convenient amount are required to fill an area of approximately 1.5$\times$1.1\,deg$^2$, which is called a ``tile". For details on the VISTA Orion survey strategy, number of observed tiles, and a description of the data sets we refer to \citet{arnaboldi10} and \citet{petr11}. 

Observations of the \so region (tile number 16 according to the nomenclature of the VISTA Orion survey) were performed between 2009 October 20 and October 28 employing the VISTA broadband filters $ZYJHK_s$. The integration times at $Z$- and $J$-band were roughly twice and 5 times, respectively, longer than what was used for the overall VISTA Orion survey, allowing us to obtain deeper images. This was crucial in order to identify the lowest mass cluster member candidates in $\sigma$~Orionis. In Table \ref{log} we provide the specific observing dates, total exposure times per filter, and seeing conditions as measured from the full width at half maximum (FWHM) of point-like sources in the stacked mosaics. 

Given the large amount of the VISTA Orion data ($\sim$559 Gb), they were processed by a dedicated pipeline run by the Cambridge Astronomical Survey Unit (CASU)\footnote{\url{http://casu.ast.cam.ac.uk}}. We used the reduced products obtained with the VIRCAM pipeline version 1.0, which delivered science-ready stacked images and mosaics, as well as photometrically and astrometrically calibrated source catalogs. Standard reduction steps included dark and flat-field corrections, and sky background subtraction. All individual images (pawprints) were aligned and stacked together, and then conveniently montaged to produce deep tiles. Automatic identification of sources and photometry were performed by the instrument pipeline in CASU. Tile images of different filters were treated individually. Aperture photometry with an aperture radius of 2\farcs0 was obtained to avoid contamination from nearby sources. Aperture corrections were applied. Instrumental $JHK_s$ magnitudes were converted into apparent calibrated magnitudes using the 2MASS photometry \citep{skrutskie06}, and the $ZY$ data were calibrated using routine observations of UKIRT photometric standard star fields \citep{casali07} carried out during the Orion survey campaign. The photometric calibration precision is typically better than 5\%\,for magnitudes brighter than the completeness magnitude. The 2MASS catalog was also used to calibrate astrometrically the VISTA Orion data to an absolute precision of 0\farcs3 (relative astrometric accuracy of 0\farcs1). Data of different filters were paired within a 2\farcs0 radius to produce a merged catalog (both non- and cross-correlated objects are included). 

We adopted the completeness magnitude of the VISTA \so survey as the faintest magnitude at which the number of sources per interval of magnitude does not deviate from an increasing distribution. This function is obtained from the fit of an exponential distribution from  bright to intermediate magnitudes in each passband. We adopted the limiting magnitude as the magnitude bin at which the total number of sources deviates by $\ge$50\%~from the prediction of the exponential law. Our determined completeness and limiting magnitudes (see Table \ref{log}) roughly correspond to source detections around the 10-$\sigma$ and 4-$\sigma$ level, respectively ($\sigma$ is the sky-subtracted background noise). At the bright end of the survey, we considered sources with $J\,\ge\,13$\,mag to ensure that the registered photon counts laid within the linearity regime of the VIRCAM detectors in all filters.

In next sections we present the cross-correlation of the VISTA \so catalog with various other catalogs compiled by us or taken from the literature. We used a cross-correlation radius of 2\farcs0 , which takes into account the astrometric uncertainties of the catalogs and possible systematic deviations. 

\subsection{Mid-infrared photometry}
\subsubsection{\textit{Spitzer}}
\begin{figure*}
\centering
\includegraphics[angle=270,scale=0.6]{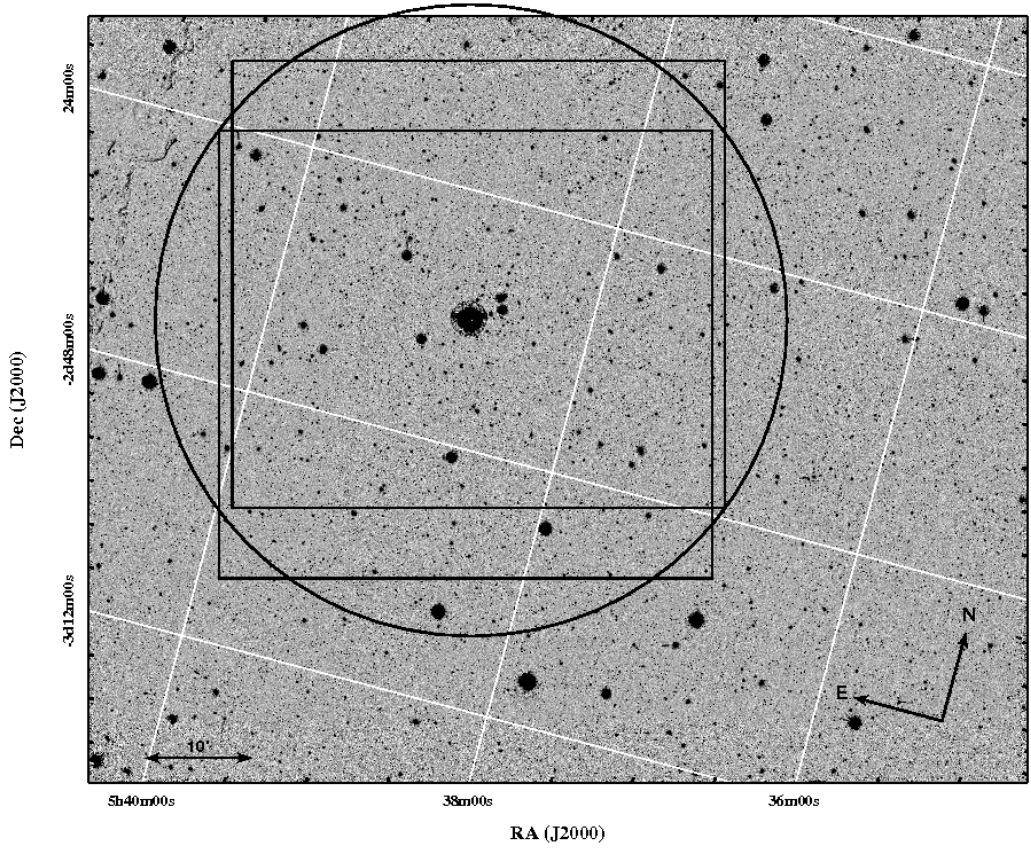}
\caption{
Mosaiced VISTA $J$-band image of size 1.2$\times$1.5 deg$^2$ (designated as tile no. 16 in the VISTA Orion survey). Our search has explored the region inside the 30\arcmin-radius circle centered on the bright, massive multiple system $\sigma$~Ori. The regions explored by {\sl Spitzer} are shown with squares, the top square corresponds to the $[3.6]$- and $[5.8]$-band images, and the bottom square to the $[4.5]$- and $[8.0]$-band images. 
\label{survey}}
\end{figure*}

To extend the VISTA wavelength coverage toward the mid-infrared wavelengths, and aimed at studying the presence of mid-infrared flux excesses in \so objects, we complemented the VISTA data set with [3.6], [4.5], [5.8], and [8.0]-band photometry acquired with the \textit{Spitzer Space Telescope} Infrared Array Camera (IRAC, \citealt{fazio04}). The four-channel {\sl Spitzer} data were collected by \citet{hernandez07} on 2004 October 9 (Guaranteed Time Observation program \#37, PI: G$.$ Fazio), and covered an area of $\sim$1739\,arcmin$^2$ in the \so cluster. These data have been widely used by different groups \citep{caballero07,zapatero07,scholz08,luhman08}. The completeness limits are 17.25, 17.0, 14.75, and 14.0\,mag at 3.6, 4.5, 5.8, and 8.0\,$\mu$m, respectively \citep{luhman08}. VISTA data (particularly the $J$-band) are sufficiently sensitive to ensure the detection of the cluster sources to the same depth (i.e., low mass objects), or even beyond, as the {\sl Spitzer} images. Here, we employed the {\sl Spitzer} photometry obtained by \citet{zapatero07}. Figure \ref{survey} shows the location of the {\sl Spitzer}/IRAC fields, which are contained in the VISTA area but cover only $\sim$62\% of the explored circular region in $\sigma$~Orionis.

\subsubsection{\textit{WISE}}
Additional mid-infrared photometry at 3.4, 4.6, 12, and 22\,$\mu$m ($W1$, $W2$, $W3$, $W4$) was provided by the \textit{Wide-field Infrared Survey Explorer (WISE)} space mission \citep{wright10}. These data were extracted from the {\sl WISE} preliminary data release\footnote{\url{http://wise2.ipac.caltech.edu/docs/release/prelim/}}, which includes the first 105 days of {\sl WISE} survey observations between 2010 January 14 and 2010 April 29. {\sl WISE} photometry is processed with initial calibrations and reduction algorithms. Astrometric accuracy of sources brighter than $W1\sim13.0$\,mag is 0\farcs2. From the on-line {\sl WISE} catalog, we retrieved magnitudes of cross-correlated VISTA sources with quoted mid-infrared photometric signal-to-noise (S/N) ratio above 4. The quoted 5-$\sigma$ {\sl WISE} detection limits  are 15.3, 14.4, 10.1 and 6.7\,mag at $W1$, $W2$, $W3$ and $W4$, respectively. In addition, we inspected {\sl WISE} images visually for all cross-correlated sources. The VISTA survey and the {\sl Spitzer} images are significantly deeper than the {\sl WISE} data. In contrast to {\sl Spitzer}, {\sl WISE} covered the full extension of the \so~cluster. The combination of VISTA and {\sl WISE} photometry (Section \ref{disksec}) will allow us to study the presence of mid-infrared flux excesses of very low-mass stars and brown dwarfs all across the cluster area.

\subsection{Optical photometry}
The combination of optical and near-infrared data is notably useful to distinguish true ultracool late-L and T-type dwarfs from photometric contaminants. Ideally, the VISTA survey should have been complemented with red optical images down to $I$\,=\,25\,mag or $Z$\,=\,24\,mag, which would have enabled us to detect the latest L- and early- to mid-T members of \so at visible wavelengths. Unfortunately, we do not have these very deep optical images. We used all optical data available to us, including published and new photometry.

The images taken with the Wide Field Camera (WFC) of the 2.5-m Isaac Newton Telescope on La Palma Island cover a total area of 970\,arcmin$^2$ around the cluster center \citep{caballero07}. We refer to this reference for further details on the WFC photometry. The authors reported a limiting magnitude of $I$\,=\,24.1\,mag.

We also employed unpublished data taken by us with the Suprime-Cam mounted on the prime focus of the 8-m Subaru telescope on Hawaii. Observations were conducted using the Sloan $z'$ and Cousins $I$ filters on 2000 December 28. These filters, centered at $\sim$800\,nm ($I_c$) and $\sim$910\,nm ($z'$), are appropriate to characterize very red and cool sources. Suprime-Cam had a mosaic of ten MIT/LL 2048\,$\times$\,4096-pixel CCDs arranged in a pattern of 5\,$\times$\,2 with interchip gaps of about 16\farcs5 in its ``old" version prior to 2008. The projection of one pixel onto the sky is 0\farcs2. At the time of the observations, the South-East CCD (number six) was dead and did not register any images. A total of five ($z'$) and ten ($I_c$) exposures of 240\,s each were acquired with dither offsets of $\sim$20\arcsec~for a proper removal of the telluric background contribution. Therefore, total integration times were 20 min ($z'$) and 40 min ($I_c$). Taking into account the overlapping area of the dithers and the size of the nine (useful) detectors, our Suprime-Cam observations provided a field of view of 570 arcmin$^2$ North of the cluster center. In Table \ref{sub_coor} we list the measured central coordinates of each of the nine detectors. Observations were carried out under good conditions of sky transparency and seeing of 0\farcs9 ($z'$ band).

Raw frames were bias-subtracted and flat-field corrected using packages running inside the {\sc iraf} environment. We constructed the flat-field images for each passband by combining all Suprime-Cam (adequately scaled) science images taken during the night. Individual processed frames were conveniently aligned and stacked together to produce deep images.

The photometric aperture and point-spread function fitting analysis was carried out using routines within {\sc daophot}, which provides image profile information needed to discriminate between stars and resolved sources. Instrumental $I_c$ magnitudes were transformed into observed magnitudes using WFC observations by \citet{bejar11}: more than 190 sources with photometric error bars $\le$0.15\,mag were in common to both WFC and Suprime-Cam searches. We found that the uncertainty in the photometric calibration of the $I_c$ band is $\sim$0.03\,mag. Unfortunately, no standard stars were observed in the Sloan $z'$ filter for a proper calibration of the data. Therefore, we have not performed an absolute calibration of this passband, but have constructed color-magnitude diagrams where calibrated $I_c$ is plotted against the semi-instrumental index $I_c-z'$. Sources with intrinsically red colors will stand out toward high values of $I_c-z'$ with respect to the great bulk of field objects. Completeness and limiting magnitudes of the Suprime-Cam survey were determined following the procedure indicated in Section \ref{limcomp}. We have estimated them to be $I_c$\,=\,22.5\,mag for the completeness (10-$\sigma$), and $I_c$\,=\,23.8\,mag for the limit (4-$\sigma$). All Suprime-Cam CCDs were astrometrically calibrated using 2MASS data to an accuracy of 0\farcs1--0\farcs3.

\section{Selection of cluster member candidates}\label{search}
\subsection{Search area definition}
We searched for \so photometric member candidates in a circular area of radius 30\arcmin~centered on the massive star of the same name. \citet{bejar04,bejar11}, \citet{caballero07a,caballero08b} and \citet{lodieu09} have discussed that the cluster sequence is clearly distinguished from the field sources up to 30\arcmin. At larger radii, the contamination by other objects (e.g., young Orion sources, foreground and background sources) increases notably, blurring the \so sequence. \citet{caballero08} describes the cluster as a dense core extending from the center to a radius of 20\arcmin~and a rarified halo at larger separations up to 30\arcmin. In addition, \citet{bejar11} observed that at radii larger than 30\arcmin~there is an overdensity of red objects to the North and West of the cluster that might be related to the young population of the Orion Belt. The extension of \so is addressed in Section \ref{spatial}. As illustrated in Figure \ref{survey}, the 30\arcmin-radius circular region of our search is contained within tile 16 of the VISTA Orion survey except for a small sector of $\sim$12\,arcmin$^2$ to the North of the cluster. This sector is covered by other tiles of the VISTA survey, but they are not as deep as tile 16 and are therefore not included. 

Within the circular area there exist massive O, B, and A-type stars strongly saturating all optical and infrared data. Their bright ``halos" extend beyond a few times the point-spread function (PSF) of the images, artificially increasing the noise of the local background level and preventing us from detecting very faint (low-mass) sources in their surroundings. We estimated that the area lost to this effect is 17.0\,arcmin$^2$. We determined the final deep area of our VISTA survey to be 2798.4\,arcmin$^2$ (0.78\,deg$^2$).

\subsection{VISTA $\bm {ZJ}$ color-magnitude diagram}\label{zjsearch}
\begin{figure*}[!ht]
\epsscale{0.7}
\plotone{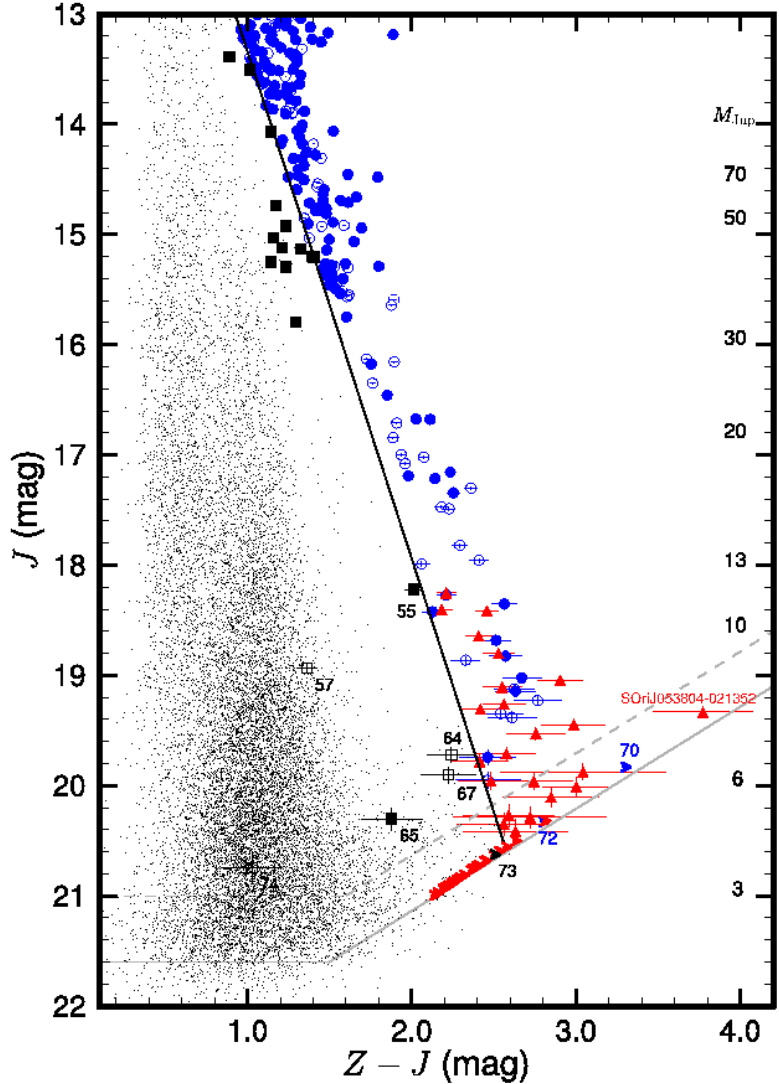}
\caption{
VISTA $J$ vs. $Z-J$ color magnitude diagram used to select \so~member candidates. Confirmed \so~members inside of the eligible zone are shown with filled blue circles, those outside of the eligible zone are in filled black squares. The straight line from $J$\,=\,13 to 20.5\,mag denotes the field-cluster separator according to our photometric criteria. Open blue circles stand for known cluster photometric candidates (not yet confirmed members) rediscovered with VISTA, those out of the eligible zone are represented with open black squares. The new discoveries ($J$\,=\,18--20.5\,mag) are plotted as red triangles. S\,Ori sources discussed in this work and the new T type candidate are labeled. The sources of the extended survey (down to $J$\,=\,21\,mag) with no $Z$ detections are shown with red arrows. The completeness- and the $\sim$4-$\sigma$ detection-limit are indicated with dashed and solid lines, respectively. Masses for an age of 3\,Myr and a distance of 352\,pc are labeled in Jupiter mass units. Field sources are plotted as black dots. A color version of this figure is available in the online journal.
\label{jzj}}
\end{figure*}

Since we sought cool objects with red $Z-J$ colors and the VISTA $Z$ and $J$-band images are quite deep,  we constructed the $J$ vs. $Z-J$ color-magnitude diagram to select photometrically \so member candidates. Previously known cluster members trace the cluster sequence from $J$\,=\,13 down to 20\,mag: there are 145 of these objects in the literature cross-correlated with VISTA and whose young age is confirmed by means of spectroscopic and mid-infrared photometric observations. As evidence for youth we considered the following signatures: the detection of the Li\,{\sc i} doublet atomic absorption at 670.8\,nm \citep{zapatero02,barrado03,kenyon05,sacco08}, broad or strong H$\alpha$ emission compatible with accretion events \citep{haromoreno53,wiramihardja89,wiramihardja91,barrado01,zapatero02,barrado02,barrado03,andrews04,weaver04,kenyon05,sacco08}, weak alkali absorption lines associated to low-gravity atmospheres \citep{bejar99,mcgovern04,kenyon05,burningham05,maxted08}, radial velocity consistent with the cluster systemic velocity, $\sim$\,30\,km\,s$^{-1}$, \citep{kenyon05,burningham05,maxted08,sacco08}, and/or infrared flux excesses indicative of the presence of surrounding disks or envelopes \citep{barrado03,jayawardhana03,oliveira06,hernandez07,caballero07,zapatero07,scholz08,luhman08}. The $ZJ$ sequence delineated by these 145 \so members is shown in Figure \ref{jzj} (blue filled circles). This population is clearly distinguishable from field sources (small dots) since, for a given magnitude, \so true members have redder $Z-J$ colors. 

To select new photometric member candidates we traced a blue envelope along the known cluster sequence. This blue envelope goes from ($Z-J$, $J$)\,=\,(0.93, 13.0) to (2.6, 20.5)\,mag and is plotted as a solid line in Figure \ref{jzj}. It separates field sources from eligible \so member candidates. The so-defined blue envelope roughly corresponds to a linear fit of the cluster $ZJ$ sequence shifted at $\sim$2-$\sigma$ (where $\sigma$ is the color dispersion of the fit) toward blue $Z-J$ indices. Of the 145 original sources, 129 are located within the eligible zone (red side of the envelope, see Table \ref{infoyoung}). Most of the cluster members that lie outside of the eligible zone have $J \le 16$\,mag, they represent about 10\%~of the confirmed young cluster members, i.e., by using the defined blue envelope we might be including the 90\%~of potential cluster members in the interval $J$\,=\,13--16\,mag. 

For the magnitude range $J$\,=\,13--20.5\,mag, where $J$ $\sim$\,20.5\,mag matches the VISTA survey limiting $Z$-band magnitude of cluster members, 219 VISTA sources comply with our $ZJ$ photometric criterion for \so membership. Of them, 129 are cluster members with confirmed features of youth (see above), 58 are reported in the literature as photometric candidates yet pending spectroscopic confirmation of their membership, and 32 are new detections. After visual inspection of the VISTA images, we rejected 12 candidates out of the 32 new sources: nine are merged bright sources, two are spurious detections, and one object is spatially resolved (its FWHM is larger than the average). The VISTA photometry of the resolved object is provided in Table\,\ref{resolved}. Here we report the finding of 207 \so members and member candidates, 20 of which are new (red symbols in Figure \ref{jzj}). All new detections have $J$-band brightness in the interval 18--20.5\,mag, which according to evolutionary models \citep{chabrier00model,baraffe03} corresponds to a mass range of $\sim$0.011--0.004\,$M_\odot$ at the adopted age and distance of the cluster, i.e., at the planet--brown dwarf boundary and below. There are no new more massive \so candidates in the VISTA survey, indicating that previous works \citep{bejar99,bejar01,bejar11,zapatero00,gonzalez06,caballero07,lodieu09,bihain09} carried out a successful census of the cluster very low-mass star and brown dwarf populations. 

One new cluster member candidate, S\,Ori\,J053804.65$-$021352.5, lies at about 1\farcs0 north of a field source. To secure the VISTA photometry, we performed PSF photometric analysis (using the {\sc daophot} package) particularly for the $Z$ band, where the candidate is significantly fainter than the field object. The $Z-J$ color of S\,Ori\,J053804.65$-$021352.5 suggests a T spectral type; this source is the only T-type candidate detected at both $Z$ and $J$ bands in the VISTA \so survey.

In Table \ref{young} we provide the VISTA, {\sl Spitzer}, and {\sl WISE} photometry of the 145 \so candidates members with youth features (129 of them within our $ZJ$ eligible zone). The observational evidence for their cluster membership along with further information, including bibliography, and an indication of whether they lie to the red side of the blue envelope in the $J$ vs. $Z-J$ diagram are given in Table \ref{infoyoung}. The VISTA, {\sl Spitzer}, and {\sl WISE} photometry of the 58 \so member candidates known in the literature that lie in the eligible zone of the $ZJ$ diagram is listed in Table \ref{known}, and Table \ref{infoknown} provides additional information for them. The data of the 20 new planetary-mass candidates reported here are given in Table \ref{new}, 90\% of them remain undetected in the {\sl WISE} dataset.

\subsection{Additional color-magnitude diagrams}
\begin{figure*}
\centering
\includegraphics[scale=0.5]{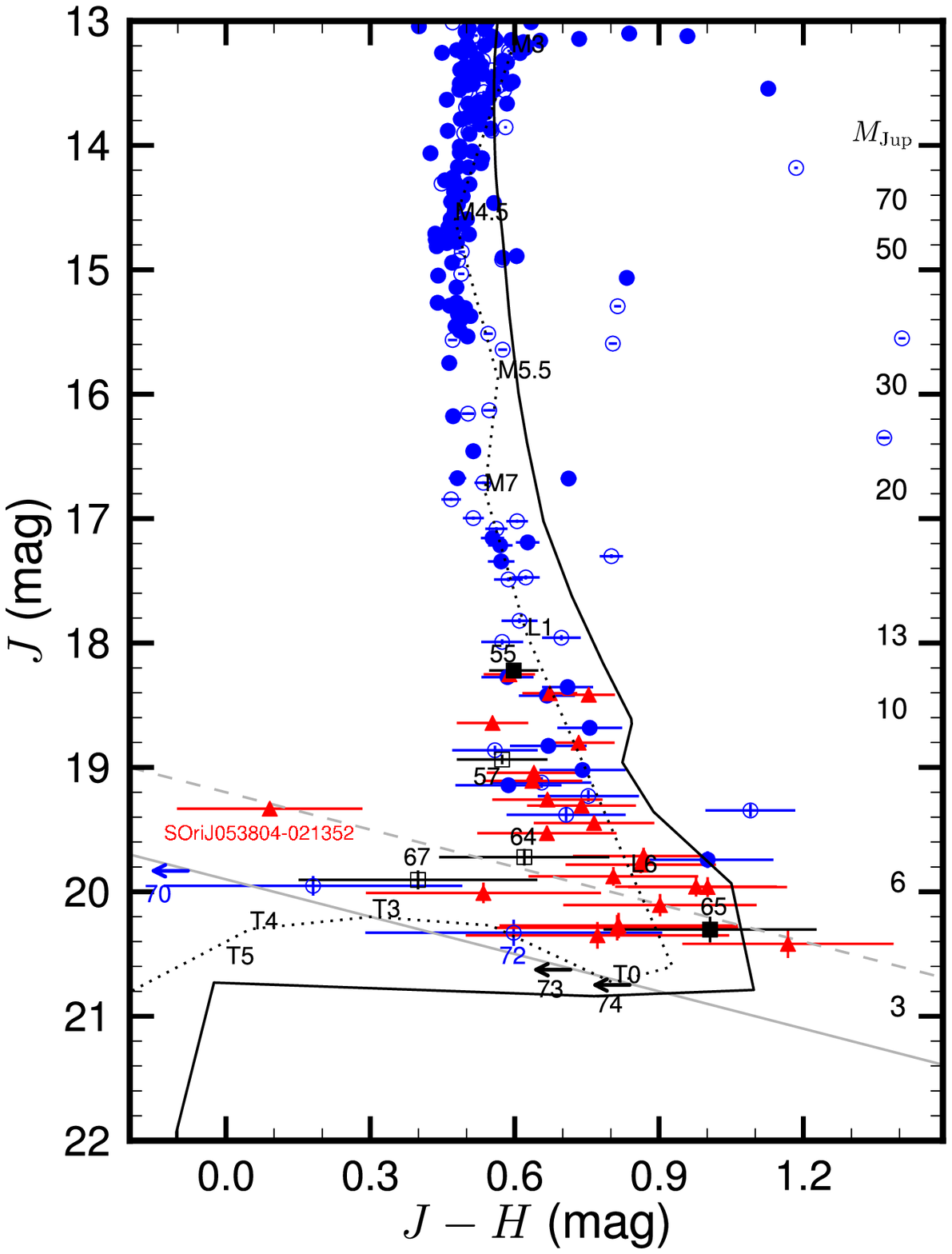}
\includegraphics[scale=0.5]{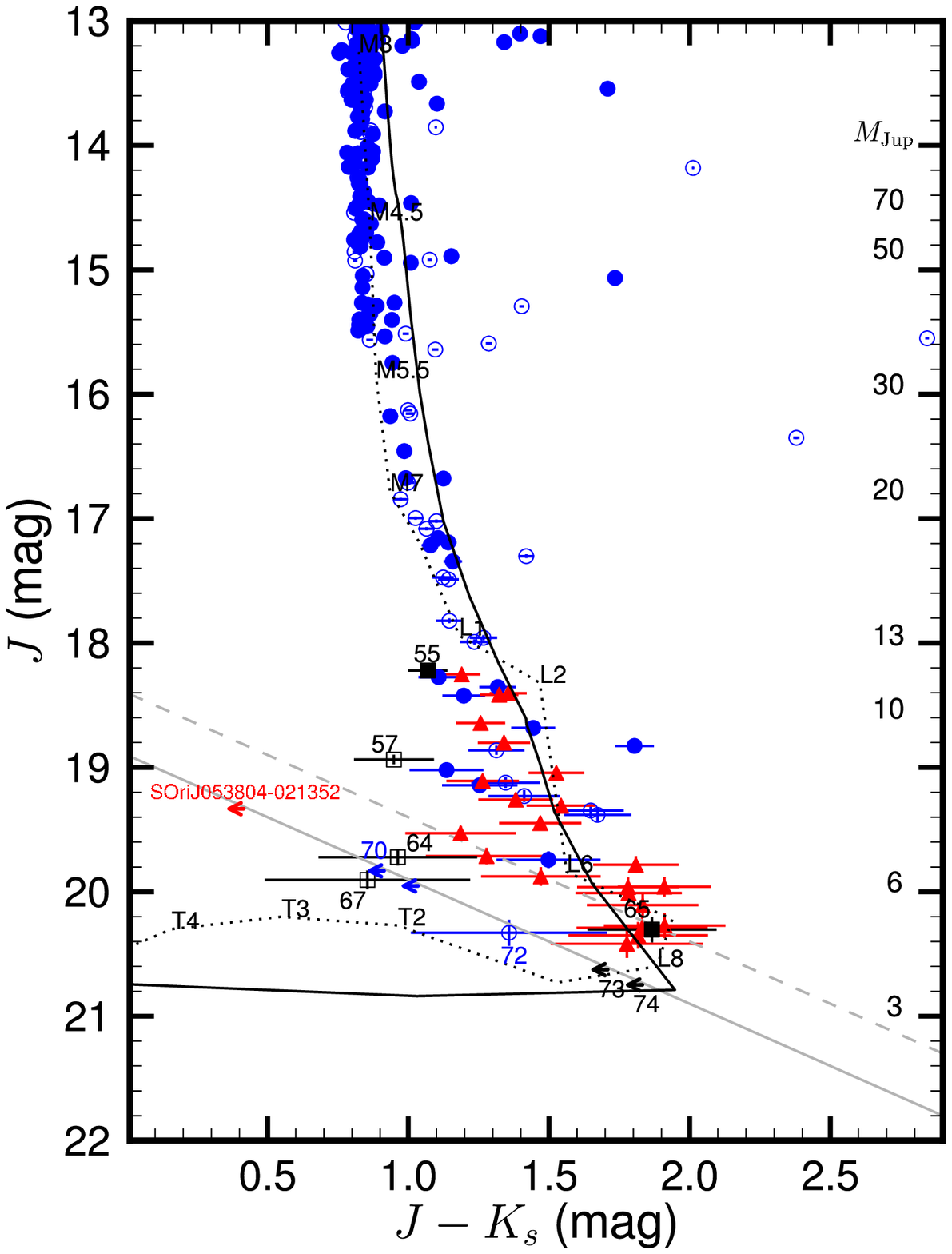}\\
\includegraphics[scale=0.5]{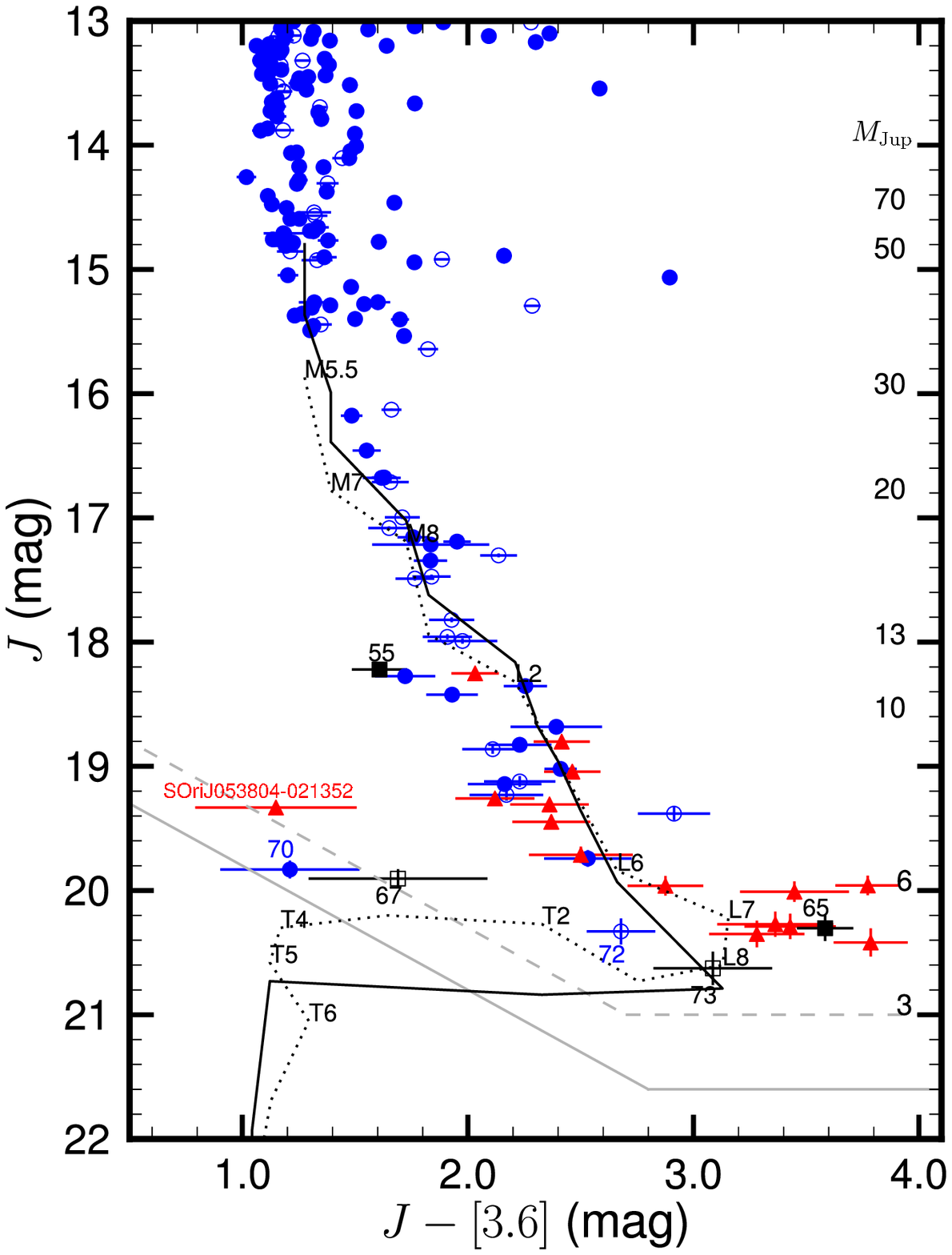}
\includegraphics[scale=0.5]{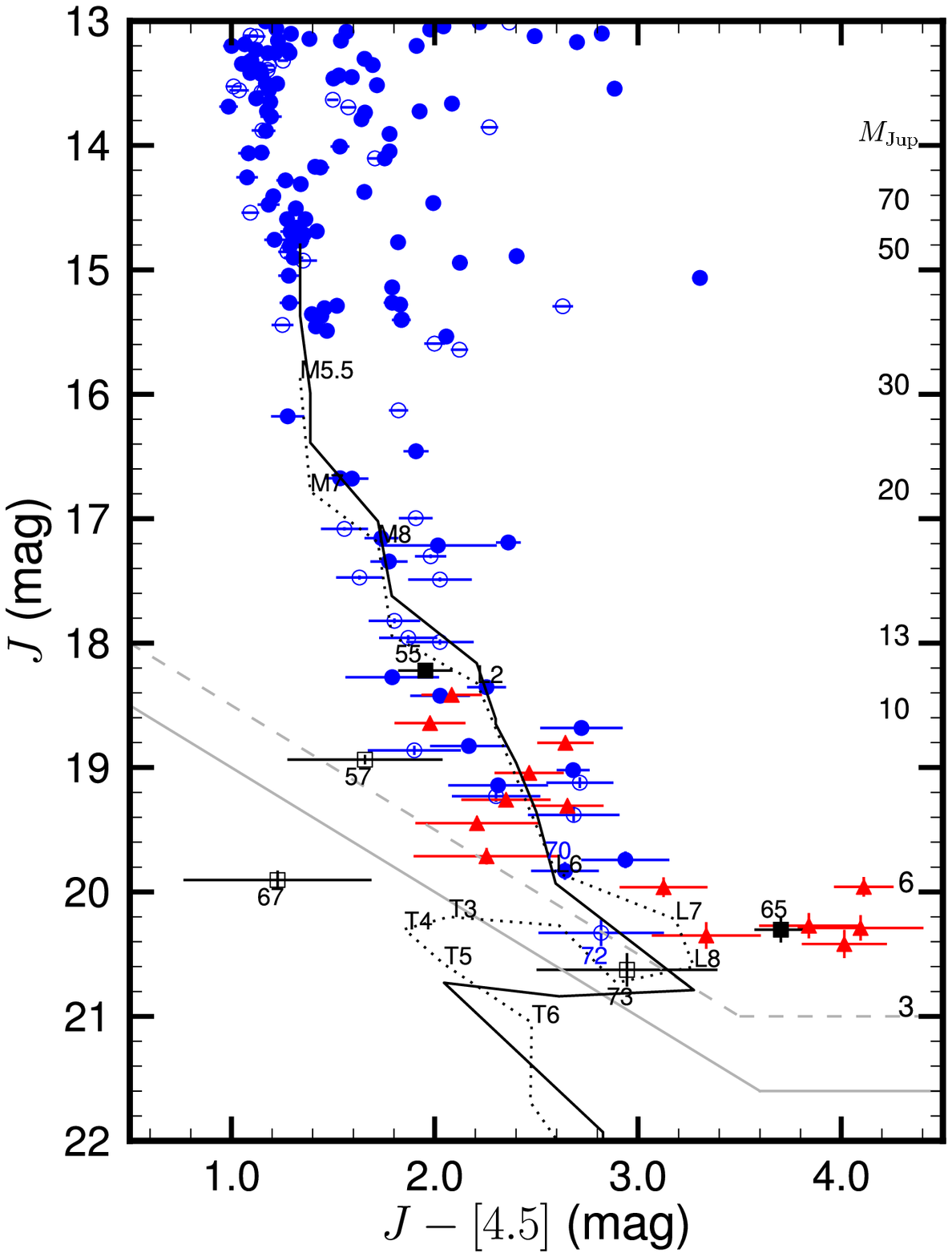}\\
\caption{
Various color-magnitude diagrams including VISTA and {\sl Spitzer} data. Symbols are as in Figure \ref{jzj}. The mean sequence of M-L-T types for field objects is overlaid (dotted line), it is normalized at the \so late-M types. The 3\,Myr evolutionary model by \citet{baraffe98,baraffe03} and \citet{chabrier00model} is included (solid line, see text). Source Mayrit\,1082188 falls out the $J$ vs. $J-[4.5]$ diagram due to its red $J-[4.5]$ color. The color version of these figures is available in the online journal.
\label{refinement}}
\end{figure*}

To test the quality of the $ZJ$ candidates as well as to delineate the \so sequence at other wavelengths, we built various color-magnitude diagrams using the VISTA data. Figure \ref{refinement} depicts the $J$ magnitudes as a function of $J-H$, and $J-K_s$. We also included the combined VISTA and {\sl Spitzer} $J$ vs. $J-[3.6]$ and $J-[4.5]$ diagrams in Figure~\ref{refinement}. For comparison purposes, we overlaid the average field sequence of mid-M to T-type objects (typical ages 1--5 Gyr) after normalization to the late-M population of the \so cluster. We constructed the average field sequence by calculating the mean magnitudes and colors for each spectral type and using the data published by \citet{hewett06}, \citet{patten06} and \citet{leggett07}.

All $ZJ$ objects nicely describe a continuous photometric sequence in the four color-magnitude diagrams of Figure \ref{refinement}. With the only exception of S\,Ori\,J053804.65$-$021352.5, none appears to deviate toward blue colors, which would have made us reject it as a cluster member candidate. According to Figure \ref{refinement}, most of the new candidates show colors typical of early-L through late-L types. S\,Ori\,J053804.65$-$021352.5 has colors (from $Z$ through [3.6]) fully compatible with T0--T4 types. Unfortunately, there is no [4.5] photometry for this source since its location was not covered by the corresponding {\sl Spitzer} image. In short, all possible color-magnitude diagrams covering the wavelength range 0.9--4.5\,$\mu$m confirm the photometric cluster member candidacy of the 207 sources originally selected from the VISTA $ZJ$ data in the interval $J$\,=\,13--20.5\,mag.

\subsection{Extended search down to $\bm J$ completeness}\label{arrowssec}
The $ZJ$ search is limited by the $Z$ band. To fully exploit the deep VISTA $J$-band data, we extended the search for \so member candidates down to $J$\,=\,21\,mag, which is the estimated completeness magnitude of the survey in this near-infrared band. According to Figure \ref{refinement}, at these faint magnitudes we would expect cluster members with colors typical of late-L and T types. The following photometric criteria, optimized for very cool objects, were applied:
\begin{itemize}
\item[-]We identified all sources with no $Z$ detections and $J$ magnitudes in the interval 19--21\,mag present in the VISTA \so catalog. This criterion would also allow us to ``recover" those objects with $Z$ magnitudes between completeness and detection limit that might have been lost by the automatic source identification of the instrument pipeline. This resulted in over a thousand potential candidates.
\item[-]After visual inspection of the $J$-band images, we were able to reject most of them since they turned out to be spurious detections or detector defects. The $Z$-band images were also inspected and we discarded those objects with $Z$-band fluxes above the 10-$\sigma$ detection (the $Z$-band automatic identification failed particularly on those objects close to bright stars). Basically, we were interested in sources with $Z-J\,\ge\,2.1$\,mag (as illustrated in Figure \ref{jzj}) since it is expected that true cluster members increase their $Z-J$ color at fainter magnitudes (or lower masses).
\item[-]Other filters were used to impose additional constraints: the VISTA $Y$-band and {\sl Spitzer} (VISTA $HK_s$ images are not deep enough for our purposes). For those objects with positive detections in any of the filters we considered the following color cuts: $Y-J\,\ge\,0.6$, $J-[3.6]\,\le\,3.5$, and $J-[4.5]\,\le\,4.5$\,mag. All color criteria have to be satisfied simultaneously. They allowed us to reject highly reddened extragalactic sources and Galactic objects earlier than M types. Figure 1 of \citet{burningham10} showed that L and T type sources have $Y-J$ colors in the range 0.5--1.7\,mag; these authors also discussed that  the $Y-J\,\ge\,0.5$\,mag criterion should not exclude any T type objects. Since the {\sl Spitzer} color cuts are based on the typical colors of L and T type sources in the field \citep{patten06}, we increased them generously to account for any possible mid-infrared flux excess present in cluster members. Those candidates from previous steps with non detections in $Y$, [3.6], and [4.5] were kept in the pool. At this stage, the number of potential candidates is reduced to 82. 
\item[-]Finally, we used the optical data set and imposed $I-J\,\ge\,3.0$\,mag. This step left us 65 new potential candidates. Of the 17 rejected objects, 12 have WFC photometry, and five have Suprime-Cam data.
\end{itemize}
After applying these criteria, we successfully recovered the known T dwarfs S\,Ori\,70 (T5.5, \citealt{zapatero02sori70}) and S\,Ori\,73 (T4, \citealt{bihain09,penaramirez11}) in the direction toward the \so cluster. S\,Ori\,70 is detected at both $Y$ and $J$, while S\,Ori\,73 is detected only in the $J$-band \citep{penaramirez11}. We are confident that our extended search should have identified all possible T-type sources in the surveyed area down to  $J$\,=\,21\,mag. However, we imposed color cuts based on those of field L and T dwarfs, which have high gravity atmospheres. If the colors of T-type sources were proven to dramatically depend on surface gravity, the selection photometric criteria should have to be revised.

Of the 65 potential candidates, 35 have visual detections in the $Z$ band. We obtained their PSF photometry and/or photometry based on the peak count ratios. All have $Z$ magnitudes between the completeness and the limiting magnitudes of the survey. Only six sources fulfilled our $Z-J$ color cut, three  of which were resolved (see Table \ref{resolved}) and the other three remain unresolved (star-like FWHM). The unresolved candidates are S\,Ori\,J054017.3$-$023623, S\,Ori\,J053923.3$-$021235 and S\,Ori\,J053716.8$-$024308, and appear as red symbols in Figures \ref{jzj} and \ref{refinement}. We provide their photometry in Table \ref{new}. All their colors (two of them have {\sl Spitzer} photometry) are compatible with the L types. 

These three sources add to the 20 new candidates found in Section \ref{zjsearch}. Therefore, in the VISTA survey we found a total of 210 (=\,207$+$3) \so member candidates with $ZJ$ detections ($J$\,=\,13--20.5\,mag), and 30 with $J$-band detection ($J$ =20.5--21\,mag)  and $Z-J\,\ge 2.1$\,mag (Table \ref{arrows}). As we will discuss in Sections \ref{cont} and \ref{spatial}, the group of 30 suffers from high object contamination, and it will not be included in the study of the cluster mass function. 

Of the 65 potential candidates, 30 have $Z$-band magnitudes beyond the limit of the survey ($Z\,\ge\,23.1$\,mag) and are not previously reported in the literature. They are listed in Table \ref{arrows} and plotted as red arrows in Figure \ref{jzj}. Of the 30, 29 have a positive detection in another filter, and their colors are consistent with the L-types. One source is detected only in the $J$ band. 

Finally, in the surveyed area of 2798.4\,arcmin$^2$ there are only three sources with colors compatible with T spectral types (i.e., significant methane absorption at near-infrared wavelengths), two of which are known (S\,Ori\,70 and 73), and the third one is a new candidate reported here (S\,Ori\,J053804.6$-$021352). In addition, there is a source detected only in the $J$-band; we cannot conclude on its likely nature.

\subsection{Known objects out of the $\bm {ZJ}$ selection}
With some exceptions, our search recovered all previously known \so confirmed members and member candidates with magnitudes in the interval $J$\,=\,16--21\,mag (well within the substellar domain and redder than our $ZJ$ blue envelope). Below we briefly discuss those exceptions; we also include S\,Ori\,70, 72, and 73, which, although found in the extended survey as $Z$-band non-detections, are not considered in the cluster mass function analysis. The photometry of the objects indicated next is given in Table \ref{young} (those with features of youth) and Table \ref{known2} (remaining sources). 
\begin{itemize}
\item[-]S\,Ori\,55 \citep{zapatero00} is an M9-type substellar member of the cluster with very intense and variable H$\alpha$ emission \citep{zapatero02}. It has a VISTA $Z-J$ color slightly bluer than our photometric color cut, and it lies very close to the blue envelope depicted in Figure \ref{jzj}.  We remark that S\,Ori\,55 follows the \so photometric sequence in all other color-magnitude diagrams shown in Figure \ref{refinement}. 
\item[-]S\,Ori\,57 \citep{zapatero00} was rejected as a cluster photometric candidate by \citet{martin01} based on near-infrared photometry. Our VISTA data confirm that this object deviates from the cluster photometric sequence in all of the color-magnitude diagrams of Figures \ref{jzj} and \ref{refinement} with the only exception of the $J$ vs. $J-H$ diagram.
\item[-]S\,Ori\,61 \citep{zapatero00} was found to have an optical spectral type inconsistent with cluster membership by \citet{barrado01}, and it was rejected as a cluster member candidate. With VISTA this object was only detected in the $Z$ band with a large error bar.
\item[-]S\,Ori\,64 \citep{zapatero00} is one of a few low-mass candidates lacking follow-up spectroscopy to assess cluster membership. It was detected in all five VISTA filters; its  $Z-J$ color is slightly bluer than our photometric color cut. 
\item[-]S\,Ori\,65 \citep{zapatero00} was assigned a spectral type of L3.5 by \citet{barrado01}. It follows the \so spectroscopic sequence. More recently, \citet{scholz08} suggested that S\,Ori\,65 has a flux excess at 8\,$\mu$m indicative of a surrounding disk. \citet{luhman08} did not reach the same conclusion arguing that the object photometric precision was insufficient for reliably detecting excess emission. S\,Ori\,65 has a slightly blue VISTA $Z-J$ color, but it nicely sits on top of the cluster photometric sequence in the color-magnitude diagrams of Figure \ref{refinement}. VISTA $J$ magnitude and the data from the literature differ by 0.4\,mag. Significant veiling at optical wavelengths, strong variability and/or underestimated photometric errors of magnitudes provided in the literature might provide an explanation for its blue VISTA $Z-J$ index.
\item[-]S\,Ori\,67 \citep{zapatero00} was typed as an L5 \so source by \citet{barrado01}. It follows the \so spectroscopic sequence. All VISTA colors locate S\,Ori\,67 very close to the photometric sequence of the cluster, but always slightly to the blue side. For this particular source, the VISTA error bars are relatively large.
\item[-]S\,Ori\,69 was the faintest candidate in \citet{zapatero00} survey. \citet{martin01} tentatively classified it as a T0 source based on a rather noisy near-infrared spectrum. \citet{caballero07} did not detect S\,Ori\,69 in their deep survey, suggesting that this object was fainter than indicated by the discovery paper. S\,Ori\,69 remains undetected in all VISTA images in agreement with the findings of these authors.
\item[-]S\,Ori\,70 \citep{zapatero02sori70} is a \so T5.5 isolated planetary-mass candidate (the spectral type was measured from low resolution spectroscopy covering the $H$ and $K$ bands). \citet{burgasser04} argued that this object is probably a foreground field T6--7 dwarf based on $J$-band near-infrared spectroscopy. The VISTA $J$-band photometry of  S\,Ori\,70 is in agreement with the values reported in the literature \citep{zapatero08}.  Its astrometric and photometric properties  have been recently discussed in \citet{scholz08}, \citet{zapatero08} and \citet{penaramirez11}.
\item[-]S\,Ori\,72 \citep{bihain09} is detected only at VISTA $JHK_s$-bands. It was recovered in the extended survey. The VISTA $J$ magnitude differs by 0.56\,mag with respect to that of the discovery paper. 
\item[-]S\,Ori\,73 \citep{bihain09} was confirmed to be a T4 source through the $H$-band methane imaging study by \citet{penaramirez11}. Its high proper motion suggests it is a likely field interloper. Its VISTA $J$ magnitude is 0.33\,mag brighter than the value given by \citet{bihain09} (see \citealt{penaramirez11}).
\item[-]S\,Ori\,74 \citep{bihain09} has VISTA colors significantly deviating from the cluster sequence in all diagrams of Figures \ref{jzj} and \ref{refinement}. VISTA data do not support the photometric candidacy of this object. 
\end{itemize}
From the discussion above, we concluded that our selection criteria are efficient in identifying \so cluster member candidates, and that only $\le$10\%~of true members might have been lost due to variability and/or photometric uncertainties.

\section{Mass estimates for $\bm \sigma$~Orionis member candidates}\label{models}
To estimate the masses of our selected photometric cluster member candidates we compared their observed $J$-band magnitudes to various evolutionary models from the Lyon group: NextGen98 \citep{baraffe98}, AMES-Dusty \citep{chabrier00model}, and AMES-Cond \citep{baraffe03}. We did not employ the absolute magnitudes provided by the models since it is known that they do not reproduce the observed spectra of cool sources convincingly. We converted theoretical luminosities and effective temperatures into observables by using the bolometric correction--temperature and color--temperature relations given in \citet{golimowski04} and \citet{hewett06}. The models in combination with the transformation equations cover the mass range $\sim$0.25--0.001\,$M_{\odot}$. The results are displayed in the various panels of Figure \ref{refinement} for the cluster age of 3\,Myr and distance of 352\,pc. Estimated masses for this canonical cluster age and distance are labeled in all color-magnitude diagrams of Figures \ref{jzj} and \ref{refinement} and will be used throughout this paper.

According to the Lyon models, our VISTA survey is complete in the mass interval 0.25--0.006\,$M_\odot$, which corresponds to $J$\,=\,13--20\,mag (or completeness given by the $Z$ band). The extended survey ($J$\,=\,21\,mag) goes down to 0.003\,$M_{\odot}$. The substellar limit is located at $J$\,=\,14.5\,mag ($\sim$0.072\,$M_{\odot}$) and the planet-brown dwarf borderline lies at $J$\,=\,18.1\,mag (13\, $M_{\rm Jup}$). Within these magnitude intervals there are 104 very low mass stars, 69 brown dwarfs, and 37 planetary-mass objects in our VISTA \so survey. Of the 37 planetary-mass candidates, 28 have magnitudes within the completeness limit of the exploration. We remark that the VISTA survey has doubled the number of \so planetary-mass candidates known to date. For a cluster distance of 300\,pc and age of 3\,Myr, our survey would be complete in the mass range 0.20-0.005\,$M_\odot$ (91 very low mass stars, 80 brown dwarfs, and 39 planetary mass objects). For the largest cluster distance of 450\,pc, the mass coverage of the VISTA exploration would shift toward higher masses particulary in the stellar regime; the survey would be complete for the interval 0.38-0.007\,$M_\odot$ (128 very low mass stars, 53 brown dwarfs, and 29 objects with planetary masses). 

\section{Contaminants}\label{cont}
In a pure photometric search without spectroscopic measurements to confirm the young nature of the candidates, in addition to possible errors in the source catalog magnitudes, the sample of cluster member candidates may be contaminated by objects displaying colors similar to those expected for members. Nevertheless, 129 out of the total of 210 candidates, i.e., 61\%, have confirmed membership in \so (see Section \ref{data}). This indeed provides secure statistics for the study of the cluster mass function. We will discuss briefly what kind of objects are expected to be contaminating a survey like ours and will try to quantify each of them as follows.

The main sources of contaminants are Galactic M, L, and T type field objects (at all magnitudes) and reddened, unresolved galaxies (particularly dominating at faint magnitudes). As shown in \citet{bihain09}, L dwarfs and red galaxies share similar near- and mid-infrared colors. However, because of their relatively blue near-infrared colors, T-type sources can be distinguished from reddened galaxies.  We derived the Galactic contribution of contaminants following the study by \citet[and references therein]{caballero08c}. In the magnitude range $J$\,=\,13--20.5\,mag (the $ZJ$ search), we estimated the contamination due to M, L, and T sources to be 31 objects. These are divided as follows: 12 M4--M6 stars in the stellar domain ($J$\,=\,13--14.5\,mag, 11.5\%), 11 M6--L0 sources in the brown dwarf regime ($J$\,=\,14.5--18.1\,mag, 15.9\%), and 7 L0--L7 plus 1 T dwarfs in the planetary-mass interval ($J$\,=\,18.1--20.5\,mag, 21.6\%). Regarding the extended survey, for the magnitude range $J$\,=\,20.5--21\,mag, the number of expected contaminants is an additional 1 L dwarf and 1 T dwarf. 

Extragalactic contamination is expected to be negligible at magnitudes brighter than $J,H$=18--19\,mag since most galaxies might have been resolved in our survey given the pixel size of the detectors and seeing conditions. We checked the FWHM of all new candidates and confirmed that they fit the stellar PSF. Resolved sources (Table \ref{resolved}) were actually rejected and considered non-photometric member candidates. Also negligible is the contamination due to background cool (M-type) giant stars all across the magnitude range of the VISTA survey. \citet{kirkpatrick94} argued that these objects are relatively numerous in areas within 10\,deg of the Galactic plane. The \so cluster is located well away from this region (at $b$\,=\,$-17.3$\,deg). 

As for the extragalactic contamination at faint magnitudes, particularly reddened galaxies and galaxies at high redshifts,  we used the GOODS--MUSIC multicolor V2.0 catalog of galaxies \citep{grazian06,santini09}, which has multi--wavelength coverage from 0.3 to 24\,$\mu$m and either spectroscopic or accurate photometric redshifts. This catalog extends beyond the limiting magnitudes of the VISTA survey, but it covers a smaller area on the sky by a factor of 1/19.5. The GOODS--MUSIC magnitudes were converted into the Vega system using the relations by \citet{hewett06} and the IRAC Data Handbook V3.0. For the magnitude interval $J$\,=\,13--20.5\,mag, we retrieved all GOOD--MUSIC sources that have $Z-J$ colors to the red side of the blue envelope depicted in Figure \ref{jzj}. We found none. The situation changes dramatically for $J$\,=\,20.5--21\,mag. With color cuts of $Z-J\,\ge\,2.1$ and $Y-J\,\ge\,0.6$\,mag, we retrieved a total of five galaxies in the GOODS--MUSIC area. These galaxies have redshifts of 1.3--1.6. By applying the additional color cuts of $J-[3.6]\,\le\,3.5$ and $J-[4.5]\,\le\,4.5$\,mag that we imposed to our extended search in the \so cluster, the number of galaxies reduces to three. All have colors ``consistent" with the field L type dwarfs, and none resembles the T dwarfs. By scaling the area of the GOODS--MUSIC and VISTA surveys, we estimated that 35--90 galaxies with redshifts $z$\,=\,1.3--1.6  may be contaminating our extended survey in the magnitude range $J$\,=\,20.5--21\,mag. This number is quite similar and even larger than the amount of \so candidates with $Z$-band non-detections found in the extended survey, suggesting that most of the 30 sources of Table \ref{arrows} could be extragalactic.

The extragalactic nature of two of these candidates is confirmed by inspecting the deep near-infrared images of \citet{penaramirez11}, particularly those with an excellent seeing of 0\farcs3. These authors covered an area of 120\,arcmin$^2$ in the \so~cluster using the $H$ and methane filters. The two candidates overlapping with that area are resolved according to the photometric analysis made by the authors. Because of the high level of extragalactic contamination expected in the magnitude bin $J$\,=\,20.5--21\,mag of our VISTA search, we shall not consider it in the mass function study.

\section{Infrared flux excesses: disks}\label{disksec}
\begin{figure*}
\epsscale{0.7}
\plotone{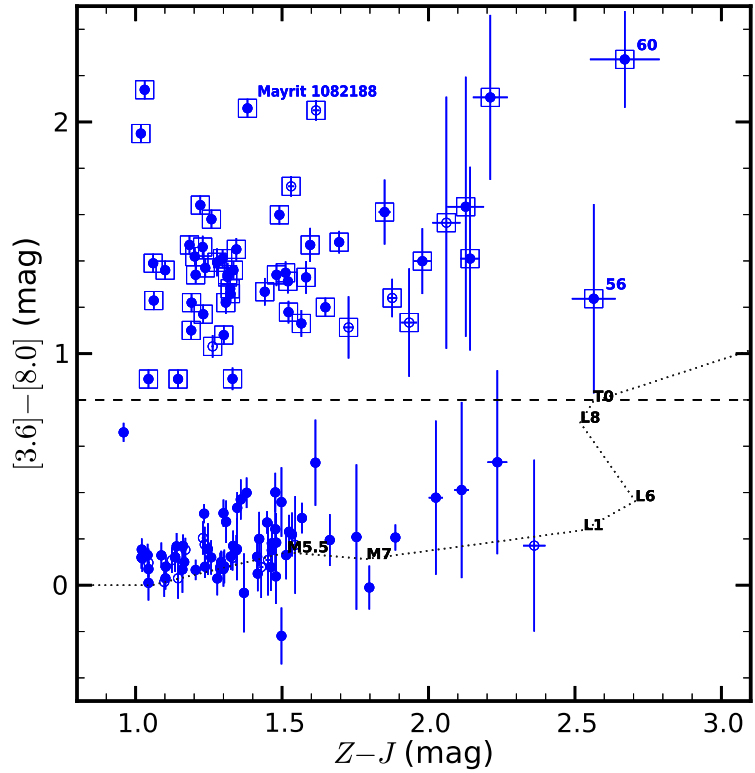}
\caption{VISTA and IRAC/\textit{Spitzer} $[3.6]-[8.0]$ vs. $Z-J$ color-color diagram of all VISTA sources with data available in the four filters. Symbols are as in Figure \ref{jzj}. Objects with significant flux excesses at 8\,$\mu$m (framed circles) lie above the horizontal dashed line defined as the separator between objects with and without surrounding disks. The average location of early-M to T0 field dwarfs is indicated with the dotted line. Spectral types are labeled. Sources without infrared flux excesses nicely follow the field sequence.\label{disk80}}
\end{figure*}

As largely discussed in the literature, infrared flux excesses of young objects are typically attributed to the presence of surrounding disks or envelopes. Figure \ref{disk80} displays the color--color diagram [3.6]--[8.0] vs. $Z-J$, which gathers 128 \so members (of the 210 selected candidates) with positive detections in IRAC/{\sl Spitzer} [3.6] and [8.0] bands. This Figure illustrates the color difference between sources with no infrared flux excesses ([3.6]--[8.0]$\,\lesssim\,0.5$\,mag) and objects that possibly host a circum(sub)stellar disk ([3.6]--[8.0]$\,\gtrsim\,0.8$\,mag, \citep{luhman06,zapatero07}. The field dwarf sequence is also plotted in Figure~\ref{disk80}; it nicely overlaps with the sequence of cluster objects with no {\sl Spitzer} excesses. The field sequence roughly traces an horizontal line from blue $Z-J$ colors to the late-M/early-L types, i.e., the [3.6]--[8.0] color is nearly constant for the color interval $Z-J$\,=\,1.0--2.5\,mag. Therefore, the distinction between objects with and without infrared excesses based on a constant color cut is valid for this color range. Two L-type sources with disks reported by \citet{zapatero00}, S\,Ori\,56 and S\,Ori\,60, have [3.6]--[8.0] colors redder than the field dwarf sequence, in agreement with the literature.  

Of the 128 sources detected in IRAC/{\sl Spitzer} [3.6] and [8.0] bands, 54 have [3.6]--[8.0]$\,\gtrsim\,0.8$\,mag (S\,Ori\,56 and S\,Ori\,60 included). We identified them as objects that probably harbor a disk. This yields a global fraction of objects with infrared excesses at 8.0\,$\mu$m of 42\,$\pm$\,6\% (Poisson uncertainty). In the \so low mass stellar regime ($J\sim$13--14.5\,mag), 31 sources out of 75 present infrared excesses at 8.0\,$\mu$m, translating into a stellar disk fraction of 41\,$\pm$\,8\%. Similarly, for cluster brown dwarfs ($J\sim$14.5--18.1\,mag), the frequency of infrared flux excesses at 8.0\,$\mu$m turns out to be 39\,$\pm$\,9\% (19 sources out of 49).  Only four planetary mass objects were detected at 3.6 and 8.0\,$\mu$m, all of them presenting flux excesses at the longest wavelength. 
 
Only two objects with disks reported in the literature do not have a color [3.6]--[8.0]$\,\gtrsim\,0.8$\,mag: S\,Ori\,7 and S\,Ori\,8. The former was studied by \citet{oliveira06}, who measured a flux excess in the $L\sp{\prime}$ band, and by \citet{hernandez07}, who classified this source as diskless based on its IRAC photometry. S\,Ori\,8 was analyzed by \citet{scholz04}. These authors found a low amplitude photometric variability, which is better explained by the existence of photospheric spots rather than by a circumstellar disk. Later, \citet{hernandez07} determined that this object may have an ``evolved disk" according to its IRAC photometry.

Of the known photometric candidates of $\sigma$~Orionis, we report here the detection of flux excesses at 8\,$\mu$m for Mayrit\,1082188 \citep{caballero08b}, see Table \ref{infoknown}. Furthermore, this source is detected at 22.0\,$\mu$m in the {\sl WISE} catalog, supporting the existence of significant flux excesses at long wavelengths.

\begin{figure*}
\epsscale{0.7}
\plotone{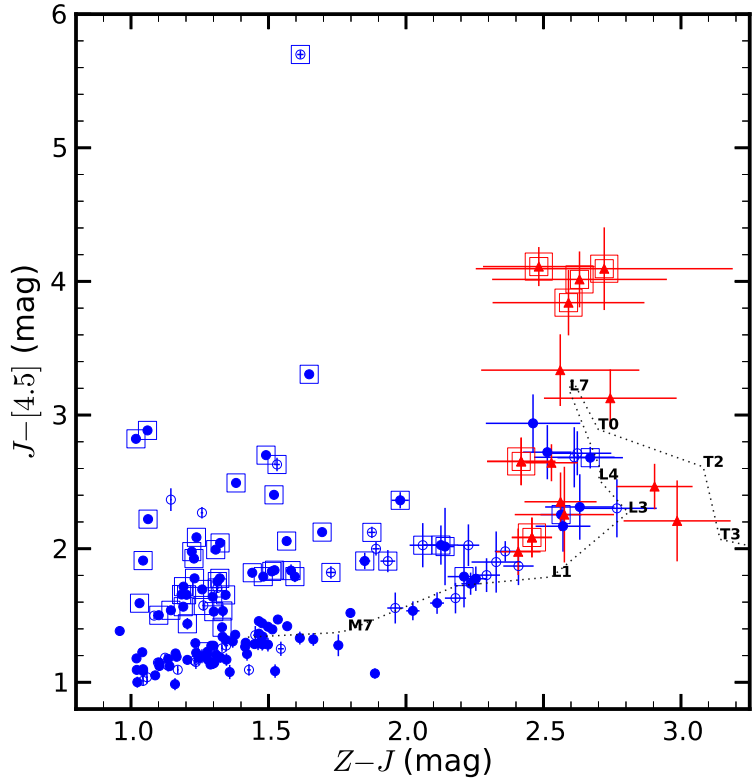}
\caption{VISTA and IRAC/\textit{Spitzer} $J-[4.5]$ vs. $Z-J$ color-color diagram of all VISTA sources with data available in the four filters. Symbols are as in Figure \ref{jzj}.  Known \so objects with flux excesses at 8\,$\mu$m have framed circles. The field sequence (early-M to early-T types) is indicated with a dotted line, and spectral types are labeled. New planetary mass candidates with $J-[4.5]$ colors indicative of flux excesses at 4.5\,$\mu$m are plotted as double framed red triangles. A color version of this figure is available in the online journal.\label{disk45}}
\end{figure*}

Because a significant number of \so sources in the planetary mass regime lacks detection at 8.0\,$\mu$m, we extended the study of infrared flux excesses to a shorter wavelength where {\sl Spitzer} was more sensitive: the IRAC/{\sl Spitzer} 4.5\,$\mu$m band. Figure \ref{disk45} depicts the $J-[4.5]$ vs. $Z-J$ color-color diagram where all sources with positive detections in all three bands are plotted. Objects with clear excesses at 8.0\,$\mu$m, as deduced from their [3.6]--[8.0] colors, typically have $J-[4.5]$ indices 0.5\,mag redder than the field dwarf sequence, which is more than twice their photometric uncertainties. Therefore, we applied the following criterion for claiming an infrared flux excess at 4.5\,$\mu$m: for a given $Z-J$ color (or spectral type), any source with a $J-[4.5]$ color redder than the field sequence by more than 2-$\sigma$ its photometric error bars may harbor a surrounding disk. According to this criterion, we derived that the global fraction of \so candidates with infrared excesses at 4.5\,$\mu$m is $\ge$\,38\,$\pm$\,5\%. As for the different mass intervals, disk frequency at 4.5\,$\mu$m is as follows: 42\,$\pm$\,7\%~(very low-mass stars), 36\,$\pm$\,8\%~(brown dwarfs), and $\ge$31\,$\pm$\,11\%~(planetary mass objects). We note that $\sim$L2--T0 objects have similar $Z-J$ color but differing $J-[4.5]$ indices (see Figure \ref{disk45}). Because of this ambiguity, without a proper spectral classification of the cluster planetary-mass objects we cannot properly assess the presence of infrared flux excesses at 4.5 $\mu$m for some of these objects based on our defined criterion. Therefore, the derived disk fraction for \so planetary-mass objects is a lower limit. 

\begin{figure*}
\epsscale{0.7}
\plotone{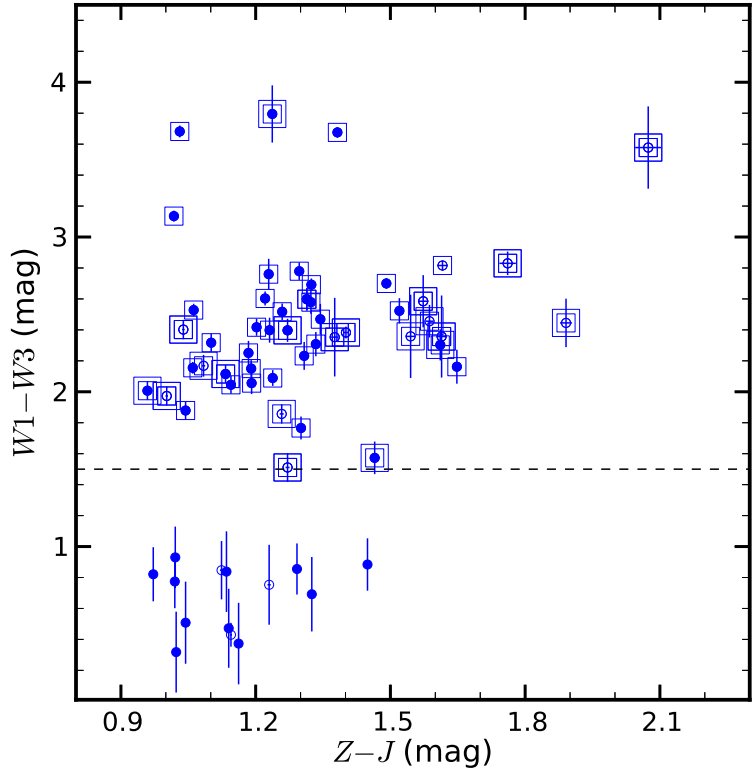}
\caption{VISTA and \textit{WISE} $Z-J$ vs. $W1-W3$ color-color diagram of all VISTA sources with data available in the four filters. Symbols are as in Figure \ref{jzj}. Sources with known IRAC/\textit{Spitzer} flux excesses at 8\,$\mu$m are plotted as framed circles. All of them lie above the dashed horizontal line, which separates objects with and without flux excesses at 12\,$\mu$m. The new \textit{WISE} detections of \so candidates that possibly harbor disks are shown as double framed cicles.\label{disk12}}
\end{figure*}

We also investigated the presence of infrared flux excesses in the \so candidates using {\sl WISE} data. Figure \ref{disk12} shows the $W1-W3$ vs. $Z-J$ color-color diagram, including a total of 65 objects with positive detections in all four bands. Sources with confirmed {\sl Spitzer} flux excesses are characterized by  having colors $W1-W3\,>\,1.5$\,mag, while objects that presumably do not have surrounding disks show $W1-W3\,<\,1$\,mag. Of the 65 {\sl WISE} detections, 51 have a large $W1-W3$ color ($W1-W3\,>\,1.5$\,mag, i.e., infrared flux excess at 12\,$\mu$m). {\sl WISE} data are biased to the detection of \so low-mass objects with infrared flux excesses. {\sl WISE} performed a shallow exploration of the sky in the $W3$ band; therefore, we expect a great majority of the \so low-mass sources without flux excesses to remain undetected at $W3$ because their purely photospheric emission lies well beyond the completeness limit of the {\sl WISE} all-sky survey. In the high mass part of the stellar regime of our VISTA survey, all cluster stars with {\sl Spitzer} flux excesses at 8.0\,$\mu$m also have excesses at 12\,$\mu$m from the {\sl WISE} data (see Figure~\ref{disk12}). Additionally, we found 20 cluster member candidates with apparent flux excesses at 12\,$\mu$m reported here for the first time (see Tables \ref{infoyoung} and \ref{infoknown}): 11 lie out of the area covered by {\sl Spitzer}, 5 do not have detections in the {\sl Spitzer} [3.6] and [8.0] bands (and were not included in our analysis above), and another 4 do not show any flux excesses at 8\,$\mu$m.

We caution that the disk fractions discussed above represent a lower value of the ``true" disk incidence in \so since, as stated in Section \ref{cont}, we expect some object contamination in our VISTA survey. Contaminants are likely diskless, because the majority of them are expected to be old late-type field objects. In addition, cold (evolved) disks and/or disks with significant interior gaps are not detected at short wavelengths ($\le$12\,$\mu$m). By using IRAC/{\sl Spitzer} and {\sl WISE} we are missing these cold disks, particularly at the lowest masses. It is expected that the derived disk frequency increases with wavelength (as an example, consider the four stars, [KJN2005]8, [SWW2004]J053926.768$-$024258.25, S\,Ori\,J053821.3$-$023336, and S\,Ori\,J053853.8$-$024459, with clear flux excesses at 12\,$\mu$m and no evidence for excesses at 8\,$\mu$m). In addition, \citet{zapatero07} pointed out that \so planetary-mass object inner disk excess emission becomes clearly detectable longward of 5 $\mu$m, e.g., S\,Ori\,60 exhibits an excess only at 8\,$\mu$m, while its emission at shorter wavelengths is mainly photospheric. 

At 4.5 and 8.0\,$\mu$m, the derived disk fractions (i.e., infrared flux excesses) for \so very low-mass stars, brown dwarfs, and planetary mass objects are quite alike, about 40\%, within error bars. This relatively high disk incidence suggests that the effect of evaporating inner disks through external ionization from massive O-type stars is not playing a major role in \so low-mass members. This fraction also agrees with the derivations available in the literature \citep{hernandez07,caballero07,zapatero07,scholz08,luhman08,bejar11}, indicating that inner/warm disks typically last for at least a few million years \citep{haisch01} for both low-mass stellar and substellar regimes.


\section{Spatial distribution}\label{spatial}
\begin{figure*}
\epsscale{0.7}
\plotone{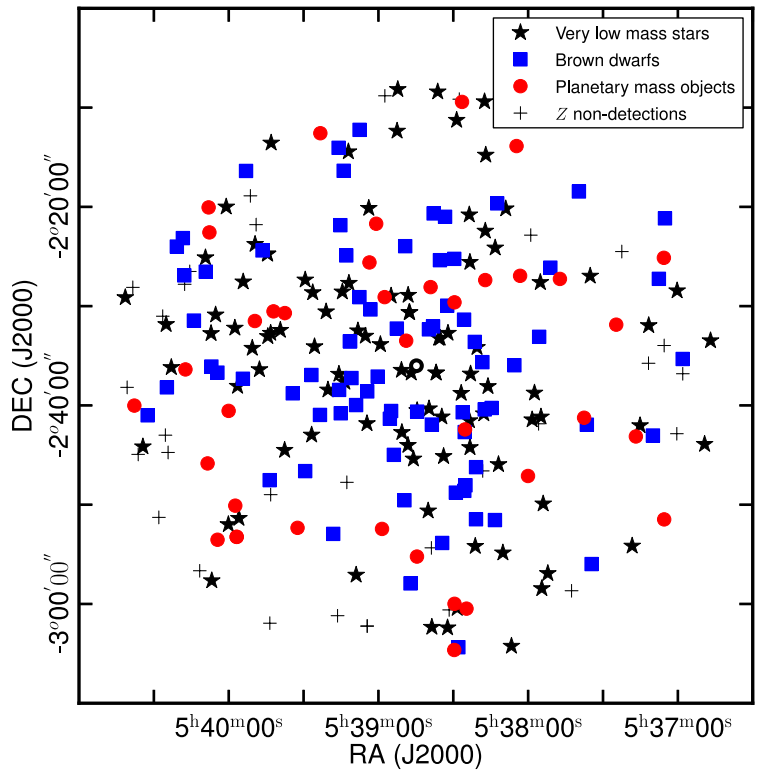}
\caption{Spatial distribution of all 210 \so member candidates and 30 $Z$-band non-detections found in the VISTA survey. The multiple, massive star $\sigma$ Ori is plotted as a central open circle. A color version of this figure is available in the online journal.
\label{spatialmag}}
\end{figure*}

\begin{figure*}
\centering
\includegraphics[angle=90,scale=0.5]{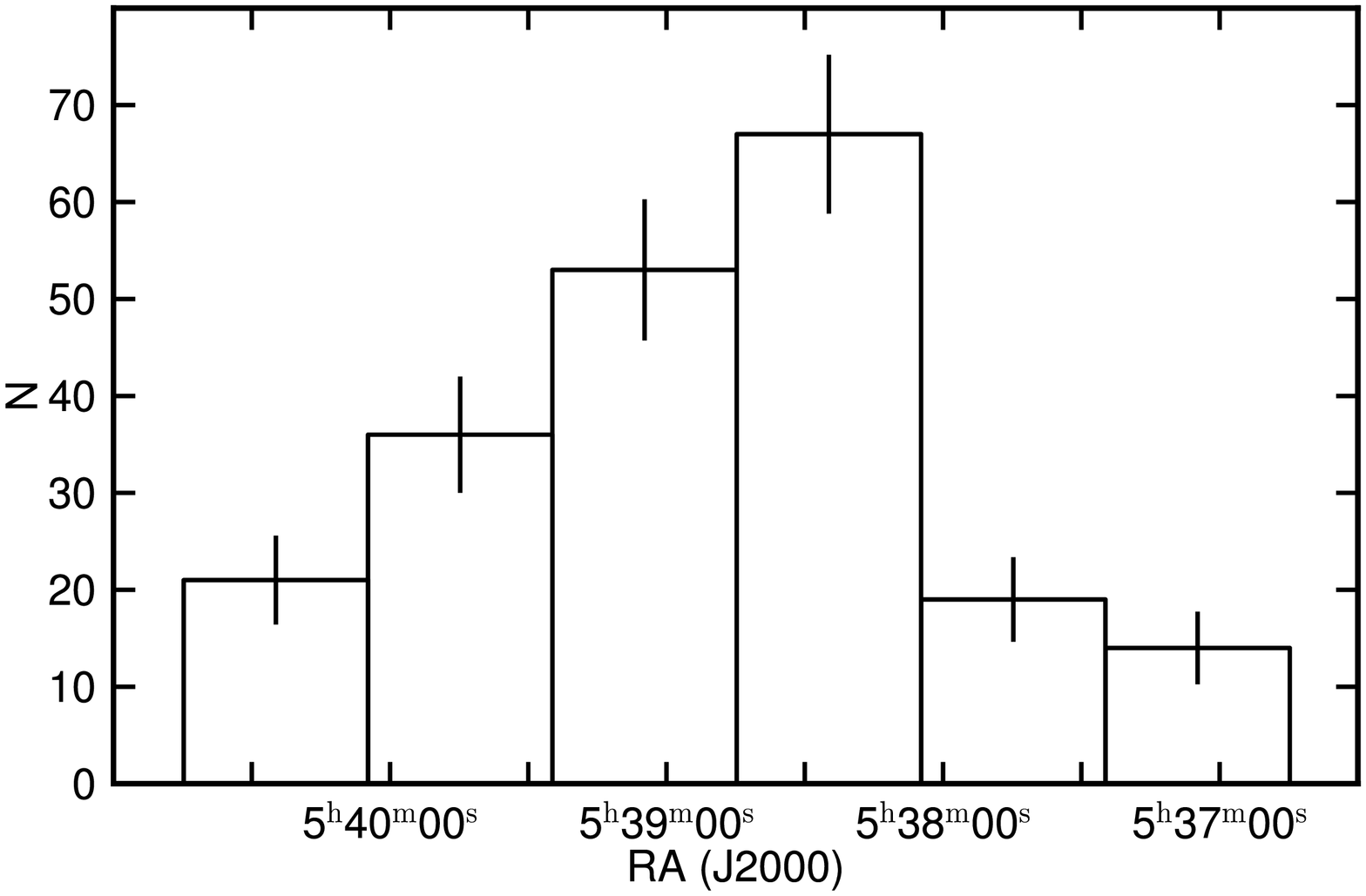}\\
\vspace{1cm}
\includegraphics[angle=90,scale=0.5]{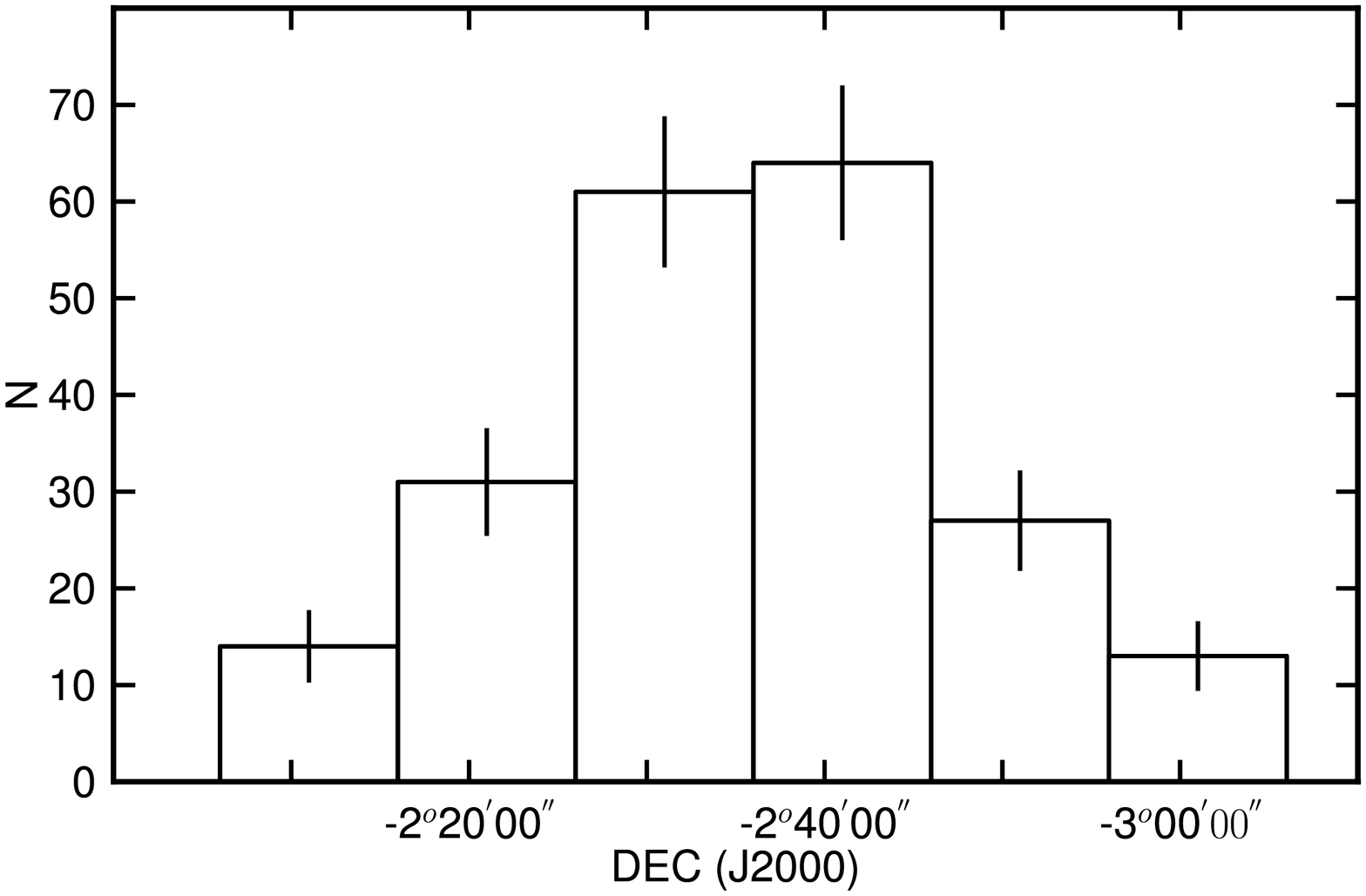}
\caption{Distribution of the 210 VISTA sources in the right ascension (RA) and declination (DEC) coordinates. The size of the bins is 10\arcmin. All panels are centered at the location of the multiple, massive star $\sigma$ Ori (plotted as a cross symbol in Figure \ref{spatialmag}.) \label{radec}}
\end{figure*}

The right ascension (RA) and declination (DEC) spatial distribution of \so sources is illustrated in Figure \ref{spatialmag}; we considered the following magnitudes intervals: very low mass stars ($J$\,=\,13--14.5\,mag), brown dwarfs ($J$\,=\,14.5--18.1\,mag), planetary mass objects ($J\,>\,18.1$\,mag), and $Z$-band non-detection candidates ($J > 20.5$\,mag). Figure \ref{radec} displays the histograms of the objects spatial distribution for the RA and DEC coordinates separately. The largest concentration of cluster member candidates lies very close to the projected location of the massive, multiple star $\sigma$ Ori\,AB. The DEC histogram is quite symmetric at both North and South directions with respect to the position of this star. From the RA histogram, it can be seen that the number of \so sources decreases smoothly toward the East, while the distribution drops sharply toward the West. \citet{caballero08} found a similar result and explained it as  an azimuthal asymmetry due to a filament-shape overdensity of objects connecting the cluster center with a part of the Horsehead Nebula. Nevertheless, the number of objects toward the West and East of the cluster center does not differ by more than 10\%, and the number of objects in each RA, DEC quadrant is coincident within 20\%, i.e., the Poisson errors. 

With respect to the $Z$-band non-detected candidates (Section \ref{arrowssec}), $\sim$50\%~of them are located in the outer 25\arcmin--30\arcmin~annulus, and none was identified in the 10\arcmin-radius central area. These objects are not included in the histogram plots of Figure \ref{radec}.

\subsection{Radial surface density profiles}
We studied the radial surface density profiles of \so sources of different mass within the circular area of radius = 30\arcmin. To extend the profiles to further distances from the cluster center, we also explored the surrounding regions covered by the VISTA survey (only tile 16, other tiles to the North of the cluster are not as deep as the observations of $\sigma$~Orionis). Within tile 16 (Figure \ref{survey}), we avoided the regions East of the cluster because of the high extinction associated to the Horsehead Nebula. We selected the area enclosed between 30\arcmin~and 50\arcmin~West of the \so cluster (1090\,arcmin$^2$). After applying the $ZJ$ photometric criteria described in Section \ref{zjsearch}, a total of 14 sources were identified with $J$ magnitudes in the interval 13--20.5\,mag (see Table \ref{out}). Based on {\sl WISE} 12\,$\mu$m data, three of them show infrared flux excesses. We also checked that all these sources follow the cluster sequence in the VISTA color magnitude diagrams of Figure \ref{refinement}, i.e., the 14 objects successfully pass all the photometric criteria for being considered as cluster member candidates, suggesting that the cluster area may extend up to or even beyond a radius of 50\arcmin~from the center. 

\begin{figure*}
\epsscale{0.7}
\plotone{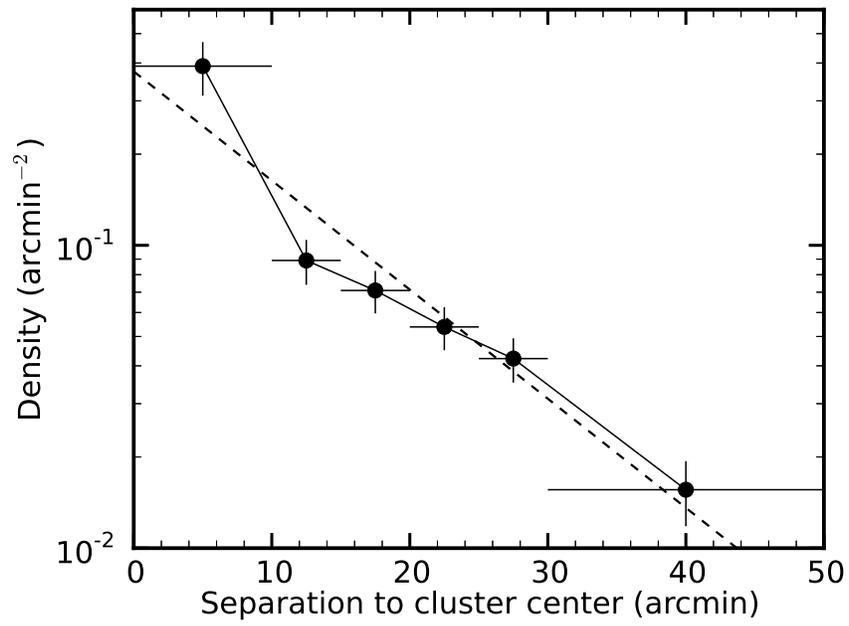}
\caption{The radial surface density profile of all 210 \so member candidates is shown with dots and solid line. The profile extends up to 50\arcmin~from the cluster center. The exponential fit is depicted with a dashed line. \label{all}}
\end{figure*}

The resulting radial surface density profile of the 210 \so member candidates is displayed in Figure \ref{all}. In general, there is a large object concentration in the central 10\arcmin~and a smooth decay by more than one order of magnitude toward the periphery. Previous studies \citep{bejar04,bejar11,caballero08,lodieu09} have found distributions similar to ours. The cluster radial surface density profile can be reproduced by an exponential law $\rho=\rho_{0}\, e^{-\frac{r}{r_0}}$ \citep{bergh60}. We found a central density $\rho_0=0.4$\,arcmin$^{-2}$ ($\rho_0\sim$36\,pc$^{-2}$) and a characteristic radius $r_0 \sim$12\arcmin~(1.2\,pc). A similar result was reported by \citet{bejar11}, who found $r_0 \sim$1.0\,pc. 

\begin{figure*}
\epsscale{0.7}
\plotone{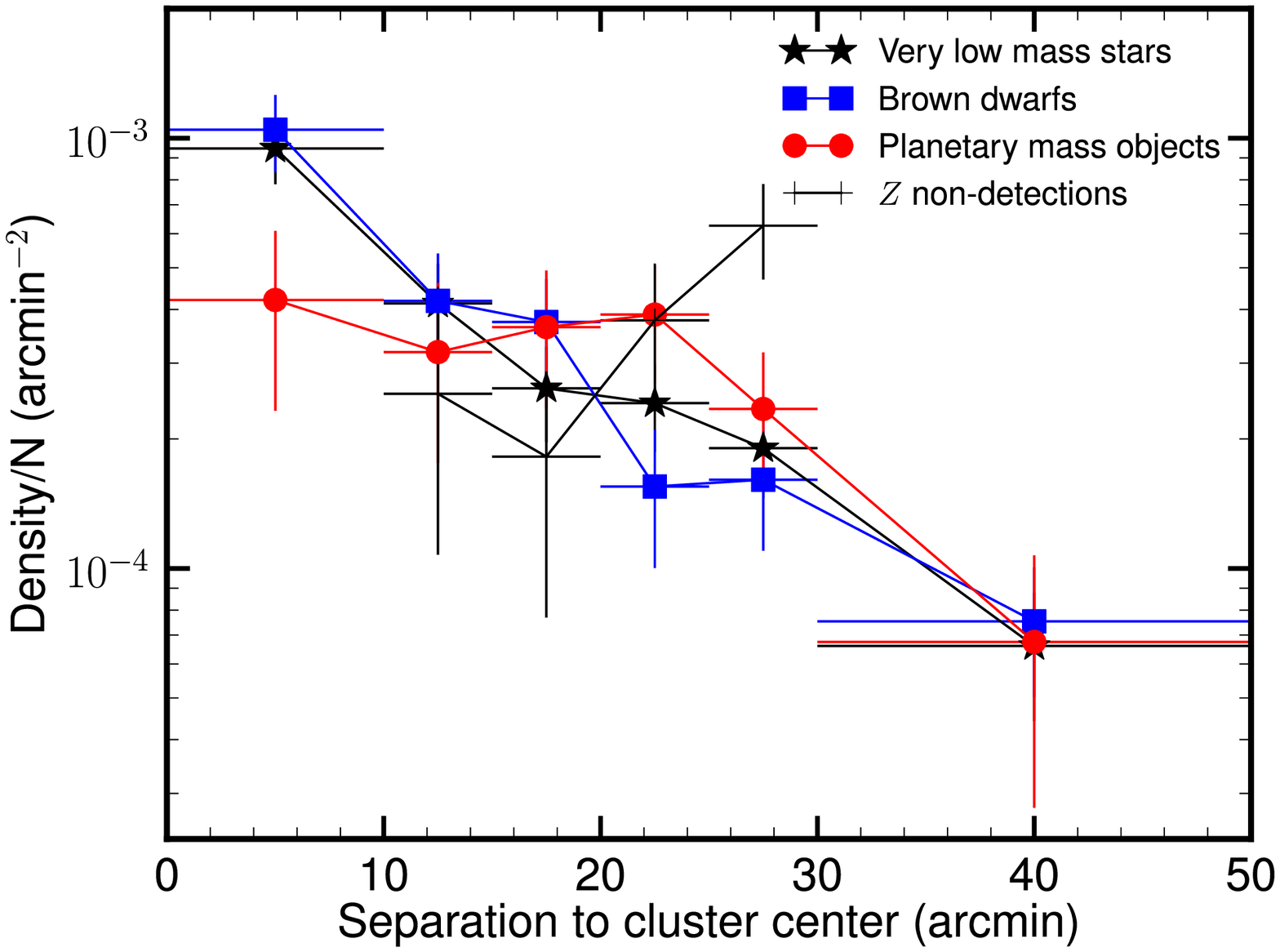}
\plotone{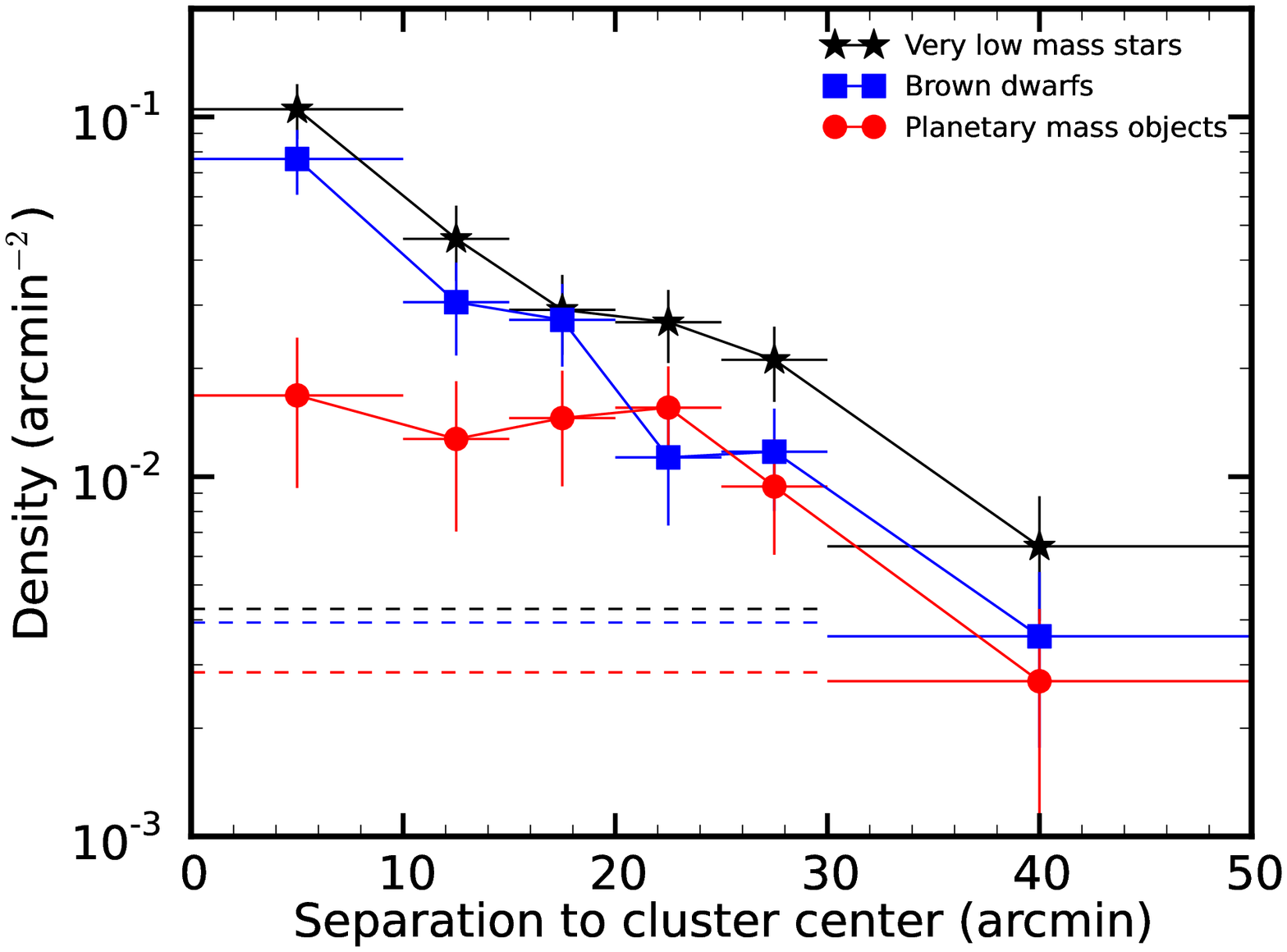}
\caption{\textit{(Upper panel)} Radial surface density profiles of the VISTA \so candidates per mass (or magnitude) interval. The profiles extend up to a separation of 50\arcmin~from the cluster center. All profiles are normalized to the total number of objects in each mass interval. The surface densities have been measured for coronal regions of the following sizes: central circle of radius = 10\arcmin, from here up to 30\arcmin~the coronal step is 5\arcmin, and the outer region goes from 30\arcmin~to 50\arcmin~(see text). The vertical error bars correspond to Poisson uncertainties while the horizontal bars account for the size of their corresponding coronal regions.  \textit{(Lower panel)} Same as previous panel but displaying the surface densities as measured (without any normalization factor). The field dwarf contamination level for each mass (magnitude) interval is plotted as dashed lines (color codified). The color version of this figure is available in the online journal.
 \label{radialprofiles}}
\end{figure*}

The radial surface density profiles for different mass intervals (stars, $J$\,=\,13--14.5\,mag; brown dwarfs, $J$\,=\,14.5--18.1\,mag; and planetary-mass objects, $J\,>\,18.1$\,mag) up to 50\arcmin~from the cluster center are shown in Figure \ref{radialprofiles}. The upper panel displays the profiles normalized to the total number of objects in each magnitude interval, and the lower panel depicts the measured densities without any normalization. Radial density profiles were normalized to the total number of sources rather than to the number of sources in the central cluster region to avoid the large uncertainties introduced by small numbers. In the very central region of the cluster ($\le$5\arcmin), we took into account the area lost (about 21\,\%) because of the presence of numerous very bright stars preventing us from detecting planetary-mass candidates, and  we thus applied an area correction only to the faintest magnitude bins. 

The radial distributions of very low mass stars and brown dwarfs do not differ significantly. The radial profile of the planetary-mass object density might deviate from those of very low-mass stars and brown dwarfs at the level of 1.5-$\sigma$, since it appears to be flatter in the central 20\arcmin. Nevertheless, at farther separations from the cluster center there is an obvious decrease in the surface planetary-mass object density. This feature can be interpreted in favor of the cluster membership of the majority of the planetary mass candidates found in our study, because if they were merely photometric contaminants they would present a flat radial profile independently of the separation from the cluster center, as expected from an homogeneous density of contaminants. Nevertheless, we argue that the observed differences between the radial profiles of very low-mass stars, brown dwarfs, and planetary-mass objects are not statistically significant, and that all three radial profiles are consistent with the hypothesis of distributions with a similar origin (after applying the Kolmogorov-Smirnov test\footnote{In a Kolmogorov-Smirnov test the P-value is an indicator of the amount of evidence against the hypothesis of distributions with the same origin. The smaller the P-value, the stronger the evidence. The reference value is 0.05 typically.}, we obtained $P > 0.93$ for the comparison of the very low-mass stars--planetary-mass objects samples and the brown dwarfs--planetary-mass objects samples). Furthermore, \citet{bouy09} studied the core of the \so cluster using adaptive optics and found a relatively large number of planetary-mass candidates; if confirmed to be true \so members, they would increase the central density of the cluster planetary mass population.  This contrasts with the underabundance of substellar objects in the central cluster region reported by \citet{caballero07b,caballero08} and \citet{lodieu09}. In the context of a dynamically evolved system, one would expect the least massive objects to be located far away from the cluster center, but this does not appear to be the case in $\sigma$~Orionis, thus confirming that \so is younger than its relaxation time.

The radial surface density profile of the $Z$-band non-detected sources (Section \ref{arrowssec}) is displayed in the upper panel of Figure \ref{radialprofiles}. In contrast to the very low-mass stars and brown dwarfs of $\sigma$~Orionis, it shows a nearly flat distribution or a moderate density increase at separations larger than 20\arcmin~from the cluster center. This feature along with the previously discussed spatial distribution may be ascribed to a faint, very low-mass \so population inhabiting the peripheral regions of the cluster or, more likely, to a population of field contaminants (also see Section \ref{cont}).

According to previous studies, over 80\% of confirmed cluster members lie within a circular area of radius\,=\,30\arcmin\, \citep{bejar04,bejar11,caballero08,lodieu09}. Using the VISTA data, we estimate that $\sim$65 potential cluster members following the cluster photometric sequence in all possible filters could be located at radial separations between 30\arcmin~and 50\arcmin~from the cluster center (14 candidates were already identified to the West of the cluster uncovering about 1/4.6 of the total coronal area). We thus estimate that the circular survey of radius\,=\,30\arcmin~accounts for more than 75\% of the \so low-mass population. Actually, this fraction is expected to be significantly larger because the number of field contaminants increases at a high rate at far separations from the \so center as compared to true  cluster members.

By taking the expected number of field dwarf contaminants per mass interval computed in Section \ref{cont}, we derived the surface density of dwarf contamination in our survey. It is plotted in the bottom panel of Figure \ref{radialprofiles}. Up to 30\arcmin--35\arcmin, the cluster surface densities are higher than the level of expected contamination for all mass intervals (very low-mass stars, brown dwarfs, and planetary-mass objects). At larger separations from the cluster center, only the stellar bin appears overnumerous with respect to the contamination level, while the substellar mass intervals reach nearly the same value as the expected contamination, indicating that (i) large observational efforts are required to distinguish true cluster members from contaminants and, (ii) any pure photometric statistical study at distances beyond $\sim$35\arcmin~from the cluster center has little significance. Therefore, we shall restrict the \so mass function analysis of the next section to the area within a radius of 30\arcmin.

\subsection{Objects with and without disks}
\begin{figure*}
\epsscale{0.7}
\plotone{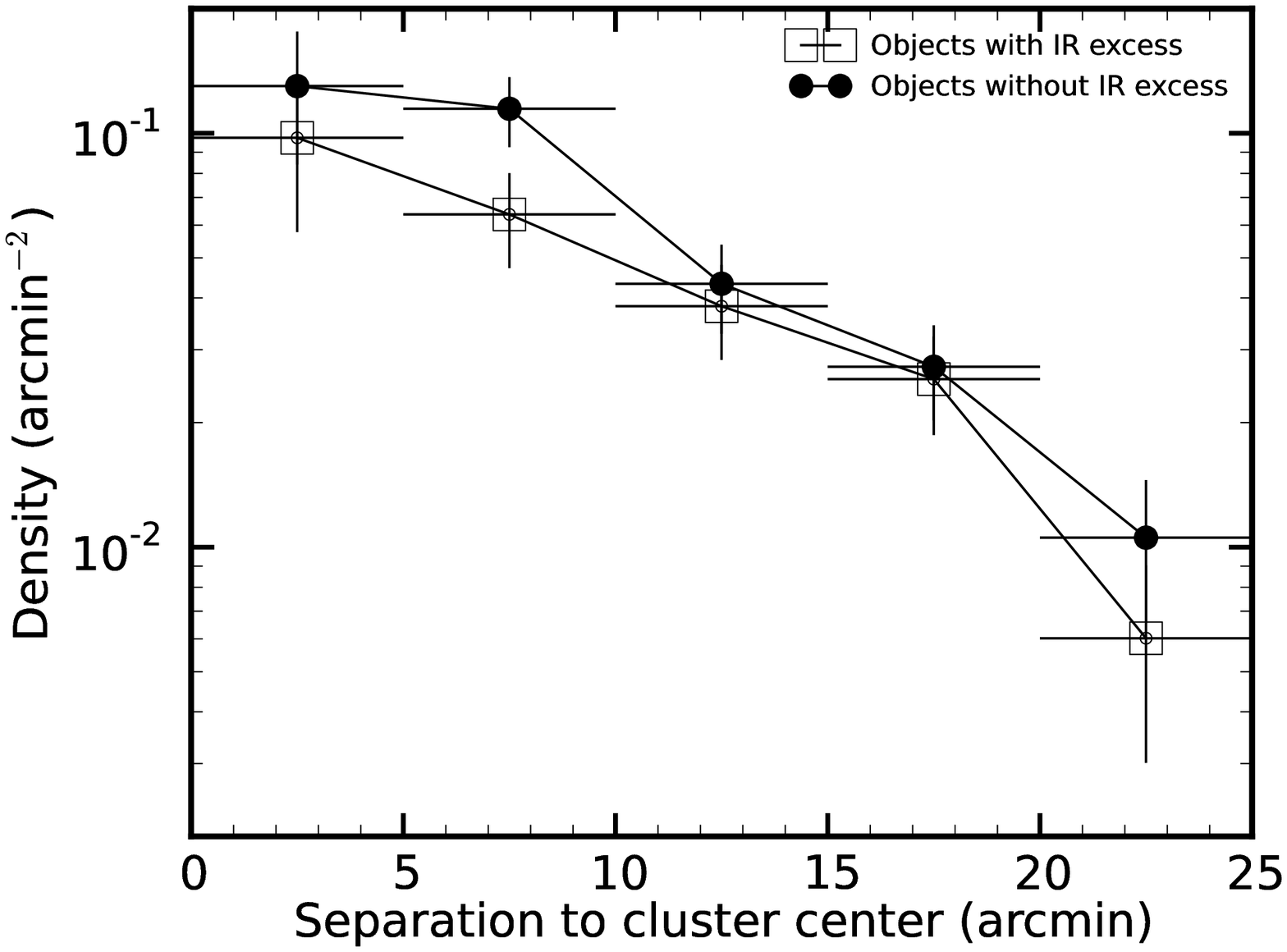}
\plotone{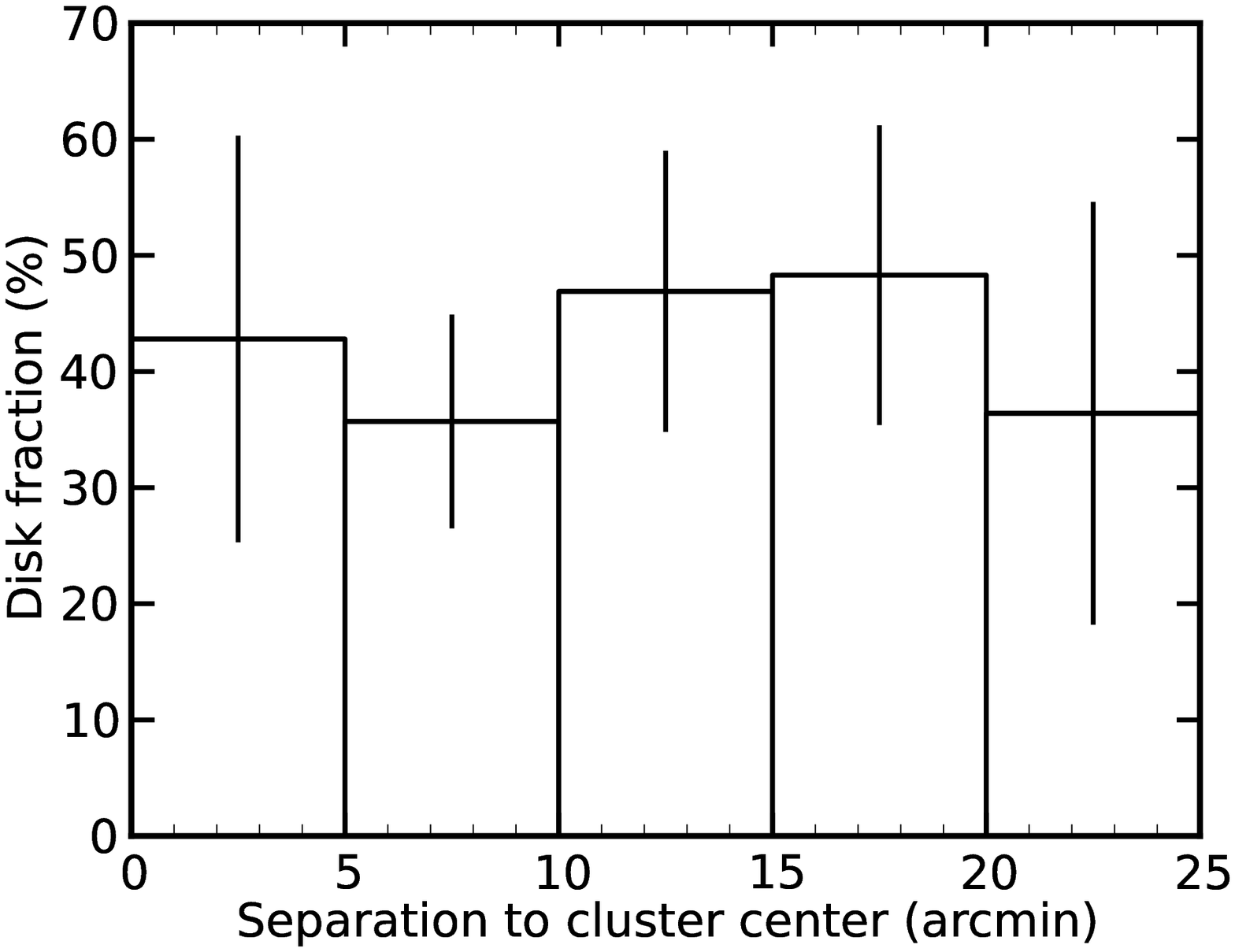}
\caption{\textit{(Upper panel)} Radial surface density profiles of \so member candidates with (squares) and without (dots) infrared flux excesses at 8\,$\mu$m. The coronal regions have a width of 5\arcmin~from the cluster center to a radial separation of 25\arcmin. Vertical error bars represent Poisson uncertainties, and horizontal bars account for the coronal size. \textit{(Lower panel)} Disk fractions based on flux excesses at 8\,$\mu$m determined at different separations from the cluster center. \label{spatialdisk}}
\end{figure*}

Radial surface density profiles of \so candidates with and without flux excesses at 8.0\,$\mu$m are shown in Figure \ref{spatialdisk}. Here, we do not distinguish different mass intervals. We counted objects inside concentric annuli of 5\arcmin~in size from the cluster center up to a separation of 25\arcmin. The lower panel of Figure \ref{spatialdisk} depicts our derivations of the disk/infrared flux excesses fractions at different separations from the cluster center. From both panels it can be seen that there is no obvious difference in the distribution or disk fraction of both \so populations; the disk fraction remains nearly constant at the 1-$\sigma$ level all across the explored cluster area. A similar property is observed when considering flux excesses at 4.5\,$\mu$m. This result is in agreement with the findings by \citet{oliveira06}, who concluded that there is no spatial segregation for low-mass objects with and without circum(sub)stellar disks.

\section{The \so mass function }\label{fmsec}
Our study allowed us to derive a reliable, comprehensive mass function for the \so cluster. First, the VISTA survey homogeneously covers a large extension of the cluster area with detection completeness in the mass interval 0.25--0.006\,$M_{\odot}$ (at the age of 3\,Myr and distance of 352\,pc). Second, we showed that our analysis of a circled area of radius of 30\arcmin~around the $\sigma$ Ori AB star uncovered $>$75\%~of all cluster members, thus giving statistical weight to the mass function derivation. Third, a large fraction of our VISTA candidates have spectral and/or photometric confirmation of their young age, and thus, they are likely \so members. As stated in Section \ref{search}, 129 objects out of the total of 210 candidates were confirmed to be young in the literature. Here, we confirmed the presence of infrared flux excesses for an additional amount of 21 objects (15 stars and 6 substellar sources). Therefore, our mass function derivation using the VISTA data relies on a sample of 210 objects, 71\%~of which have a confirmed young age fully compatible with cluster membership.

\subsection{Substellar mass spectrum}\label{fmsub}
\begin{figure*}
\plotone{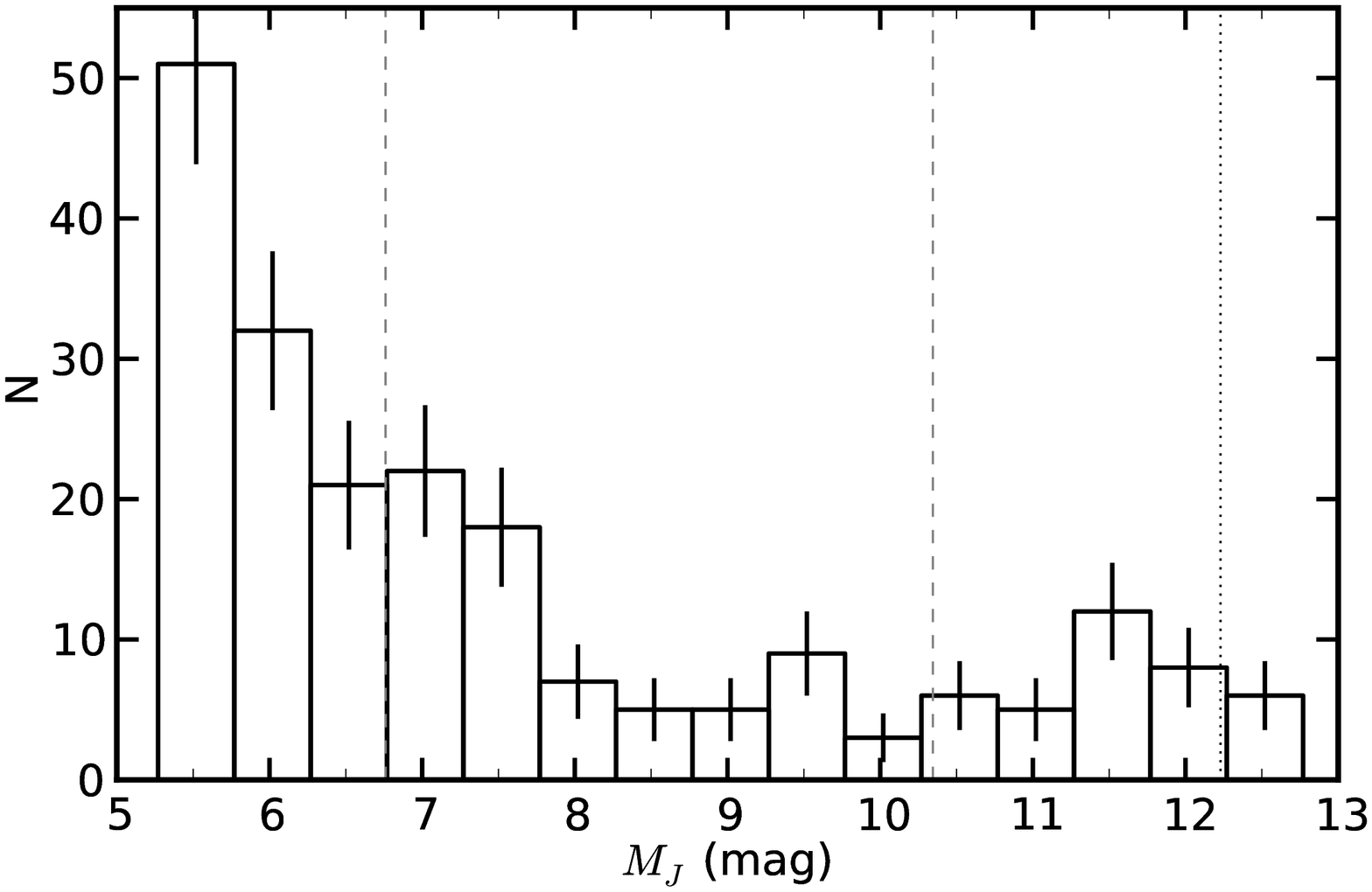}
\plotone{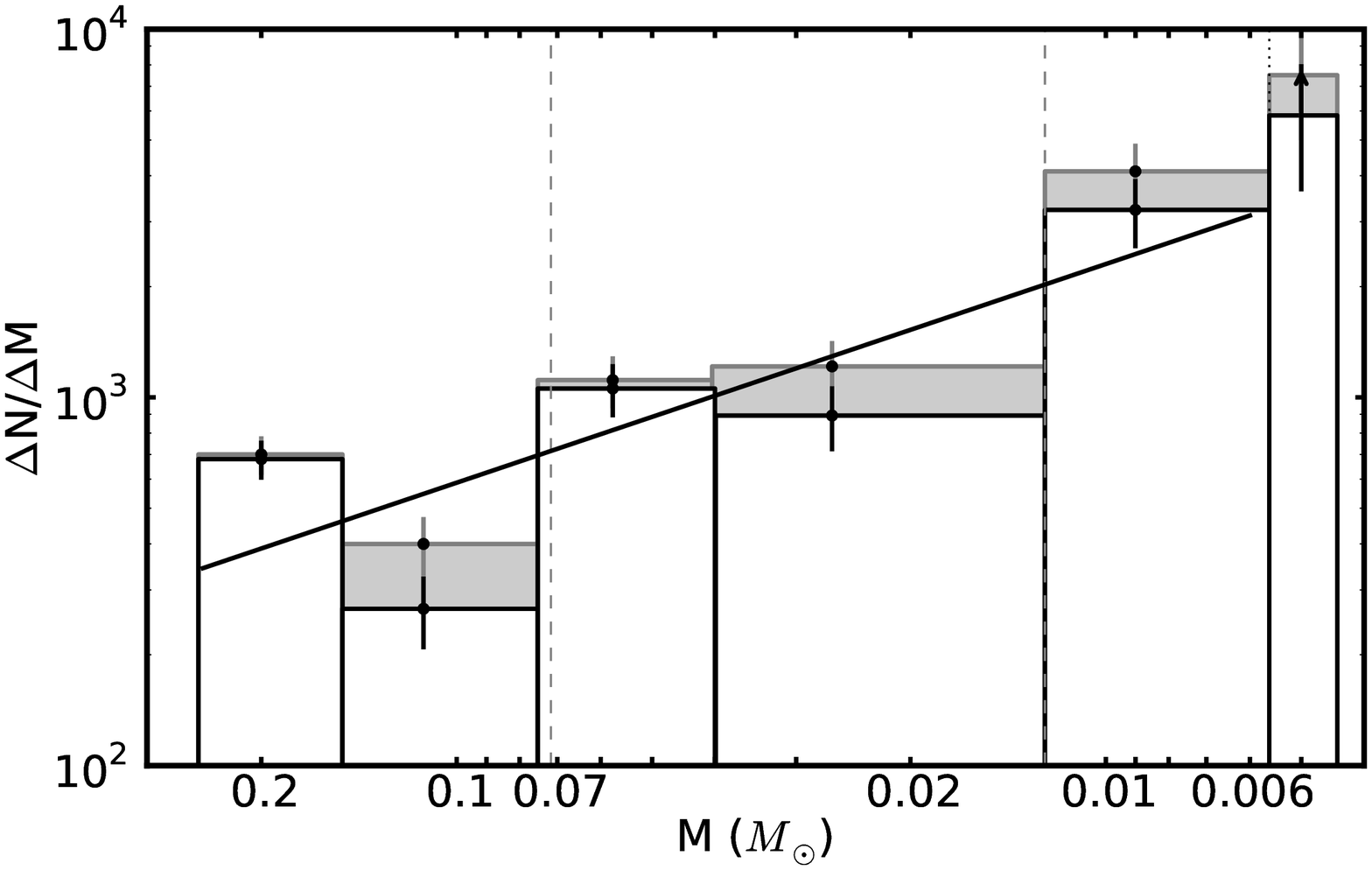}
\caption{\textit{(Upper panel)} The $J$-band luminosity function obtained from the 210 VISTA \so candidates. From left to right, the vertical dashed lines and the dotted line denote the location of the substellar limit, the brown dwarf--planet borderline, and the completeness limit of the VISTA $ZJ$ survey. Distance modulus is $\sim$7.73\,mag for a cluster distance of 352\,pc. \textit{(Lower panel)} The \so low-mass spectrum ($\Delta N/\Delta M$) derived for the mass interval 0.25--0.004\,$M_{\odot}$. The mass spectrum free of field contaminants is plotted as the open histogram, while the mass spectrum as measured is indicated by the grey upward extension of the mass bins. Vertical error bars stand for the Poisson count statistics. Vertical lines as in the upper panel of this Figure. The best power-law fit to the decontaminated mass spectrum is shown with a continuos line ($\Delta N/\Delta M \sim M^{-0.6\pm0.2}$, obtained for the complete mass interval 0.25--0.006\,$M_{\odot}$). \label{fm}}
\end{figure*}

Using the VISTA $ZJ$ selected cluster candidates in the circular area of radius\,=\,30\arcmin~around the cluster center, we built the \so luminosity function (number of objects per $J$-band magnitude interval) depicted in Figure \ref{fm}. In the range of absolute magnitudes $M_J$\,=\,5--8\,mag, the cluster luminosity function steadily declines, while it remains nearly constant from $M_J$\,=\,8 through 13\,mag. The marked drop at $M_J$\,$\sim8$\,mag corresponds to spectral types mid- to late-M, a feature attributed to the appearance of dust formation in cool atmospheres \citep{dobbie02}. Nevertheless, there is no definitive observational or theoretical explanation for it. 

Figure \ref{fm} also displays the \so low-mass stellar and substellar-mass spectrum (i.e., number of objects per mass interval, $\Delta N/\Delta M$) obtained from the cluster luminosity function and the 3\,Myr theoretical mass-luminosity relations indicated in Section \ref{models}. The stellar, brown dwarf, and planetary-mass regimes were divided into two bins each. We corrected the cluster mass spectrum from contamination by field dwarfs as described in Section \ref{cont}. In the following, we shall consider only the decontaminated mass spectrum. The ``VISTA"  mass spectrum is complete in the mass interval 0.25--0.006\,$M_\odot$, where completeness is driven by the $Z$-band observations.

The mass spectrum smoothly increases toward smaller mass. There is a continuous transition from the low-mass stellar domain to the brown dwarf regime, and from here to the planetary-mass objects with no apparent change in the slope of the mass spectrum. It can be fit by a power-law function $\Delta N/\Delta M \propto M^{-\alpha}$, for which we measured an index $\alpha\,=\,0.6\pm0.2$ in the mass interval  0.25--0.006\,$M_{\odot}$. This result is in agreement with the derivations found in the literature (\citealt{caballero11} and references therein). For example, the first substellar mass spectrum obtained for the mass range 0.2--0.013\,$M_{\odot}$ and an area of 847\,arcmin$^2$ by \citet{bejar01} was fit with $\alpha\,=\,0.8\pm0.4$. \citet{caballero07} measured $\alpha\,=\,0.6\pm0.2$ for the mass interval 0.11--0.006\,$M_{\odot}$ and an area of 790\,arcmin$^2$. The similarity of our measurement and that of \citet{caballero07} is remarkable since both working mass intervals are alike, but our explored area is 3.5 times larger. The recent works by \citet{lodieu09} and \citet{bejar11} covered an area similar to our VISTA survey, and they found $\alpha\,=\,0.5\pm0.2$ (0.5--0.01\,$M_{\odot}$) and $\alpha\,=\,0.7\pm0.3$ (0.1--0.013\,$M_{\odot}$), respectively.  As discussed in \citet{bejar11}, the various brown dwarf mass spectrum studies carried out in different star-forming regions and associations with ages  below 100--200\,Myr have yielded rising functions with typical power-law indices between 0.4 and 1.0, e.g., $\rho$ Ophiuchi \citep{luhman00}, IC 348 \citep{najita00,luhman03}, the Trapezium \citep{lucas00,hillenbrand00,luhman00}, Chamaeleon \citep{lopez04,luhman07}, $\lambda$ Orionis \citep{barrado04}, Upper Scorpius \citep{lodieu07}, $\alpha$ Persei \citep{barrado02b}, Blanco 1 \citep{moraux07}, and the Pleiades cluster \citep{moraux03,bihain06,lodieu07}. 

The extrapolation of the \so mass spectrum between the $Z$-band completeness and the $Z$-band detection limit (or  $J$\,=\,19.95--20.5\,mag) yields about five cluster planetary-mass objects in the range 0.006--0.004\,$M_{\odot}$ with expected spectral types from mid- to late-L based on their predicted surface temperatures at the age of 3\,Myr. This number estimate is consistent with our measurement free of contaminants (7$\pm$2 objects) at the 1-$\sigma$ level (Figure \ref{fm}). The \so mass function appears to extend from high mass stars down to planetary-mass objects with 0.004\,$M_{\odot}$. For smaller masses and the magnitude interval $J$\,=\,20.5--21\,mag (completeness magnitude of the $J$-band data), and based on the theoretical 3\,Myr model and the normalized sequence of field L and T dwarfs depicted in Figures \ref{jzj} and \ref{refinement}, the VISTA $J$-band data are sensitive to \so sources warmer than about T6. In our study we identified 3--4 T-type objects: the previously known S\,Ori\,70 (T5.5, \citealt{zapatero02sori70}) and S\,Ori\,73 (T4,  \citealt{bihain09,penaramirez11}), a new object S\,Ori\,J053804.65$-$021352.5 (see Table \ref{new}) with optical and infrared colors consistent with an ultracool atmosphere, and one source detected only in the $J$-band (see Table \ref{arrows} and Section \ref{arrowssec}). The cluster membership of the two former objects was recently discussed by \citet{penaramirez11}, who found that both T dwarfs have proper motions higher ($>$ 4-$\sigma$) than that of the \so cluster. The extrapolation of the \so mass spectrum down to 0.003\,$M_{\odot}$ or $J$\,=\,21\,mag yields a population of T sources that clearly outnumbers our findings. Furthermore, the expected contamination of $\sim$2 field T-type objects (Section \ref{cont}) contributes to increase the observed discrepancy. There are two possible explanations to account for this difference: (i) the evolutionary models may make wrong predictions particularly at these low masses and/or our conversions from theoretical parameters to observables based on field calibrations are not valid at very low masses and young ages; (ii) and/or the \so mass spectrum (or mass function) may have a turn over (and possibly a mass cut-off) below 0.004\,$M_{\odot}$ or $J$\,=\,20.5\,mag (in agreement with \citealt{bihain09}). Deeper images are required for a robust claim on the detection of a low-mass turn over or cut-off for the \so mass function, which has large impact in the theory of substellar formation processes. If none of the 3--4 T-type objects found in the VISTA \so data turned out to be a cluster member, we estimate the field T dwarf density at factors of 1.5--7.5 higher than the density measurements available in the literature \citep{burgasser07,metchev08,reyle10}.

\subsection{From high mass stars to planets}\label{total}
The mass spectrum presented in the previous section is focused on the low mass stellar and substellar domains of $\sigma$~Orionis. To determine the cluster mass function covering a wider mass range we need to expand our study to more massive cluster members. The Mayrit catalog \citep{caballero08b} presents a compilation of 338 \so star and brown dwarf member candidates, 241 (71\%) of which have known features of youth. Interestingly, the area covered by the Mayrit catalog fully overlaps with our VISTA explored region. Therefore, both the Mayrit catalog and our VISTA survey were merged to build the complete \so mass function from the most massive stars ($\sim$19\,$M_{\odot}$) to the planetary-mass objects of 0.006\,$M_{\odot}$. Our VISTA data and the Mayrit catalog have a common magnitude interval where both surveys are complete: $J$\,=\,13--14.6\,mag, where the number of detections is quite similar for the two catalogs. For $J < 13$\,mag, there are 189 Mayrit sources, which we have added to our 210 VISTA objects. Of the 189 Mayrit sources,  85 (45$\pm$6\%) have infrared flux excesses at 8.0\,$\mu$m, and more than the 70\%~show spectroscopic features of youth or X-ray emission. For both samples, the number of confirmed young objects and likely \so members is similar (71\%) and statistically significant, rendering our cluster mass function reliable.

\begin{figure*}
\epsscale{0.7}
\plotone{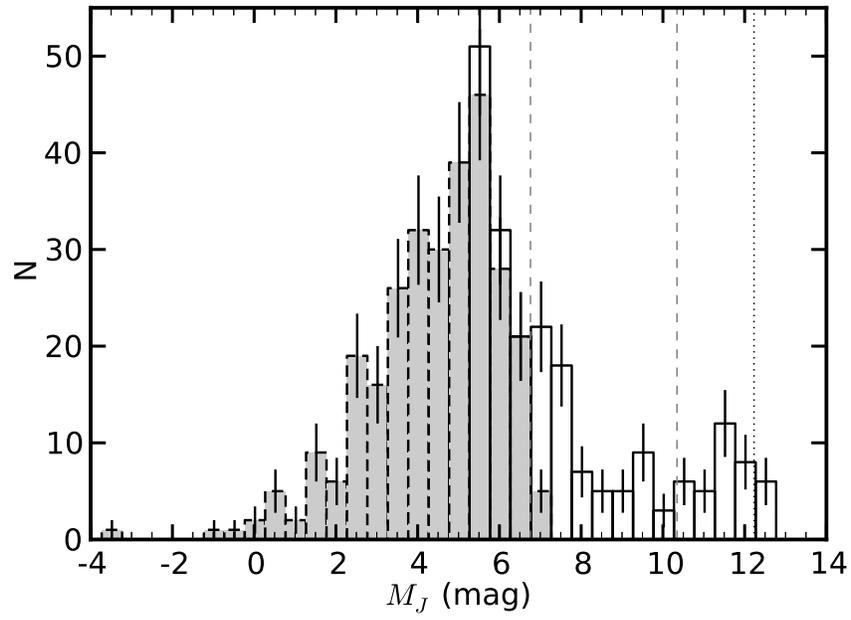}
\caption{Combined Mayrit (grey dashed histogram) and VISTA (open histogram) \so $J$-band luminosity function. From left to right, vertical lines as in Figure \ref{fm}. Vertical error bars stand for the Poisson count statistics. The faintest bin of the Mayrit catalog remains largely incomplete. \label{lum}}
 \end{figure*}

The combined Mayrit and VISTA $J$-band luminosity function is shown in Figure \ref{lum}. The observed differences between the Mayrit catalog and the VISTA data in the overlapping magnitude range $J$\,=\,13--14.6\,mag ($M_J$\,=\,5.3--6.9\,mag) are within the Poisson errors. This adds support to the combination of both catalogs. 

\begin{figure*}
\epsscale{0.8}
\plotone{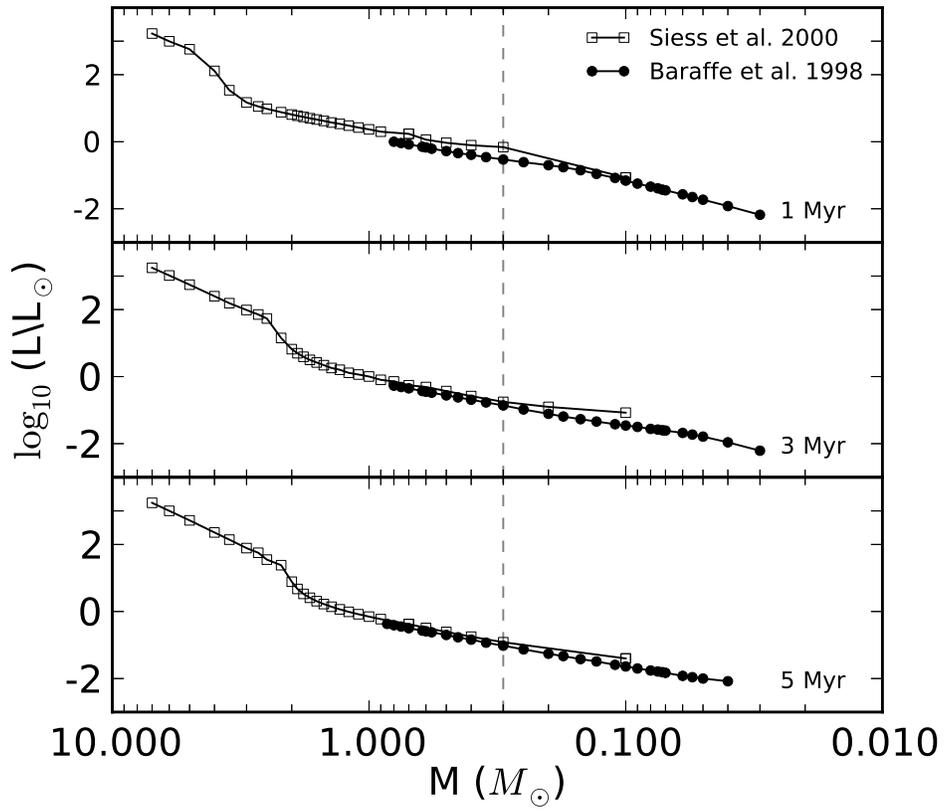}
\caption{Theoretical mass--luminosity relations for 1, 3, and 5\,Myr given by \citet{siess00} and \citet{baraffe98}. Solar metallicity is adopted. The vertical dashed line lies at 0.3\,$M_\odot$.  \label{comp}}
 \end{figure*}

To build the mass function we used the stellar mass-luminosity relationships corresponding to the 3\,Myr, solar metallicity isochrones from \citet{siess00} and \citet{baraffe98} (NextGen model). Theoretical luminosities and effective temperatures were transformed to observable magnitudes and colors using the relations given in \citet{worthey11}. As illustrated in Figure \ref{comp}, the mass--luminosity relationships used are quite similar for a wide range of masses, and only differ significantly between $\sim$0.1 and 0.3\,$M_{\odot}$. Actually, this difference becomes larger for a broader mass interval at ages younger than 3\,Myr, and both models converge toward older ages (see Figure \ref{comp}), indicating that the choice of the zero age for the theoretical computations may account for the discrepancies. For the age of 3\,Myr, the mass derivations obtained from the NextGen and \citet{siess00} models do not differ by more than 20\%. The masses of the most luminous $\sigma$ Ori stars  ($\sigma$ Ori Aa, Ab, B, D and E ) were taken from \citet{simon11}. These authors reported the detection of a third massive star component in the \so AB system, traditionally considered to be a binary system. \citet{simon11} employed a distance of 385\,pc to the cluster. We checked that by adopting the distance of 352\,pc the most massive bin is not changed in our study.

\begin{figure*}
\epsscale{0.65}
\plotone{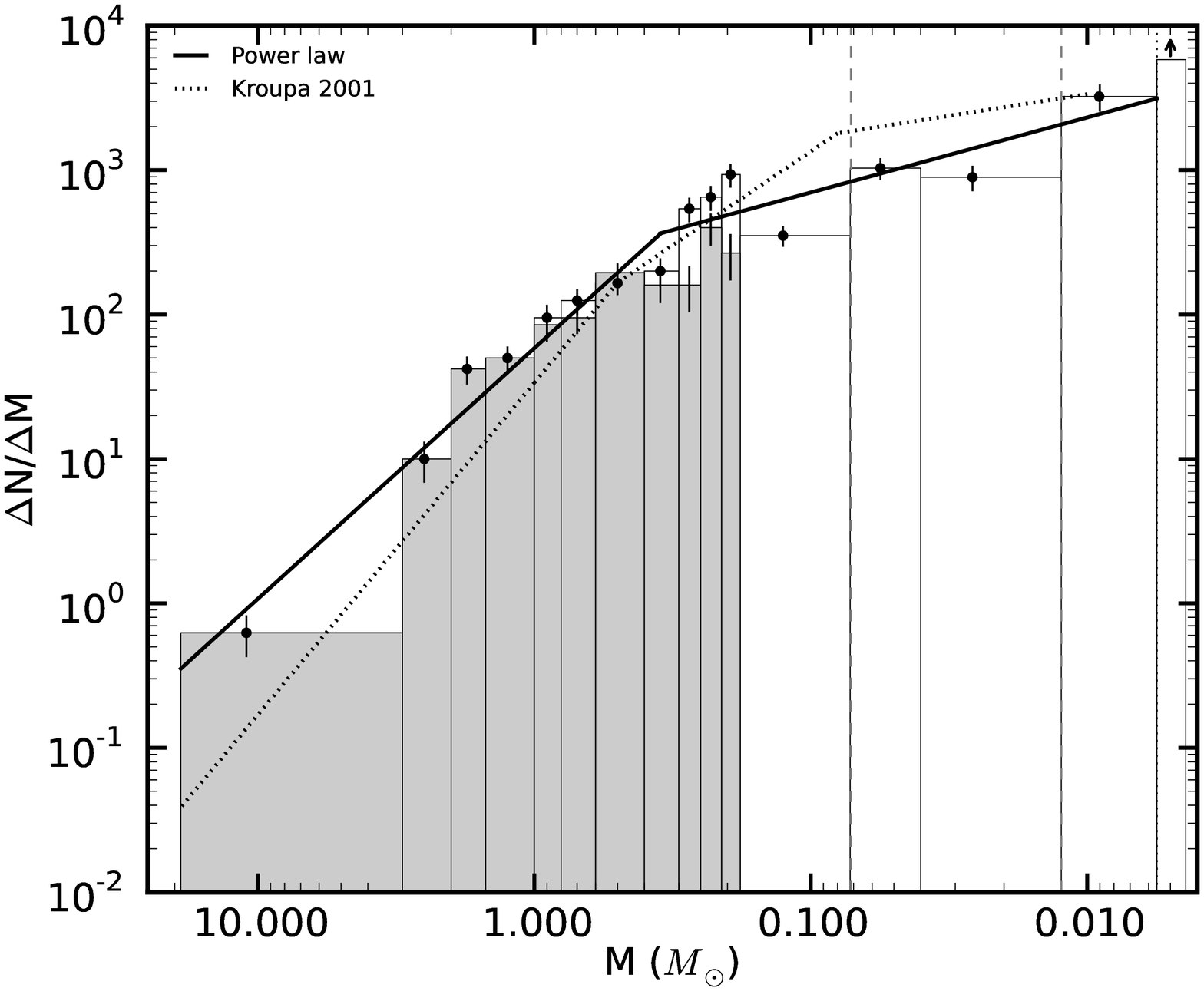}
\plotone{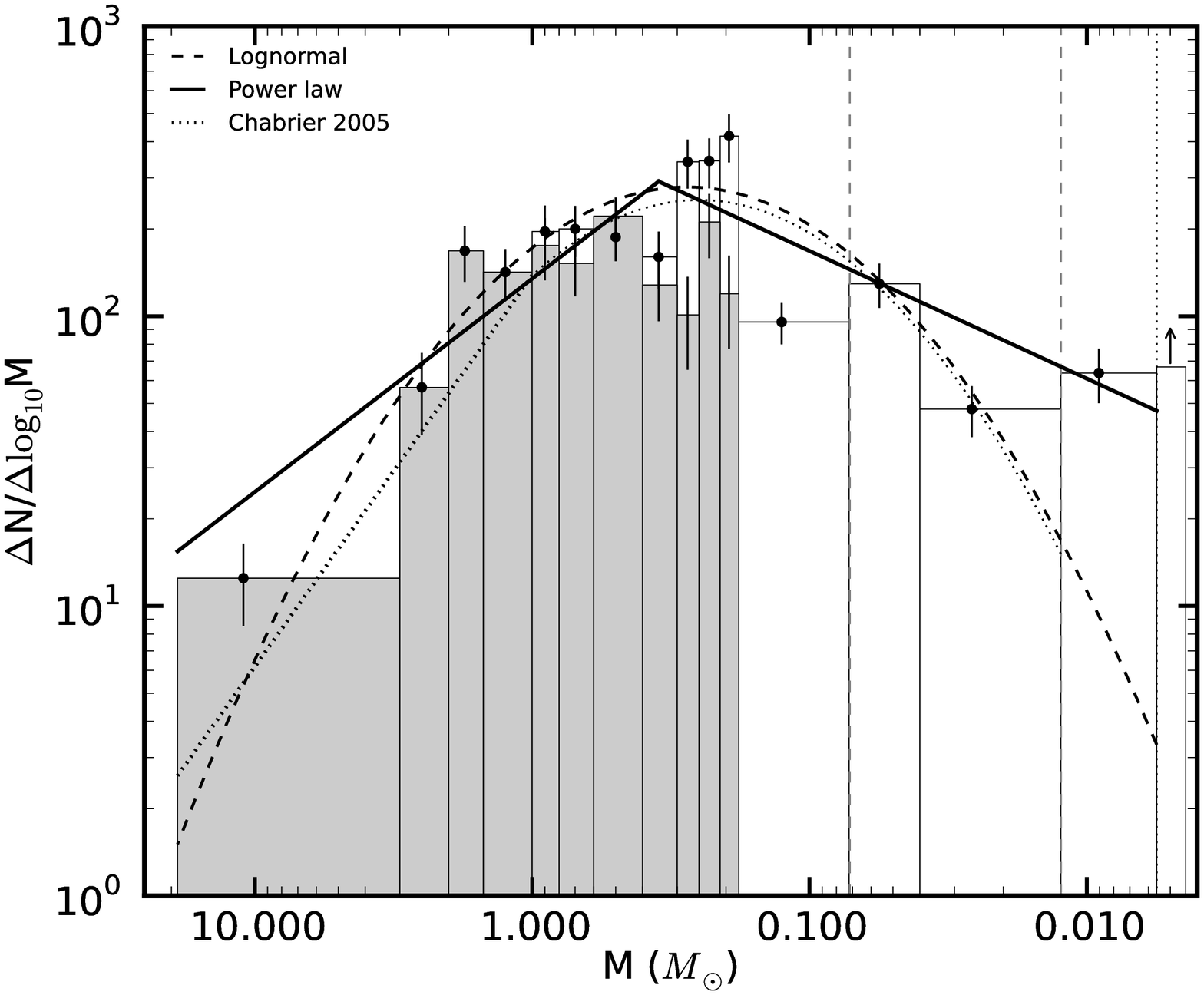}
\caption{\textit{(Upper panel)} \so mass spectrum from 19 through 0.004\,$M_{\odot}$ using the \citet{siess00} models (grey area) and the Lyon models (open histogram, \citealt{baraffe98,chabrier00model,baraffe03}). The best power-law fits to the mass domains $M\,>\,0.35$\,$M_\odot$ and $M\,<\,0.35$\,$M_\odot$ (using the Lyon models) are shown with continous lines. For comparison purposes, the mass spectrum derived by \citet{kroupa01} is overlaid normalized to the total number of objects. \textit{(Lower panel)} The \so mass function using the \citet{siess00} models (grey area) and the Lyon models (open histogram). The best log-normal fit to our data is depicted with a dashed line, the double power-law fit from our mass spectrum is illustrated with a continuos line, and the \citet{chabrier05} mass function normalized to the total number of sources is plotted as a dotted line. In both panels, from left to right, vertical lines as in Figure \ref{fm}. Vertical error bars stand for the Poisson uncertainties.
\label{fmall}}
\end{figure*}

The combined \so stellar and substellar mass spectrum is depicted in Figure \ref{fmall}. We plotted the derivations using the NextGen and \citet{siess00} models. To give a similar statistical weight to each mass bin, we considered mass intervals with similar amounts of \so member candidates (20--30), thus the mass bin size is not constant along the mass spectrum. Only the incomplete bin in the planetary-mass regime and the complete most massive bin contain less number of objects. We remark the good match between the Mayrit and the VISTA catalogs since there is no obvious discontinuity at the mass both samples were merged (0.25\,$M_\odot$). 

The \so mass spectrum is a rising function, i.e., the number of \so members increases toward low masses from $\sim$19 to 0.006\,$M_\odot$. It can be fit by two power-law equations of different exponents. In Table \ref{fits} we summarize the $\alpha$ parameters found for the mass intervals 19--0.35\,$M_{\odot}$ and 0.35--0.006\,$M_{\odot}$. To take into account the differences that may arise from the two employed theoretical mass-luminosity relations, we also provide fits for the following cases: case I where the Lyon- and \citet{siess00}-based masses cover the intervals 0.006--1\,$M_\odot$ and 1--19\,$M_\odot$, respectively, and case II where the \citet{siess00}-based masses cover 0.3--19\,$M_\odot$ and the smaller masses are derived using the Lyon isochrone. The fits shown in Figure \ref{fmall} correspond only to case I (for the clarity of the Figure). Despite the differences in the models (e.g., from Figure \ref{fmall} it can be seen that some mass bins differ by more than the Poisson uncertainty), the derived $\alpha$ values are consistent with each other at the 1-$\sigma$ level. The $\alpha$ exponent obtained for the mass range 19--0.35\,$M_{\odot}$ ($\alpha \sim 1.7$) appears to be slightly smaller than the Salpeter value of 2.35 \citep{salpeter55}. 

Figure \ref{fmall} also illustrates the comparison of the \so mass spectrum with the multiple segment power-law function proposed by \citet{kroupa01}. This author defined an average or Galactic-field mass spectrum with changes in the power-law index at two masses: $\sim$0.5\,$M_{\odot}$ and $\sim$0.08\,$M_{\odot}$. The Kroupa functional form normalized to the total number of \so objects is shown in Figure \ref{fmall}. According to this function, the number of substellar objects relative to the population of stars appears to be larger than what is observed from our \so mass spectrum. However, we remark that the \so mass spectrum and \citet{kroupa01} function are consistent within the uncertainties quoted for both.  We found that a two segment power-law function ($\alpha = 1.7 \pm 0.2$ for $M > 0.35$\,$M_{\odot}$, and $\alpha = 0.6 \pm 0.2$ for $M < 0.35$\,$M_{\odot}$) provides a better fit to the cluster data.

To compare the \so mass spectrum with known mass functions in other star forming regions and the field, we built the cluster mass function as originally defined by \citet{salpeter55}, i.e., $\xi(\log M)=\Delta N/\Delta \log M$. It is displayed in Figure \ref{fmall}. We performed a lognormal fit to the data by finding the characteristic mass ($M_c$, the mass at which the mass function reaches its maximum) and the variance ($\sigma$) of the distribution:
\begin{equation*}
\xi(\log M) \sim \exp\Bigg(-\frac{(\log M - \log M_c)^2}{2\sigma^2}\Bigg)
\end{equation*}
The derived parameters for cases I and II are listed in Table \ref{fits}. The characteristic mass was found at 0.27\,$M_{\odot}$ for both cases, the variance of the functions ranged from  0.6 to 0.7. This is in agreement with other works on open clusters, e.g., Pleiades \citep{moraux03}, $\rho$ Ophiuchi, IC\,348 and the Trapezium \citep{luhman00}, Chamaeleon I \citep{luhman07}, and the Galactic field \citep{chabrier05}.

In Figure \ref{fmall}, we also compared the \so mass function with the functional form of \citet{chabrier05}, the latter being normalized to the total number of \so sources. \citet{chabrier05} function consists of a combined Salpeter's power-law for $M\,>\,1$\,$M_{\odot}$ and a lognormal parametrization for lower masses, where $M_c$\,=\,0.25\,$M_{\odot}$ and variance $\sigma\,=\,0.55$. The similarity between this functional form and the pure lognormal one is noteworthy for the stellar mass regime (see Table \ref{fits}). However, the \so planetary-mass interval clearly outnumbers the prediction of the \citet{chabrier05} function. This author did not cover such low masses in his study. By adopting \citet{chabrier05} functional form and fitting the characteristic mass and variance of the lognormal equation from 1 down to 0.006\,$M_{\odot}$, we found $M_c \sim$ 0.25\,$M_{\odot}$ and $\sigma\,=\,$0.5--0.6 for $\sigma$~Orionis; the planetary-mass bins are still not well reproduced. As for the cluster mass spectrum, we conclude that two power-law expressions appear to fit better the \so mass function, from the very high masses down to the free-floating planetary-mass objects; these power-laws are illustrated in Figure \ref{fmall}. 


\citet{sumi11} reported the discovery of a large population of unbound planetary mass objects located at wide separations from their parent stars by the gravitational microlensing technique. In their work, the authors claimed that the number of brown dwarfs and planetary mass objects derived from microlensing events is $\sim$0.7 and $\sim$1.8 times the number of main sequence stars with masses in the interval 0.08--1.0\,$M_{\odot}$. Our results for the \so cluster indicate that brown dwarfs represent about one third of the population of sub-solar mass stars (0.08--1.0\,$M_{\odot}$), while planetary mass objects down to $\sim$6\, $M_{\rm Jup}$ represent a fraction of only $\sim$0.2. Our study found a relative number of substellar objects smaller than that obtained by \citet{sumi11}. However, the alternative model presented by the authors, where they used a discontinuous substellar mass function (model 3 from \citealt{sumi11}), suggests a fraction of brown dwarfs relative to the number of stars less massive than the Sun of $\sim$0.2, which is consistent with our results. Because our search is complete down to $\sim$6\, $M_{\rm Jup}$, objects of 1\,$M_{\rm Jup}$ are not covered by the VISTA data, then we cannot compare directly to the results of \citet{sumi11} for this mass regime.

\section{Conclusions}\label{final}
Using VISTA Orion data obtained in the $ZYJHK_s$ filters, we performed a photometric exploration of a circular area of radius 30\arcmin~around the multiple, O-type star $\sigma$ Ori. The survey uncovered $>$75\%~of the \so cluster population with masses ranging from 0.25\,$M_{\odot}$ ($J$ = 13\,mag)  to 0.006\,$M_{\odot}$ ($Z$-band completeness at 22.6\,mag) and 0.003\,$M_{\odot}$ ($J$-band completeness at 21\,mag). These mass estimates are based on theoretical isochrones of 3\,Myr, solar metallicity, and shifted to the distance of 352\,pc, which are the canonical parameters of the \so cluster. The VISTA data were combined with optical ($I$-band) and mid-infrared {\sl WISE} and {\sl Spitzer} photometry to identify a total of 210 \so candidates detected at both $Z$- and $J$-passbands that comply with all photometric criteria expected for bona-fide cluster members. Of the 210, 23 are new discoveries with mass estimates in the planetary-mass interval 0.011--0.004\,$M_{\odot}$ and likely L spectral types. One of them, S\,Ori\,J053804.65$-$021352.5, has VISTA and {\sl Spitzer} colors compatible with a ``methane" spectral type T0--T4. Our findings double the number of \so planetary-mass candidates known to date. 

Based on infrared flux excesses at 8.0\,$\mu$m, we derived a disk frequency of 41\,\,$\pm$\,\,8\,\%~and 39\,\,$\pm$\,\,9\,\%~for very low-mass stars (0.25--0.072\,$M_{\odot}$) and brown dwarfs (0.072--0.012\,$M_{\odot}$). As for the planetary domain (0.012--0.004\,$M_{\odot}$), the study of the presence of infrared flux excesses was based on the deeper {\sl Spitzer} 4.5\,$\mu$m data, from which we measured that $\ge$31\,$\pm$\,11\%~of the \so planetary-mass objects harbor warm inner disks.

The combination of an homogeneous survey depth, wide spatial and mass coverages, and high percentage of sources with confirmed membership in \so ($\sim$70\%) has allowed us to confidently study the cluster spatial distribution and mass function. Low-mass stars, brown dwarfs, and planetary-mass objects in \so are spatially concentrated with an effective radius of 12\arcmin~(1.2\,pc) around the multiple star $\sigma$ Ori. They have a central density of $\rho_0\sim0.4$\,arcmin$^{-2}$ (36\,pc$^{-2}$). The radial density distributions of all three mass intervals are alike, suggesting little dynamical evolution of the cluster. Furthermore, we find no significant evidence for spatial distribution differences between the disk-bearing and diskless cluster members, suggesting that there is no spatial segregation between these two type of populations. 

The \so low-mass spectrum is a smooth rising function that can be fit with a single power-law expression ($\Delta N / \Delta M \sim M^{-\alpha}$), where $\alpha = 0.6 \pm 0.2$ for the mass interval 0.25--0.006\,$M_{\odot}$ ($Z$-band completeness or $J$=20\,mag). The \so mass spectrum extends well into the planetary-mass regime down to 0.004\,$M_{\odot}$. Using the Mayrit catalog by  \citet{caballero08b}, we were able to expand the \so mass function and the mass spectrum up to the most massive, O-type stars of the cluster, thus covering the wide mass range  $\sim$19--0.006\,$M_{\odot}$. We compared the \so mass function to various functional forms discussed in the literature finding a characteristic mass of 0.27\,$\pm$\,0.10\,$M_{\odot}$ and a variance ranging from 0.5 to 0.7 in the case of a lognormal mass function. However, this functional form appears to underestimate the number of cluster planetary-mass objects. The entire \so mass function and mass spectrum are reasonably described by two power-law expressions with mass spectrum indices of $\alpha = 1.7 \pm 0.2$ for $M > 0.35$\,$M_{\odot}$, and $\alpha = 0.6 \pm 0.2$ for $M < 0.35 $\,$M_{\odot}$. The low number of T-type objects found in the VISTA \so survey (two previously known ones, S\,Ori\,70 and 73, and a new one reported here) as compared to the prediction made by the extrapolation of the \so mass function between 0.004 and 0.003\,$M_{\odot}$ (or between $J$=20.5\,mag and $J$-band completeness) seems to indicate that either the cluster substellar mass function has a turnover (and possibly a mass cut-off) at around 0.004\,$M_{\odot}$, field T dwarf densities are underestimated in the literature, and/or cluster T-type objects are fainter than predicted by theoretical models.

\acknowledgments
We thank the referee, Jos\'e A. Caballero, for constructive comments. Based on observations made with ESO Te\-les\-co\-pes at Cerro Paranal Observatory during VISTA science verification, under the program ID 60.A-9285(B). This work is in part based on observations made with the \textit{Spitzer Space Telescope}, which is operated by the Jet Propulsion Laboratory, California Institute of Technology under a contract with NASA. This publication makes use of data products from the \textit{Wide-field Infrared Survey Explorer}, which is a joint project of the University of California, Los Angeles, and the Jet Propulsion Laboratory/California Institute of Technology, funded by the National Aeronautics and Space Administration. IRAF is distributed by National Optical Astronomy Observatory, which is operated by the Association of Universities for Research in Astronomy, Inc., under contract with the National Science Foundation. This research has been supported by the Spanish Ministry of Economics and Competitiveness under the projects AYA2010-21308-C3-02 and AYA2010-20535.


\clearpage

\bibliography{pena_ramirez_et_al}

\begin{thebibliography}{121}
\expandafter\ifx\csname natexlab\endcsname\relax\def\natexlab#1{#1}\fi

\bibitem[{{Andrews} {et~al.}(2004){Andrews}, {Reipurth}, {Bally}, \&
  {Heathcote}}]{andrews04}
{Andrews}, S.~M., {Reipurth}, B., {Bally}, J., \& {Heathcote}, S.~R. 2004,
  \apj, 606, 353

\bibitem[{{Arnaboldi} {et~al.}(2010){Arnaboldi}, {Petr-Gotzens}, {Rejkuba},
  {Neeser}, {Szeifert}, {Ivanov}, {Hummel}, {Hilker}, {Neumayer}, {M{\o}ller},
  {Nilsson}, {Venemans}, {Hatziminaoglou}, {Hussain}, {Stanke}, {Teixeira},
  {Ramsay}, {Retzlaff}, {Slijkhuis}, {Comer{\'o}n}, {Melnick}, {Romaniello},
  {Emerson}, {Sutherland}, {Irwin}, {Lewis}, {Hodgkin}, \&
  {Gonzales-Solares}}]{arnaboldi10}
{Arnaboldi}, M., {Petr-Gotzens}, M., {Rejkuba}, M., {Neeser}, M., {Szeifert},
  T., {Ivanov}, V.~D., {Hummel}, W., {Hilker}, M., {Neumayer}, N., {M{\o}ller},
  P., {Nilsson}, K., {Venemans}, B., {Hatziminaoglou}, E., {Hussain}, G.,
  {Stanke}, T., {Teixeira}, P., {Ramsay}, S., {Retzlaff}, J., {Slijkhuis}, R.,
  {Comer{\'o}n}, F., {Melnick}, J., {Romaniello}, M., {Emerson}, J.,
  {Sutherland}, W., {Irwin}, M., {Lewis}, J., {Hodgkin}, S., \&
  {Gonzales-Solares}, E. 2010, The Messenger, 139, 6

\bibitem[{{Baraffe} {et~al.}(1998){Baraffe}, {Chabrier}, {Allard}, \&
  {Hauschildt}}]{baraffe98}
{Baraffe}, I., {Chabrier}, G., {Allard}, F., \& {Hauschildt}, P.~H. 1998, \aap,
  337, 403

\bibitem[{{Baraffe} {et~al.}(2003){Baraffe}, {Chabrier}, {Barman}, {Allard}, \&
  {Hauschildt}}]{baraffe03}
{Baraffe}, I., {Chabrier}, G., {Barman}, T.~S., {Allard}, F., \& {Hauschildt},
  P.~H. 2003, \aap, 402, 701

\bibitem[{{Barrado y Navascu{\'e}s} {et~al.}(2003){Barrado y Navascu{\'e}s},
  {B{\'e}jar}, {Mundt}, {Mart{\'{\i}}n}, {Rebolo}, {Zapatero Osorio}, \&
  {Bailer-Jones}}]{barrado03}
{Barrado y Navascu{\'e}s}, D., {B{\'e}jar}, V.~J.~S., {Mundt}, R.,
  {Mart{\'{\i}}n}, E.~L., {Rebolo}, R., {Zapatero Osorio}, M.~R., \&
  {Bailer-Jones}, C.~A.~L. 2003, \aap, 404, 171

\bibitem[{{Barrado y Navascu{\'e}s} {et~al.}(2002{\natexlab{a}}){Barrado y
  Navascu{\'e}s}, {Bouvier}, {Stauffer}, {Lodieu}, \&
  {McCaughrean}}]{barrado02b}
{Barrado y Navascu{\'e}s}, D., {Bouvier}, J., {Stauffer}, J.~R., {Lodieu}, N.,
  \& {McCaughrean}, M.~J. 2002{\natexlab{a}}, \aap, 395, 813

\bibitem[{{Barrado y Navascu{\'e}s} {et~al.}(2004){Barrado y Navascu{\'e}s},
  {Stauffer}, {Bouvier}, {Jayawardhana}, \& {Cuillandre}}]{barrado04}
{Barrado y Navascu{\'e}s}, D., {Stauffer}, J.~R., {Bouvier}, J.,
  {Jayawardhana}, R., \& {Cuillandre}, J.-C. 2004, \apj, 610, 1064

\bibitem[{{Barrado y Navascu{\'e}s} {et~al.}(2001){Barrado y Navascu{\'e}s},
  {Zapatero Osorio}, {B{\'e}jar}, {Rebolo}, {Mart{\'{\i}}n}, {Mundt}, \&
  {Bailer-Jones}}]{barrado01}
{Barrado y Navascu{\'e}s}, D., {Zapatero Osorio}, M.~R., {B{\'e}jar}, V.~J.~S.,
  {Rebolo}, R., {Mart{\'{\i}}n}, E.~L., {Mundt}, R., \& {Bailer-Jones},
  C.~A.~L. 2001, \aap, 377, L9

\bibitem[{{Barrado y Navascu{\'e}s} {et~al.}(2002{\natexlab{b}}){Barrado y
  Navascu{\'e}s}, {Zapatero Osorio}, {Mart{\'{\i}}n}, {B{\'e}jar}, {Rebolo}, \&
  {Mundt}}]{barrado02}
{Barrado y Navascu{\'e}s}, D., {Zapatero Osorio}, M.~R., {Mart{\'{\i}}n},
  E.~L., {B{\'e}jar}, V.~J.~S., {Rebolo}, R., \& {Mundt}, R.
  2002{\natexlab{b}}, \aap, 393, L85

\bibitem[{{Bastian} {et~al.}(2010){Bastian}, {Covey}, \& {Meyer}}]{bastian10}
{Bastian}, N., {Covey}, K.~R., \& {Meyer}, M.~R. 2010, \araa, 48, 339

\bibitem[{{B{\'e}jar}(2001)}]{tesisbejar}
{B{\'e}jar}, V.~J.~S. 2001, PhD thesis, Instituto de Astrof\'isica de Canarias

\bibitem[{{B{\'e}jar} {et~al.}(2004{\natexlab{a}}){B{\'e}jar}, {Caballero},
  {Rebolo}, {Zapatero Osorio}, \& {Barrado y Navascu{\'e}s}}]{bejar04}
{B{\'e}jar}, V.~J.~S., {Caballero}, J.~A., {Rebolo}, R., {Zapatero Osorio},
  M.~R., \& {Barrado y Navascu{\'e}s}, D. 2004{\natexlab{a}}, \apss, 292, 339

\bibitem[{{B{\'e}jar} {et~al.}(2001){B{\'e}jar}, {Mart{\'{\i}}n}, {Zapatero
  Osorio}, {Rebolo}, {Barrado y Navascu{\'e}s}, {Bailer-Jones}, {Mundt},
  {Baraffe}, {Chabrier}, \& {Allard}}]{bejar01}
{B{\'e}jar}, V.~J.~S., {Mart{\'{\i}}n}, E.~L., {Zapatero Osorio}, M.~R.,
  {Rebolo}, R., {Barrado y Navascu{\'e}s}, D., {Bailer-Jones}, C.~A.~L.,
  {Mundt}, R., {Baraffe}, I., {Chabrier}, C., \& {Allard}, F. 2001, \apj, 556,
  830

\bibitem[{{B{\'e}jar} {et~al.}(1999){B{\'e}jar}, {Zapatero Osorio}, \&
  {Rebolo}}]{bejar99}
{B{\'e}jar}, V.~J.~S., {Zapatero Osorio}, M.~R., \& {Rebolo}, R. 1999, \apj,
  521, 671

\bibitem[{{B{\'e}jar} {et~al.}(2004{\natexlab{b}}){B{\'e}jar}, {Zapatero
  Osorio}, \& {Rebolo}}]{bejar04a}
---. 2004{\natexlab{b}}, AN, 325, 705

\bibitem[{{B{\'e}jar} {et~al.}(2011){B{\'e}jar}, {Zapatero Osorio}, {Rebolo},
  {Caballero}, {Barrado}, {Mart{\'{\i}}n}, {Mundt}, \&
  {Bailer-Jones}}]{bejar11}
{B{\'e}jar}, V.~J.~S., {Zapatero Osorio}, M.~R., {Rebolo}, R., {Caballero},
  J.~A., {Barrado}, D., {Mart{\'{\i}}n}, E.~L., {Mundt}, R., \& {Bailer-Jones},
  C.~A.~L. 2011, \apj, 743, 64

\bibitem[{{Bihain} {et~al.}(2006){Bihain}, {Rebolo}, {B{\'e}jar}, {Caballero},
  {Bailer-Jones}, {Mundt}, {Acosta-Pulido}, \& {Manchado Torres}}]{bihain06}
{Bihain}, G., {Rebolo}, R., {B{\'e}jar}, V.~J.~S., {Caballero}, J.~A.,
  {Bailer-Jones}, C.~A.~L., {Mundt}, R., {Acosta-Pulido}, J.~A., \& {Manchado
  Torres}, A. 2006, \aap, 458, 805

\bibitem[{{Bihain} {et~al.}(2009){Bihain}, {Rebolo}, {Zapatero Osorio},
  {B{\'e}jar}, {Vill{\'o}-P{\'e}rez}, {D{\'{\i}}az-S{\'a}nchez},
  {P{\'e}rez-Garrido}, {Caballero}, {Bailer-Jones}, {Barrado y Navascu{\'e}s},
  {Eisl{\"o}ffel}, {Forveille}, {Goldman}, {Henning}, {Mart{\'{\i}}n}, \&
  {Mundt}}]{bihain09}
{Bihain}, G., {Rebolo}, R., {Zapatero Osorio}, M.~R., {B{\'e}jar}, V.~J.~S.,
  {Vill{\'o}-P{\'e}rez}, I., {D{\'{\i}}az-S{\'a}nchez}, A.,
  {P{\'e}rez-Garrido}, A., {Caballero}, J.~A., {Bailer-Jones}, C.~A.~L.,
  {Barrado y Navascu{\'e}s}, D., {Eisl{\"o}ffel}, J., {Forveille}, T.,
  {Goldman}, B., {Henning}, T., {Mart{\'{\i}}n}, E.~L., \& {Mundt}, R. 2009,
  \aap, 506, 1169

\bibitem[{{Bouvier} {et~al.}(1998){Bouvier}, {Stauffer}, {Mart{\'i}n}, {Barrado
  y Navascu{\'e}s}, {Wallace}, \& {B{\'e}jar}}]{bouvier98}
{Bouvier}, J., {Stauffer}, J.~R., {Mart{\'i}n}, E.~L., {Barrado y
  Navascu{\'e}s}, D., {Wallace}, B., \& {B{\'e}jar}, V.~J.~S. 1998, \aap, 336,
  490

\bibitem[{{Bouy} {et~al.}(2009){Bouy}, {Hu{\'e}lamo}, {Mart{\'{\i}}n},
  {Marchis}, {Barrado Y Navascu{\'e}s}, {Kolb}, {Marchetti}, {Petr-Gotzens},
  {Sterzik}, {Ivanov}, {K{\"o}hler}, \& {N{\"u}rnberger}}]{bouy09}
{Bouy}, H., {Hu{\'e}lamo}, N., {Mart{\'{\i}}n}, E.~L., {Marchis}, F., {Barrado
  Y Navascu{\'e}s}, D., {Kolb}, J., {Marchetti}, E., {Petr-Gotzens}, M.~G.,
  {Sterzik}, M., {Ivanov}, V.~D., {K{\"o}hler}, R., \& {N{\"u}rnberger}, D.
  2009, \aap, 493, 931

\bibitem[{{Brown} {et~al.}(1994){Brown}, {de Geus}, \& {de Zeeuw}}]{brown94}
{Brown}, A.~G.~A., {de Geus}, E.~J., \& {de Zeeuw}, P.~T. 1994, \aap, 289, 101

\bibitem[{{Burgasser}(2007)}]{burgasser07}
{Burgasser}, A.~J. 2007, \apj, 659, 655

\bibitem[{{Burgasser} {et~al.}(2004){Burgasser}, {Kirkpatrick}, {McGovern},
  {McLean}, {Prato}, \& {Reid}}]{burgasser04}
{Burgasser}, A.~J., {Kirkpatrick}, J.~D., {McGovern}, M.~R., {McLean}, I.~S.,
  {Prato}, L., \& {Reid}, I.~N. 2004, \apj, 604, 827

\bibitem[{{Burningham} {et~al.}(2005){Burningham}, {Naylor}, {Littlefair}, \&
  {Jeffries}}]{burningham05}
{Burningham}, B., {Naylor}, T., {Littlefair}, S.~P., \& {Jeffries}, R.~D. 2005,
  \mnras, 356, 1583

\bibitem[{{Burningham} {et~al.}(2010){Burningham}, {Pinfield}, {Lucas},
  {Leggett}, {Deacon}, {Tamura}, {Tinney}, {Lodieu}, {Zhang}, {Hu{\'e}lamo},
  {Jones}, {Murray}, {Mortlock}, {Patel}, {Barrado y Navascu{\'e}s}, {Zapatero
  Osorio}, {Ishii}, {Kuzuhara}, \& {Smart}}]{burningham10}
{Burningham}, B., {Pinfield}, D.~J., {Lucas}, P.~W., {Leggett}, S.~K.,
  {Deacon}, N.~R., {Tamura}, M., {Tinney}, C.~G., {Lodieu}, N., {Zhang}, Z.~H.,
  {Hu{\'e}lamo}, N., {Jones}, H.~R.~A., {Murray}, D.~N., {Mortlock}, D.~J.,
  {Patel}, M., {Barrado y Navascu{\'e}s}, D., {Zapatero Osorio}, M.~R.,
  {Ishii}, M., {Kuzuhara}, M., \& {Smart}, R.~L. 2010, \mnras, 406, 1885

\bibitem[{{Caballero}(2006)}]{tesiscaballero}
{Caballero}, J.~A. 2006, PhD thesis, Instituto de Astrof\'isica de Canarias

\bibitem[{{Caballero}(2007{\natexlab{a}})}]{caballero07b}
---. 2007{\natexlab{a}}, AN, 328, 917

\bibitem[{{Caballero}(2007{\natexlab{b}})}]{caballero07a}
---. 2007{\natexlab{b}}, \aap, 466, 917

\bibitem[{{Caballero}(2008{\natexlab{a}})}]{caballero08a}
---. 2008{\natexlab{a}}, \mnras, 383, 750

\bibitem[{{Caballero}(2008{\natexlab{b}})}]{caballero08}
---. 2008{\natexlab{b}}, \mnras, 383, 375

\bibitem[{{Caballero}(2008{\natexlab{c}})}]{caballero08b}
---. 2008{\natexlab{c}}, \aap, 478, 667

\bibitem[{{Caballero}(2011)}]{caballero11}
{Caballero}, J.~A. 2011, in Stellar Clusters \& Associations: A RIA Workshop on
  Gaia, 108

\bibitem[{{Caballero} {et~al.}(2007){Caballero}, {B{\'e}jar}, {Rebolo},
  {Eisl{\"o}ffel}, {Zapatero Osorio}, {Mundt}, {Barrado Y Navascu{\'e}s},
  {Bihain}, {Bailer-Jones}, {Forveille}, \& {Mart{\'{\i}}n}}]{caballero07}
{Caballero}, J.~A., {B{\'e}jar}, V.~J.~S., {Rebolo}, R., {Eisl{\"o}ffel}, J.,
  {Zapatero Osorio}, M.~R., {Mundt}, R., {Barrado Y Navascu{\'e}s}, D.,
  {Bihain}, G., {Bailer-Jones}, C.~A.~L., {Forveille}, T., \& {Mart{\'{\i}}n},
  E.~L. 2007, \aap, 470, 903

\bibitem[{{Caballero} {et~al.}(2004){Caballero}, {B{\'e}jar}, {Rebolo}, \&
  {Zapatero Osorio}}]{caballero04}
{Caballero}, J.~A., {B{\'e}jar}, V.~J.~S., {Rebolo}, R., \& {Zapatero Osorio},
  M.~R. 2004, \aap, 424, 857

\bibitem[{{Caballero} {et~al.}(2008{\natexlab{a}}){Caballero}, {Burgasser}, \&
  {Klement}}]{caballero08c}
{Caballero}, J.~A., {Burgasser}, A.~J., \& {Klement}, R. 2008{\natexlab{a}},
  \aap, 488, 181

\bibitem[{{Caballero} {et~al.}(2006){Caballero}, {Mart{\'{\i}}n}, {Dobbie}, \&
  {Barrado Y Navascu{\'e}s}}]{caballero06}
{Caballero}, J.~A., {Mart{\'{\i}}n}, E.~L., {Dobbie}, P.~D., \& {Barrado Y
  Navascu{\'e}s}, D. 2006, \aap, 460, 635

\bibitem[{{Caballero} {et~al.}(2008{\natexlab{b}}){Caballero}, {Valdivielso},
  {Mart{\'{\i}}n}, {Montes}, {Pascual}, \&
  {P{\'e}rez-Gonz{\'a}lez}}]{caballero08d}
{Caballero}, J.~A., {Valdivielso}, L., {Mart{\'{\i}}n}, E.~L., {Montes}, D.,
  {Pascual}, S., \& {P{\'e}rez-Gonz{\'a}lez}, P.~G. 2008{\natexlab{b}}, \aap,
  491, 515

\bibitem[{{Casali} {et~al.}(2007){Casali}, {Adamson}, {Alves de Oliveira},
  {Almaini}, {Burch}, {Chuter}, {Elliot}, {Folger}, {Foucaud}, {Hambly},
  {Hastie}, {Henry}, {Hirst}, {Irwin}, {Ives}, {Lawrence}, {Laidlaw}, {Lee},
  {Lewis}, {Lunney}, {McLay}, {Montgomery}, {Pickup}, {Read}, {Rees}, {Robson},
  {Sekiguchi}, {Vick}, {Warren}, \& {Woodward}}]{casali07}
{Casali}, M., {Adamson}, A., {Alves de Oliveira}, C., {Almaini}, O., {Burch},
  K., {Chuter}, T., {Elliot}, J., {Folger}, M., {Foucaud}, S., {Hambly}, N.,
  {Hastie}, M., {Henry}, D., {Hirst}, P., {Irwin}, M., {Ives}, D., {Lawrence},
  A., {Laidlaw}, K., {Lee}, D., {Lewis}, J., {Lunney}, D., {McLay}, S.,
  {Montgomery}, D., {Pickup}, A., {Read}, M., {Rees}, N., {Robson}, I.,
  {Sekiguchi}, K., {Vick}, A., {Warren}, S., \& {Woodward}, B. 2007, \aap, 467,
  777

\bibitem[{{Chabrier}(2005)}]{chabrier05}
{Chabrier}, G. 2005, in Astrophysics and Space Science Library, Vol. 327, The
  Initial Mass Function 50 Years Later, eds. {E.~Corbelli, F.~Palla, \&
  H.~Zinnecker}, 41

\bibitem[{{Chabrier} {et~al.}(2000){Chabrier}, {Baraffe}, {Allard}, \&
  {Hauschildt}}]{chabrier00model}
{Chabrier}, G., {Baraffe}, I., {Allard}, F., \& {Hauschildt}, P. 2000, \apj,
  542, 464

\bibitem[{{Dalton} {et~al.}(2006){Dalton}, {Caldwell}, {Ward}, {Whalley},
  {Woodhouse}, {Edeson}, {Clark}, {Beard}, {Gallie}, {Todd}, {Strachan},
  {Bezawada}, {Sutherland}, \& {Emerson}}]{vista_1}
{Dalton}, G.~B., {Caldwell}, M., {Ward}, A.~K., {Whalley}, M.~S., {Woodhouse},
  G., {Edeson}, R.~L., {Clark}, P., {Beard}, S.~M., {Gallie}, A.~M., {Todd},
  S.~P., {Strachan}, J.~M.~D., {Bezawada}, N.~N., {Sutherland}, W.~J., \&
  {Emerson}, J.~P. 2006, in Society of Photo-Optical Instrumentation Engineers
  (SPIE) Conference, Vol. 6269, Society of Photo-Optical Instrumentation
  Engineers (SPIE) Conference Series

\bibitem[{{de Wit} {et~al.}(2006){de Wit}, {Bouvier}, {Palla}, {Cuillandre},
  {James}, {Kendall}, {Lodieu}, {McCaughrean}, {Moraux}, {Randich}, \&
  {Testi}}]{dewit06}
{de Wit}, W.~J., {Bouvier}, J., {Palla}, F., {Cuillandre}, J.-C., {James},
  D.~J., {Kendall}, T.~R., {Lodieu}, N., {McCaughrean}, M.~J., {Moraux}, E.,
  {Randich}, S., \& {Testi}, L. 2006, \aap, 448, 189

\bibitem[{{Dobbie} {et~al.}(2002){Dobbie}, {Pinfield}, {Jameson}, \&
  {Hodgkin}}]{dobbie02}
{Dobbie}, P.~D., {Pinfield}, D.~J., {Jameson}, R.~F., \& {Hodgkin}, S.~T. 2002,
  \mnras, 335, L79

\bibitem[{{Emerson} {et~al.}(2006){Emerson}, {McPherson}, \&
  {Sutherland}}]{vista_2}
{Emerson}, J., {McPherson}, A., \& {Sutherland}, W. 2006, The Messenger, 126,
  41

\bibitem[{{Fazio} {et~al.}(2004){Fazio}, {Hora}, {Allen}, {Ashby}, {Barmby},
  {Deutsch}, {Huang}, {Kleiner}, {Marengo}, {Megeath}, {Melnick}, {Pahre},
  {Patten}, {Polizotti}, {Smith}, {Taylor}, {Wang}, {Willner}, {Hoffmann},
  {Pipher}, {Forrest}, {McMurty}, {McCreight}, {McKelvey}, {McMurray}, {Koch},
  {Moseley}, {Arendt}, {Mentzell}, {Marx}, {Losch}, {Mayman}, {Eichhorn},
  {Krebs}, {Jhabvala}, {Gezari}, {Fixsen}, {Flores}, {Shakoorzadeh}, {Jungo},
  {Hakun}, {Workman}, {Karpati}, {Kichak}, {Whitley}, {Mann}, {Tollestrup},
  {Eisenhardt}, {Stern}, {Gorjian}, {Bhattacharya}, {Carey}, {Nelson},
  {Glaccum}, {Lacy}, {Lowrance}, {Laine}, {Reach}, {Stauffer}, {Surace},
  {Wilson}, {Wright}, {Hoffman}, {Domingo}, \& {Cohen}}]{fazio04}
{Fazio}, G.~G., {Hora}, J.~L., {Allen}, L.~E., {Ashby}, M.~L.~N., {Barmby}, P.,
  {Deutsch}, L.~K., {Huang}, J.-S., {Kleiner}, S., {Marengo}, M., {Megeath},
  S.~T., {Melnick}, G.~J., {Pahre}, M.~A., {Patten}, B.~M., {Polizotti}, J.,
  {Smith}, H.~A., {Taylor}, R.~S., {Wang}, Z., {Willner}, S.~P., {Hoffmann},
  W.~F., {Pipher}, J.~L., {Forrest}, W.~J., {McMurty}, C.~W., {McCreight},
  C.~R., {McKelvey}, M.~E., {McMurray}, R.~E., {Koch}, D.~G., {Moseley}, S.~H.,
  {Arendt}, R.~G., {Mentzell}, J.~E., {Marx}, C.~T., {Losch}, P., {Mayman}, P.,
  {Eichhorn}, W., {Krebs}, D., {Jhabvala}, M., {Gezari}, D.~Y., {Fixsen},
  D.~J., {Flores}, J., {Shakoorzadeh}, K., {Jungo}, R., {Hakun}, C., {Workman},
  L., {Karpati}, G., {Kichak}, R., {Whitley}, R., {Mann}, S., {Tollestrup},
  E.~V., {Eisenhardt}, P., {Stern}, D., {Gorjian}, V., {Bhattacharya}, B.,
  {Carey}, S., {Nelson}, B.~O., {Glaccum}, W.~J., {Lacy}, M., {Lowrance},
  P.~J., {Laine}, S., {Reach}, W.~T., {Stauffer}, J.~A., {Surace}, J.~A.,
  {Wilson}, G., {Wright}, E.~L., {Hoffman}, A., {Domingo}, G., \& {Cohen}, M.
  2004, \apjs, 154, 10

\bibitem[{{Franciosini} {et~al.}(2006){Franciosini}, {Pallavicini}, \&
  {Sanz-Forcada}}]{franciosini06}
{Franciosini}, E., {Pallavicini}, R., \& {Sanz-Forcada}, J. 2006, \aap, 446,
  501

\bibitem[{{Garrison}(1967)}]{garrison67}
{Garrison}, R.~F. 1967, \pasp, 79, 433

\bibitem[{{Golimowski} {et~al.}(2004){Golimowski}, {Leggett}, {Marley}, {Fan},
  {Geballe}, {Knapp}, {Vrba}, {Henden}, {Luginbuhl}, {Guetter}, {Munn},
  {Canzian}, {Zheng}, {Tsvetanov}, {Chiu}, {Glazebrook}, {Hoversten},
  {Schneider}, \& {Brinkmann}}]{golimowski04}
{Golimowski}, D.~A., {Leggett}, S.~K., {Marley}, M.~S., {Fan}, X., {Geballe},
  T.~R., {Knapp}, G.~R., {Vrba}, F.~J., {Henden}, A.~A., {Luginbuhl}, C.~B.,
  {Guetter}, H.~H., {Munn}, J.~A., {Canzian}, B., {Zheng}, W., {Tsvetanov},
  Z.~I., {Chiu}, K., {Glazebrook}, K., {Hoversten}, E.~A., {Schneider}, D.~P.,
  \& {Brinkmann}, J. 2004, \aj, 127, 3516

\bibitem[{{Gonz{\'a}lez-Garc{\'{\i}}a}
  {et~al.}(2006){Gonz{\'a}lez-Garc{\'{\i}}a}, {Zapatero Osorio}, {B{\'e}jar},
  {Bihain}, {Barrado y Navascu{\'e}s}, {Caballero}, \&
  {Morales-Calder{\'o}n}}]{gonzalez06}
{Gonz{\'a}lez-Garc{\'{\i}}a}, B.~M., {Zapatero Osorio}, M.~R., {B{\'e}jar},
  V.~J.~S., {Bihain}, G., {Barrado y Navascu{\'e}s}, D., {Caballero}, J.~A., \&
  {Morales-Calder{\'o}n}, M. 2006, \aap, 460, 799

\bibitem[{{Gonz{\'a}lez Hern{\'a}ndez} {et~al.}(2008){Gonz{\'a}lez
  Hern{\'a}ndez}, {Caballero}, {Rebolo}, {B{\'e}jar}, {Barrado y
  Navascu{\'e}s}, {Mart{\'{\i}}n}, \& {Zapatero Osorio}}]{hernandez08}
{Gonz{\'a}lez Hern{\'a}ndez}, J.~I., {Caballero}, J.~A., {Rebolo}, R.,
  {B{\'e}jar}, V.~J.~S., {Barrado y Navascu{\'e}s}, D., {Mart{\'{\i}}n}, E.~L.,
  \& {Zapatero Osorio}, M.~R. 2008, \aap, 490, 1135

\bibitem[{{Grazian} {et~al.}(2006){Grazian}, {Fontana}, {de Santis}, {Nonino},
  {Salimbeni}, {Giallongo}, {Cristiani}, {Gallozzi}, \& {Vanzella}}]{grazian06}
{Grazian}, A., {Fontana}, A., {de Santis}, C., {Nonino}, M., {Salimbeni}, S.,
  {Giallongo}, E., {Cristiani}, S., {Gallozzi}, S., \& {Vanzella}, E. 2006,
  \aap, 449, 951

\bibitem[{{Haisch} {et~al.}(2001){Haisch}, {Lada}, \& {Lada}}]{haisch01}
{Haisch}, Jr., K.~E., {Lada}, E.~A., \& {Lada}, C.~J. 2001, \apjl, 553, L153

\bibitem[{{Haro} \& {Moreno}(1953)}]{haromoreno53}
{Haro}, G. \& {Moreno}, A. 1953, Bolet{\'{\i}}n de los Observatorios
  Tonantzintla y Tacubaya, 1, 11

\bibitem[{{Hern{\'a}ndez} {et~al.}(2007){Hern{\'a}ndez}, {Hartmann}, {Megeath},
  {Gutermuth}, {Muzerolle}, {Calvet}, {Vivas}, {Brice{\~n}o}, {Allen},
  {Stauffer}, {Young}, \& {Fazio}}]{hernandez07}
{Hern{\'a}ndez}, J., {Hartmann}, L., {Megeath}, T., {Gutermuth}, R.,
  {Muzerolle}, J., {Calvet}, N., {Vivas}, A.~K., {Brice{\~n}o}, C., {Allen},
  L., {Stauffer}, J., {Young}, E., \& {Fazio}, G. 2007, \apj, 662, 1067

\bibitem[{{Hewett} {et~al.}(2006){Hewett}, {Warren}, {Leggett}, \&
  {Hodgkin}}]{hewett06}
{Hewett}, P.~C., {Warren}, S.~J., {Leggett}, S.~K., \& {Hodgkin}, S.~T. 2006,
  \mnras, 367, 454

\bibitem[{{Hillenbrand} \& {Carpenter}(2000)}]{hillenbrand00}
{Hillenbrand}, L.~A. \& {Carpenter}, J.~M. 2000, \apj, 540, 236

\bibitem[{{Jayawardhana} {et~al.}(2003){Jayawardhana}, {Ardila}, {Stelzer}, \&
  {Haisch}}]{jayawardhana03}
{Jayawardhana}, R., {Ardila}, D.~R., {Stelzer}, B., \& {Haisch}, Jr., K.~E.
  2003, \aj, 126, 1515

\bibitem[{{Jeffries} {et~al.}(2006){Jeffries}, {Maxted}, {Oliveira}, \&
  {Naylor}}]{jeffries06}
{Jeffries}, R.~D., {Maxted}, P.~F.~L., {Oliveira}, J.~M., \& {Naylor}, T. 2006,
  \mnras, 371, L6

\bibitem[{{Kenyon} {et~al.}(2005){Kenyon}, {Jeffries}, {Naylor}, {Oliveira}, \&
  {Maxted}}]{kenyon05}
{Kenyon}, M.~J., {Jeffries}, R.~D., {Naylor}, T., {Oliveira}, J.~M., \&
  {Maxted}, P.~F.~L. 2005, \mnras, 356, 89

\bibitem[{{Kirkpatrick} {et~al.}(1994){Kirkpatrick}, {McGraw}, {Hess},
  {Liebert}, \& {McCarthy}}]{kirkpatrick94}
{Kirkpatrick}, J.~D., {McGraw}, J.~T., {Hess}, T.~R., {Liebert}, J., \&
  {McCarthy}, Jr., D.~W. 1994, \apjs, 94, 749

\bibitem[{{Kroupa}(2001)}]{kroupa01}
{Kroupa}, P. 2001, \mnras, 322, 231

\bibitem[{{Lee}(1968)}]{lee68}
{Lee}, T.~A. 1968, \apj, 152, 913

\bibitem[{{Leggett} {et~al.}(2007){Leggett}, {Marley}, {Freedman}, {Saumon},
  {Liu}, {Geballe}, {Golimowski}, \& {Stephens}}]{leggett07}
{Leggett}, S.~K., {Marley}, M.~S., {Freedman}, R., {Saumon}, D., {Liu}, M.~C.,
  {Geballe}, T.~R., {Golimowski}, D.~A., \& {Stephens}, D.~C. 2007, \apj, 667,
  537

\bibitem[{{Lodieu} {et~al.}(2009){Lodieu}, {Burningham}, {Hambly}, \&
  {Pinfield}}]{lodieu09}
{Lodieu}, N., {Burningham}, B., {Hambly}, N.~C., \& {Pinfield}, D.~J. 2009,
  \mnras, 397, 258

\bibitem[{{Lodieu} {et~al.}(2011){Lodieu}, {de Wit}, {Carraro}, {Moraux},
  {Bouvier}, \& {Hambly}}]{lodieu11}
{Lodieu}, N., {de Wit}, W.-J., {Carraro}, G., {Moraux}, E., {Bouvier}, J., \&
  {Hambly}, N.~C. 2011, \aap, 532, A103

\bibitem[{{Lodieu} {et~al.}(2007{\natexlab{a}}){Lodieu}, {Dobbie}, {Deacon},
  {Hodgkin}, {Hambly}, \& {Jameson}}]{lodieu07}
{Lodieu}, N., {Dobbie}, P.~D., {Deacon}, N.~R., {Hodgkin}, S.~T., {Hambly},
  N.~C., \& {Jameson}, R.~F. 2007{\natexlab{a}}, \mnras, 380, 712

\bibitem[{{Lodieu} {et~al.}(2007{\natexlab{b}}){Lodieu}, {Hambly}, {Jameson},
  {Hodgkin}, {Carraro}, \& {Kendall}}]{lodieu07b}
{Lodieu}, N., {Hambly}, N.~C., {Jameson}, R.~F., {Hodgkin}, S.~T., {Carraro},
  G., \& {Kendall}, T.~R. 2007{\natexlab{b}}, \mnras, 374, 372

\bibitem[{{L{\'o}pez-Mart{\'{\i}}} {et~al.}(2004){L{\'o}pez-Mart{\'{\i}}},
  {Eisl{\"o}ffel}, {Scholz}, \& {Mundt}}]{lopez04}
{L{\'o}pez-Mart{\'{\i}}}, B., {Eisl{\"o}ffel}, J., {Scholz}, A., \& {Mundt}, R.
  2004, \aap, 416, 555

\bibitem[{{Lucas} \& {Roche}(2000)}]{lucas00}
{Lucas}, P.~W. \& {Roche}, P.~F. 2000, \mnras, 314, 858

\bibitem[{{Luhman}(2007)}]{luhman07}
{Luhman}, K.~L. 2007, \apjs, 173, 104

\bibitem[{{Luhman} {et~al.}(2008){Luhman}, {Hern{\'a}ndez}, {Downes},
  {Hartmann}, \& {Brice{\~n}o}}]{luhman08}
{Luhman}, K.~L., {Hern{\'a}ndez}, J., {Downes}, J.~J., {Hartmann}, L., \&
  {Brice{\~n}o}, C. 2008, \apj, 688, 362

\bibitem[{{Luhman} {et~al.}(2000){Luhman}, {Rieke}, {Young}, {Cotera}, {Chen},
  {Rieke}, {Schneider}, \& {Thompson}}]{luhman00}
{Luhman}, K.~L., {Rieke}, G.~H., {Young}, E.~T., {Cotera}, A.~S., {Chen}, H.,
  {Rieke}, M.~J., {Schneider}, G., \& {Thompson}, R.~I. 2000, \apj, 540, 1016

\bibitem[{{Luhman} {et~al.}(2003){Luhman}, {Stauffer}, {Muench}, {Rieke},
  {Lada}, {Bouvier}, \& {Lada}}]{luhman03}
{Luhman}, K.~L., {Stauffer}, J.~R., {Muench}, A.~A., {Rieke}, G.~H., {Lada},
  E.~A., {Bouvier}, J., \& {Lada}, C.~J. 2003, \apj, 593, 1093

\bibitem[{{Luhman} {et~al.}(2006){Luhman}, {Whitney}, {Meade}, {Babler},
  {Indebetouw}, {Bracker}, \& {Churchwell}}]{luhman06}
{Luhman}, K.~L., {Whitney}, B.~A., {Meade}, M.~R., {Babler}, B.~L.,
  {Indebetouw}, R., {Bracker}, S., \& {Churchwell}, E.~B. 2006, \apj, 647, 1180

\bibitem[{{Lyng{\aa}}(1981)}]{lynga81}
{Lyng{\aa}}, G. 1981, Astronomical Data Center Bulletin, 1, 90

\bibitem[{{Lyng{\aa}}(1983)}]{lynga83}
{Lyng{\aa}}, G. 1983, in Star Clusters and Associations and their Relation to
  the Evolution of the Galaxy, ed. {J.~Ruprecht \& J.~Palous}, 13--27

\bibitem[{{Mart{\'{\i}}n} {et~al.}(2000){Mart{\'{\i}}n}, {Brandner}, {Bouvier},
  {Luhman}, {Stauffer}, {Basri}, {Zapatero Osorio}, \& {Barrado y
  Navascu{\'e}s}}]{martin00}
{Mart{\'{\i}}n}, E.~L., {Brandner}, W., {Bouvier}, J., {Luhman}, K.~L.,
  {Stauffer}, J., {Basri}, G., {Zapatero Osorio}, M.~R., \& {Barrado y
  Navascu{\'e}s}, D. 2000, \apj, 543, 299

\bibitem[{{Mart{\'{\i}}n} {et~al.}(2001){Mart{\'{\i}}n}, {Zapatero Osorio},
  {Barrado y Navascu{\'e}s}, {B{\'e}jar}, \& {Rebolo}}]{martin01}
{Mart{\'{\i}}n}, E.~L., {Zapatero Osorio}, M.~R., {Barrado y Navascu{\'e}s},
  D., {B{\'e}jar}, V.~J.~S., \& {Rebolo}, R. 2001, \apjl, 558, L117

\bibitem[{{Maxted} {et~al.}(2008){Maxted}, {Jeffries}, {Oliveira}, {Naylor}, \&
  {Jackson}}]{maxted08}
{Maxted}, P.~F.~L., {Jeffries}, R.~D., {Oliveira}, J.~M., {Naylor}, T., \&
  {Jackson}, R.~J. 2008, \mnras, 385, 2210

\bibitem[{{Mayne} \& {Naylor}(2008)}]{mayne08}
{Mayne}, N.~J. \& {Naylor}, T. 2008, \mnras, 386, 261

\bibitem[{{McGovern} {et~al.}(2004){McGovern}, {Kirkpatrick}, {McLean},
  {Burgasser}, {Prato}, \& {Lowrance}}]{mcgovern04}
{McGovern}, M.~R., {Kirkpatrick}, J.~D., {McLean}, I.~S., {Burgasser}, A.~J.,
  {Prato}, L., \& {Lowrance}, P.~J. 2004, \apj, 600, 1020

\bibitem[{{Metchev} {et~al.}(2008){Metchev}, {Kirkpatrick}, {Berriman}, \&
  {Looper}}]{metchev08}
{Metchev}, S.~A., {Kirkpatrick}, J.~D., {Berriman}, G.~B., \& {Looper}, D.
  2008, \apj, 676, 1281

\bibitem[{{Moraux} {et~al.}(2007){Moraux}, {Bouvier}, {Stauffer}, {Barrado y
  Navascu{\'e}s}, \& {Cuillandre}}]{moraux07}
{Moraux}, E., {Bouvier}, J., {Stauffer}, J.~R., {Barrado y Navascu{\'e}s}, D.,
  \& {Cuillandre}, J.-C. 2007, \aap, 471, 499

\bibitem[{{Moraux} {et~al.}(2003){Moraux}, {Bouvier}, {Stauffer}, \&
  {Cuillandre}}]{moraux03}
{Moraux}, E., {Bouvier}, J., {Stauffer}, J.~R., \& {Cuillandre}, J.-C. 2003,
  \aap, 400, 891

\bibitem[{{Muench} {et~al.}(2002){Muench}, {Lada}, {Lada}, \&
  {Alves}}]{muench02}
{Muench}, A.~A., {Lada}, E.~A., {Lada}, C.~J., \& {Alves}, J. 2002, \apj, 573,
  366

\bibitem[{{Najita} {et~al.}(2000){Najita}, {Tiede}, \& {Carr}}]{najita00}
{Najita}, J.~R., {Tiede}, G.~P., \& {Carr}, J.~S. 2000, \apj, 541, 977

\bibitem[{{Oliveira} {et~al.}(2006){Oliveira}, {Jeffries}, {van Loon}, \&
  {Rushton}}]{oliveira06}
{Oliveira}, J.~M., {Jeffries}, R.~D., {van Loon}, J.~T., \& {Rushton}, M.~T.
  2006, \mnras, 369, 272

\bibitem[{{Patten} {et~al.}(2006){Patten}, {Stauffer}, {Burrows}, {Marengo},
  {Hora}, {Luhman}, {Sonnett}, {Henry}, {Raghavan}, {Megeath}, {Liebert}, \&
  {Fazio}}]{patten06}
{Patten}, B.~M., {Stauffer}, J.~R., {Burrows}, A., {Marengo}, M., {Hora},
  J.~L., {Luhman}, K.~L., {Sonnett}, S.~M., {Henry}, T.~J., {Raghavan}, D.,
  {Megeath}, S.~T., {Liebert}, J., \& {Fazio}, G.~G. 2006, \apj, 651, 502

\bibitem[{{Pe{\~n}a Ram{\'{\i}}rez} {et~al.}(2011){Pe{\~n}a Ram{\'{\i}}rez},
  {Zapatero Osorio}, {B{\'e}jar}, {Rebolo}, \& {Bihain}}]{penaramirez11}
{Pe{\~n}a Ram{\'{\i}}rez}, K., {Zapatero Osorio}, M.~R., {B{\'e}jar}, V.~J.~S.,
  {Rebolo}, R., \& {Bihain}, G. 2011, \aap, 532, A42

\bibitem[{{Perryman} {et~al.}(1997){Perryman}, {Lindegren}, {Kovalevsky},
  {Hoeg}, {Bastian}, {Bernacca}, {Cr{\'e}z{\'e}}, {Donati}, {Grenon}, {van
  Leeuwen}, {van der Marel}, {Mignard}, {Murray}, {Le Poole}, {Schrijver},
  {Turon}, {Arenou}, {Froeschl{\'e}}, \& {Petersen}}]{perryman97}
{Perryman}, M.~A.~C., {Lindegren}, L., {Kovalevsky}, J., {Hoeg}, E., {Bastian},
  U., {Bernacca}, P.~L., {Cr{\'e}z{\'e}}, M., {Donati}, F., {Grenon}, M., {van
  Leeuwen}, F., {van der Marel}, H., {Mignard}, F., {Murray}, C.~A., {Le
  Poole}, R.~S., {Schrijver}, H., {Turon}, C., {Arenou}, F., {Froeschl{\'e}},
  M., \& {Petersen}, C.~S. 1997, \aap, 323, L49

\bibitem[{{Petr-Gotzens} {et~al.}(2011){Petr-Gotzens}, {Alcal{\'a}},
  {Brice{\~n}o}, {Gonz{\'a}lez-Solares}, {Spezzi}, {Teixeira}, {Zapatero
  Osorio}, {Comer{\'o}n}, {Emerson}, {Hodgkin}, {Hussain}, {McCaughrean},
  {Melnick}, {Oliveira}, {Ramsay}, {Stanke}, {Winston}, \&
  {Zinnecker}}]{petr11}
{Petr-Gotzens}, M., {Alcal{\'a}}, J.~M., {Brice{\~n}o}, C.,
  {Gonz{\'a}lez-Solares}, E., {Spezzi}, L., {Teixeira}, P., {Zapatero Osorio},
  M.~R., {Comer{\'o}n}, F., {Emerson}, J., {Hodgkin}, S., {Hussain}, G.,
  {McCaughrean}, M., {Melnick}, J., {Oliveira}, J., {Ramsay}, S., {Stanke}, T.,
  {Winston}, E., \& {Zinnecker}, H. 2011, The Messenger, 145, 29

\bibitem[{{Reipurth} {et~al.}(1998){Reipurth}, {Bally}, {Fesen}, \&
  {Devine}}]{reipurth98}
{Reipurth}, B., {Bally}, J., {Fesen}, R.~A., \& {Devine}, D. 1998, \nat, 396,
  343

\bibitem[{{Reyl{\'e}} {et~al.}(2010){Reyl{\'e}}, {Delorme}, {Willott},
  {Albert}, {Delfosse}, {Forveille}, {Artigau}, {Malo}, {Hill}, \&
  {Doyon}}]{reyle10}
{Reyl{\'e}}, C., {Delorme}, P., {Willott}, C.~J., {Albert}, L., {Delfosse}, X.,
  {Forveille}, T., {Artigau}, E., {Malo}, L., {Hill}, G.~J., \& {Doyon}, R.
  2010, \aap, 522, A112

\bibitem[{{Rigliaco} {et~al.}(2009){Rigliaco}, {Natta}, {Randich}, \&
  {Sacco}}]{rigliaco09}
{Rigliaco}, E., {Natta}, A., {Randich}, S., \& {Sacco}, G. 2009, \aap, 495, L13

\bibitem[{{Sacco} {et~al.}(2008){Sacco}, {Franciosini}, {Randich}, \&
  {Pallavicini}}]{sacco08}
{Sacco}, G.~G., {Franciosini}, E., {Randich}, S., \& {Pallavicini}, R. 2008,
  \aap, 488, 167

\bibitem[{{Sacco} {et~al.}(2007){Sacco}, {Randich}, {Franciosini},
  {Pallavicini}, \& {Palla}}]{sacco07}
{Sacco}, G.~G., {Randich}, S., {Franciosini}, E., {Pallavicini}, R., \&
  {Palla}, F. 2007, \aap, 462, L23

\bibitem[{{Salpeter}(1955)}]{salpeter55}
{Salpeter}, E.~E. 1955, \apj, 121, 161

\bibitem[{{Santini} {et~al.}(2009){Santini}, {Fontana}, {Grazian}, {Salimbeni},
  {Fiore}, {Fontanot}, {Boutsia}, {Castellano}, {Cristiani}, {de Santis},
  {Gallozzi}, {Giallongo}, {Menci}, {Nonino}, {Paris}, {Pentericci}, \&
  {Vanzella}}]{santini09}
{Santini}, P., {Fontana}, A., {Grazian}, A., {Salimbeni}, S., {Fiore}, F.,
  {Fontanot}, F., {Boutsia}, K., {Castellano}, M., {Cristiani}, S., {de
  Santis}, C., {Gallozzi}, S., {Giallongo}, E., {Menci}, N., {Nonino}, M.,
  {Paris}, D., {Pentericci}, L., \& {Vanzella}, E. 2009, \aap, 504, 751

\bibitem[{{Scholz} \& {Eisl{\"o}ffel}(2004)}]{scholz04}
{Scholz}, A. \& {Eisl{\"o}ffel}, J. 2004, \aap, 419, 249

\bibitem[{{Scholz} \& {Jayawardhana}(2008)}]{scholz08}
{Scholz}, A. \& {Jayawardhana}, R. 2008, \apjl, 672, L49

\bibitem[{{Sherry} {et~al.}(2004){Sherry}, {Walter}, \& {Wolk}}]{sherry04}
{Sherry}, W.~H., {Walter}, F.~M., \& {Wolk}, S.~J. 2004, \aj, 128, 2316

\bibitem[{{Sherry} {et~al.}(2008){Sherry}, {Walter}, {Wolk}, \&
  {Adams}}]{sherry08}
{Sherry}, W.~H., {Walter}, F.~M., {Wolk}, S.~J., \& {Adams}, N.~R. 2008, \aj,
  135, 1616

\bibitem[{{Siess} {et~al.}(2000){Siess}, {Dufour}, \& {Forestini}}]{siess00}
{Siess}, L., {Dufour}, E., \& {Forestini}, M. 2000, \aap, 358, 593

\bibitem[{{Sim{\'o}n-D{\'{\i}}az} {et~al.}(2011){Sim{\'o}n-D{\'{\i}}az},
  {Caballero}, \& {Lorenzo}}]{simon11}
{Sim{\'o}n-D{\'{\i}}az}, S., {Caballero}, J.~A., \& {Lorenzo}, J. 2011, \apj,
  742, 55

\bibitem[{{Skrutskie} {et~al.}(2006){Skrutskie}, {Cutri}, {Stiening},
  {Weinberg}, {Schneider}, {Carpenter}, {Beichman}, {Capps}, {Chester},
  {Elias}, {Huchra}, {Liebert}, {Lonsdale}, {Monet}, {Price}, {Seitzer},
  {Jarrett}, {Kirkpatrick}, {Gizis}, {Howard}, {Evans}, {Fowler}, {Fullmer},
  {Hurt}, {Light}, {Kopan}, {Marsh}, {McCallon}, {Tam}, {Van Dyk}, \&
  {Wheelock}}]{skrutskie06}
{Skrutskie}, M.~F., {Cutri}, R.~M., {Stiening}, R., {Weinberg}, M.~D.,
  {Schneider}, S., {Carpenter}, J.~M., {Beichman}, C., {Capps}, R., {Chester},
  T., {Elias}, J., {Huchra}, J., {Liebert}, J., {Lonsdale}, C., {Monet}, D.~G.,
  {Price}, S., {Seitzer}, P., {Jarrett}, T., {Kirkpatrick}, J.~D., {Gizis},
  J.~E., {Howard}, E., {Evans}, T., {Fowler}, J., {Fullmer}, L., {Hurt}, R.,
  {Light}, R., {Kopan}, E.~L., {Marsh}, K.~A., {McCallon}, H.~L., {Tam}, R.,
  {Van Dyk}, S., \& {Wheelock}, S. 2006, \aj, 131, 1163

\bibitem[{{Sumi} {et~al.}(2011){Sumi}, {Kamiya}, {Bennett}, {Bond}, {Abe},
  {Botzler}, {Fukui}, {Furusawa}, {Hearnshaw}, {Itow}, {Kilmartin}, {Korpela},
  {Lin}, {Ling}, {Masuda}, {Matsubara}, {Miyake}, {Motomura}, {Muraki},
  {Nagaya}, {Nakamura}, {Ohnishi}, {Okumura}, {Perrott}, {Rattenbury}, {Saito},
  {Sako}, {Sullivan}, {Sweatman}, {Tristram}, {Udalski}, {Szyma{\'n}ski},
  {Kubiak}, {Pietrzy{\'n}ski}, {Poleski}, {Soszy{\'n}ski}, {Wyrzykowski},
  {Ulaczyk}, \& {Microlensing Observations in Astrophysics (MOA)
  Collaboration}}]{sumi11}
{Sumi}, T., {Kamiya}, K., {Bennett}, D.~P., {Bond}, I.~A., {Abe}, F.,
  {Botzler}, C.~S., {Fukui}, A., {Furusawa}, K., {Hearnshaw}, J.~B., {Itow},
  Y., {Kilmartin}, P.~M., {Korpela}, A., {Lin}, W., {Ling}, C.~H., {Masuda},
  K., {Matsubara}, Y., {Miyake}, N., {Motomura}, M., {Muraki}, Y., {Nagaya},
  M., {Nakamura}, S., {Ohnishi}, K., {Okumura}, T., {Perrott}, Y.~C.,
  {Rattenbury}, N., {Saito}, T., {Sako}, T., {Sullivan}, D.~J., {Sweatman},
  W.~L., {Tristram}, P.~J., {Udalski}, A., {Szyma{\'n}ski}, M.~K., {Kubiak},
  M., {Pietrzy{\'n}ski}, G., {Poleski}, R., {Soszy{\'n}ski}, I., {Wyrzykowski},
  {\L}., {Ulaczyk}, K., \& {Microlensing Observations in Astrophysics (MOA)
  Collaboration}. 2011, \nat, 473, 349

\bibitem[{{van den Bergh} \& {Sher}(1960)}]{bergh60}
{van den Bergh}, S. \& {Sher}, D. 1960, Publications of the David Dunlap
  Observatory, 2, 203

\bibitem[{{Walter} {et~al.}(1997){Walter}, {Wolk}, {Freyberg}, \&
  {Schmitt}}]{walter97}
{Walter}, F.~M., {Wolk}, S.~J., {Freyberg}, M., \& {Schmitt}, J.~H.~M.~M. 1997,
  \memsai, 68, 1081

\bibitem[{{Weaver} \& {Babcock}(2004)}]{weaver04}
{Weaver}, W.~B. \& {Babcock}, A. 2004, \pasp, 116, 1035

\bibitem[{{Wiramihardja} {et~al.}(1991){Wiramihardja}, {Kogure}, {Yoshida},
  {Nakano}, {Ogura}, \& {Iwata}}]{wiramihardja91}
{Wiramihardja}, S.~D., {Kogure}, T., {Yoshida}, S., {Nakano}, M., {Ogura}, K.,
  \& {Iwata}, T. 1991, \pasj, 43, 27

\bibitem[{{Wiramihardja} {et~al.}(1989){Wiramihardja}, {Kogure}, {Yoshida},
  {Ogura}, \& {Nakano}}]{wiramihardja89}
{Wiramihardja}, S.~D., {Kogure}, T., {Yoshida}, S., {Ogura}, K., \& {Nakano},
  M. 1989, \pasj, 41, 155

\bibitem[{{Wolk}(1996)}]{wolk96}
{Wolk}, S.~J. 1996, PhD thesis, SUNY Stony Brook, New York, USA

\bibitem[{{Worthey} \& {Lee}(2011)}]{worthey11}
{Worthey}, G. \& {Lee}, H.-c. 2011, \apjs, 193, 1

\bibitem[{{Wright} {et~al.}(2010){Wright}, {Eisenhardt}, {Mainzer}, {Ressler},
  {Cutri}, {Jarrett}, {Kirkpatrick}, {Padgett}, {McMillan}, {Skrutskie},
  {Stanford}, {Cohen}, {Walker}, {Mather}, {Leisawitz}, {Gautier}, {McLean},
  {Benford}, {Lonsdale}, {Blain}, {M{\'e}ndez}, {Irace}, {Duval}, {Liu},
  {Royer}, {Heinrichsen}, {Howard}, {Shannon}, {Kendall}, {Walsh}, {Larsen},
  {Cardon}, {Schick}, {Schwalm}, {Abid}, {Fabinsky}, {Naes}, \&
  {Tsai}}]{wright10}
{Wright}, E.~L., {Eisenhardt}, P.~R.~M., {Mainzer}, A.~K., {Ressler}, M.~E.,
  {Cutri}, R.~M., {Jarrett}, T., {Kirkpatrick}, J.~D., {Padgett}, D.,
  {McMillan}, R.~S., {Skrutskie}, M., {Stanford}, S.~A., {Cohen}, M., {Walker},
  R.~G., {Mather}, J.~C., {Leisawitz}, D., {Gautier}, III, T.~N., {McLean}, I.,
  {Benford}, D., {Lonsdale}, C.~J., {Blain}, A., {M{\'e}ndez}, B., {Irace},
  W.~R., {Duval}, V., {Liu}, F., {Royer}, D., {Heinrichsen}, I., {Howard}, J.,
  {Shannon}, M., {Kendall}, M., {Walsh}, A.~L., {Larsen}, M., {Cardon}, J.~G.,
  {Schick}, S., {Schwalm}, M., {Abid}, M., {Fabinsky}, B., {Naes}, L., \&
  {Tsai}, C.-W. 2010, \aj, 140, 1868

\bibitem[{{Zapatero Osorio} {et~al.}(2008){Zapatero Osorio}, {B{\'e}jar},
  {Bihain}, {Mart{\'{\i}}n}, {Rebolo}, {Vill{\'o}-P{\'e}rez},
  {D{\'{\i}}az-S{\'a}nchez}, {P{\'e}rez Garrido}, {Caballero}, {Henning},
  {Mundt}, {Barrado y Navascu{\'e}s}, \& {Bailer-Jones}}]{zapatero08}
{Zapatero Osorio}, M.~R., {B{\'e}jar}, V.~J.~S., {Bihain}, G., {Mart{\'{\i}}n},
  E.~L., {Rebolo}, R., {Vill{\'o}-P{\'e}rez}, I., {D{\'{\i}}az-S{\'a}nchez},
  A., {P{\'e}rez Garrido}, A., {Caballero}, J.~A., {Henning}, T., {Mundt}, R.,
  {Barrado y Navascu{\'e}s}, D., \& {Bailer-Jones}, C.~A.~L. 2008, \aap, 477,
  895

\bibitem[{{Zapatero Osorio} {et~al.}(2000){Zapatero Osorio}, {B{\'e}jar},
  {Mart{\'{\i}}n}, {Rebolo}, {Barrado y Navascu{\'e}s}, {Bailer-Jones}, \&
  {Mundt}}]{zapatero00}
{Zapatero Osorio}, M.~R., {B{\'e}jar}, V.~J.~S., {Mart{\'{\i}}n}, E.~L.,
  {Rebolo}, R., {Barrado y Navascu{\'e}s}, D., {Bailer-Jones}, C.~A.~L., \&
  {Mundt}, R. 2000, Science, 290, 103

\bibitem[{{Zapatero Osorio} {et~al.}(2002{\natexlab{a}}){Zapatero Osorio},
  {B{\'e}jar}, {Mart{\'{\i}}n}, {Rebolo}, {Barrado y Navascu{\'e}s}, {Mundt},
  {Eisl{\"o}ffel}, \& {Caballero}}]{zapatero02sori70}
{Zapatero Osorio}, M.~R., {B{\'e}jar}, V.~J.~S., {Mart{\'{\i}}n}, E.~L.,
  {Rebolo}, R., {Barrado y Navascu{\'e}s}, D., {Mundt}, R., {Eisl{\"o}ffel},
  J., \& {Caballero}, J.~A. 2002{\natexlab{a}}, \apj, 578, 536

\bibitem[{{Zapatero Osorio} {et~al.}(2002{\natexlab{b}}){Zapatero Osorio},
  {B{\'e}jar}, {Pavlenko}, {Rebolo}, {Allende Prieto}, {Mart{\'{\i}}n}, \&
  {Garc{\'{\i}}a L{\'o}pez}}]{zapatero02}
{Zapatero Osorio}, M.~R., {B{\'e}jar}, V.~J.~S., {Pavlenko}, Y., {Rebolo}, R.,
  {Allende Prieto}, C., {Mart{\'{\i}}n}, E.~L., \& {Garc{\'{\i}}a L{\'o}pez},
  R.~J. 2002{\natexlab{b}}, \aap, 384, 937

\bibitem[{{Zapatero Osorio} {et~al.}(1999){Zapatero Osorio}, {B{\'e}jar},
  {Rebolo}, {Mart{\'{\i}}n}, \& {Basri}}]{zapatero99}
{Zapatero Osorio}, M.~R., {B{\'e}jar}, V.~J.~S., {Rebolo}, R., {Mart{\'{\i}}n},
  E.~L., \& {Basri}, G. 1999, \apjl, 524, L115

\bibitem[{{Zapatero Osorio} {et~al.}(2007){Zapatero Osorio}, {Caballero},
  {B{\'e}jar}, {Rebolo}, {Barrado Y Navascu{\'e}s}, {Bihain}, {Eisl{\"o}ffel},
  {Mart{\'{\i}}n}, {Bailer-Jones}, {Mundt}, {Forveille}, \&
  {Bouy}}]{zapatero07}
{Zapatero Osorio}, M.~R., {Caballero}, J.~A., {B{\'e}jar}, V.~J.~S., {Rebolo},
  R., {Barrado Y Navascu{\'e}s}, D., {Bihain}, G., {Eisl{\"o}ffel}, J.,
  {Mart{\'{\i}}n}, E.~L., {Bailer-Jones}, C.~A.~L., {Mundt}, R., {Forveille},
  T., \& {Bouy}, H. 2007, \aap, 472, L9

\bibitem[{{Zapatero Osorio} {et~al.}(1997){Zapatero Osorio}, {Rebolo},
  {Mart{\'i}n}, {Basri}, {Magazz{\`u}}, {Hodgkin}, {Jameson}, \&
  {Cossburn}}]{zapatero97}
{Zapatero Osorio}, M.~R., {Rebolo}, R., {Mart{\'i}n}, E.~L., {Basri}, G.,
  {Magazz{\`u}}, A., {Hodgkin}, S.~T., {Jameson}, R.~F., \& {Cossburn}, M.~R.
  1997, \apjl, 491, L81

\end{thebibliography}



\clearpage

\clearpage

\clearpage

\clearpage

\clearpage

\clearpage

\clearpage

\clearpage

\clearpage

\clearpage

\clearpage


\clearpage

\begin{deluxetable}{cclcccc}\label{log}
\tablecaption{Log of VISTA observational data on the \so cluster\label{log}}
\tablewidth{-12pt}
\tabletypesize{\tiny}
\tablehead{
\colhead{RA~~(J2000)~~~DEC} &
\colhead{Filter } &
\colhead{Observing date } &
\colhead{$t_{\rm exp}$\tablenotemark{a}}&
\colhead{Seeing } &
\colhead{Completeness } &
\colhead{4-$\sigma$ limit }\\
\colhead{($^{\rm h~~~m~~~s}$)  ~~~~($^{\rm o~~~'~~~''}$)} &
\colhead{} &
\colhead{(UT)} &
\colhead{(s)}&
\colhead{(arcsec)} &
\colhead{(mag)} &
\colhead{(mag)}}

\startdata

05 38 20.6 ~~~$-$02 47 46.0& $Z$ & 2009 October 21,22 & 6084 & 0.8 & 22.6 & 23.1\\
&$Y$ & 2009 October 21 & 1008 &0.9 & 21.0 & 21.6\\
&$J$ & 2009 October 20,21,24,28 & 2112 & 0.9 & 21.0 & 21.6\\
&$H$ & 2009 October 21 & 288 &0.9 & 19.2 & 19.9\\
&$K_s$ & 2009 October 21 & 288 & 0.7 & 18.4 & 18.9\\

\enddata
\tablenotetext{a}{Given is the total time used to observe Tile 16. Because of the observing strategy (see text), the total exposure time per pixel is typically a factor of one third of the exposure time listed.}
\end{deluxetable}


\begin{deluxetable}{cl}\label{sub_coor}
\tablecaption{Central coordinates of the Suprime-Cam CCDs \label{sub_coor}}
\tablewidth{-12pt}
\tablehead{
\colhead{CCD} &
\colhead{RA~~(J2000)~~~DEC} \\
\colhead{} &
\colhead{($^{\rm h~~~m~~~s}$)  ~~~~($^{\rm o~~~'~~~''}$)}}

\startdata

0 & 05 39 42.2~~~ $-$02 11 08.0   \\
1 & 05 39 14.1~~~ $-$02 11 06.5 \\
2 & 05 38 45.4~~~ $-$02 24 56.0  \\
3 & 05 38 16.9~~~ $-$02 24 57.0  \\
4 & 05 38 16.8~~~ $-$02 11 06.0   \\
5 & 05 38 44.4~~~ $-$02 11 05.7 \\
7 & 05 39 12.9~~~ $-$02 24 56.6 \\
8 & 05 37 47.7~~~ $-$02 24 54.0   \\
9 & 05 37 48.8~~~ $-$02 11 09.0   \\

\enddata
\end{deluxetable}

\clearpage


\setlength{\tabcolsep}{0.03in}
\addtolength{\topmargin}{0.1in}
\begin{landscape}
\begin{deluxetable}{llcclr}
\tablecaption{Properties of \so candidate members with youth features \label{infoyoung}}
\tabletypesize{\scriptsize}
\tablewidth{600.0pt}
\tablehead{
\colhead{$\alpha$~~(J2000)~~~$\delta$} &
\colhead{Youth features\tablenotemark{a}} &
\colhead{Envelope\tablenotemark{b}} &
\colhead{Infrared excess\tablenotemark{c}} &
\colhead{Alt. names} &
\colhead{References} \\
\colhead{($^{\rm h~~~m~~~s}$)  ~~~~($^{\rm o~~~'~~~''}$)} &
\colhead{} &
\colhead{} &
\colhead{($\mu$m)} &
\colhead{} &
\colhead{} 
}
\startdata

05 38 36.88 $-$02 36 43.2  &  RV,g  &   $\checkmark$  &   N  &  [SWW2004]J053836.759-023643.68,[MJO2008]J053836.8-023643  &  22,35\\   
05 39 43.18 $-$02 32 43.3  &  d  &      $\checkmark$  &  Y(4.5,8.0,12.0)  &    S\,Ori\,J053943.2-023243,[HHM2007]1182   &  20,22,31\\  
05 38 21.20 $-$02 54 11.1  &  d  &      $\checkmark$  &   Y(4.5,8.0,12.0) &   [SWW2004]J053821.194-025411.03,[HHM2007]467  &  22,31\\    
05 38 50.78 $-$02 36 26.7  &  Li,g,RV,d  &  $\checkmark$  &   Y(12.0) &  [KJN2005]8,[OJV2006]22  &  24,28\\    
05 40 09.32 $-$02 25 06.7  &  Li,g,RV,d  &  $\checkmark$  &  Y(4.5,8.0,12.0) &    S\,Ori\,J054009.3-022507,[KJN2005]10,[HHM2007]1362   &  20,24,31\\  
05 38 17.78 $-$02 40 50.1  &  d  &      $\checkmark$  &  Y(4.5,8.0,12.0)  &    S\,Ori\,J053817.8-024050,[HHM2007]435   &  20,22,31\\  
05 39 29.35 $-$02 27 21.0  &   H$\alpha$,d   &  $\checkmark$  &  Y(4.5,8.0,12.0)  &    Kiso\,A-0976\,351\,B,[HHM2007]1075   &  2,31\\    
05 40 01.01 $-$02 19 59.8  &  Li,g,RV,d  &  $\checkmark$  &   N  &  [SE2004]122,[KJN2005]6,[HHM2007]1323  &  19,24,31\\
\enddata
\tablenotetext{a}{``RV" stands for radial velocity consistent with the systemic cluster radial velocity; ``Li" denotes lithium absorption at 607.82\,nm; ``g" indicates spectroscopic evidence for a low-gravity atmosphere; ``H$\alpha$" refers to strong, broad  H$\alpha$ emission likely related to accretion/jet events; ``d" stands for the presence of a disk.}
\tablenotetext{b}{There is a $\checkmark$ symbol for those objects located in the $ZJ$ eligible zone of cluster members according to our photometric criteria.}
\tablenotetext{c}{Infrared flux excesses (``Y") based on {\sl Spitzer} (4.5 and 8.0\,$\mu$m) and {\sl WISE} (12.0\,$\mu$m) data. No flux excesses are denoted with ``N". Objects off limits of the {\sl Spitzer} images are indicated with ``OL".}
\tablerefs{
(1) \citet{haromoreno53}; (2) \citet{wiramihardja91};  (3) \citet{wolk96}; (4) \citet{reipurth98};  (5) \citet{bejar99};  (6) \citet{zapatero99};  (7) \citet{zapatero00};  (8) \citet{tesisbejar};  (9) \citet{bejar01};  (10) \citet{martin01};  (11) \citet{barrado01};  (12) \citet{zapatero02};  (13) \citet{barrado02};  (14) \citet{zapatero02sori70};  (15) \citet{barrado03};  (16) \citet{jayawardhana03};  (17) \citet{mcgovern04}; (18) \citet{andrews04};  (19) \citet{scholz04};   (20) \citet{bejar04a};  (21) \citet{caballero04};  (22) \citet{sherry04};  (23) \citet{weaver04};  (24) \citet{kenyon05};  (25) \citet{burningham05};  (26) \citet{tesiscaballero};  (27) \citet{franciosini06};  (28) \citet{oliveira06};   (29) \citet{caballero06}; (30) \citet{gonzalez06};  (31) \citet{hernandez07};  (32) \citet{caballero07};  (33) \citet{zapatero07};  (34) \citet{scholz08};  (35) \citet{maxted08};  (36) \citet{caballero08b};   (37) \citet{sacco08};   (38) \citet{luhman08};  (39) \citet{lodieu09};  (40) \citet{bihain09};   (41) \citet{penaramirez11}; (42) \citet{bejar11}.
}
\end{deluxetable}
\end{landscape}
\clearpage


\begin{deluxetable}{cccccccccccccc}
\tablecaption{VISTA photometry of resolved sources
\label{resolved}}
\tabletypesize{\tiny}
\tablewidth{-12pt}
\tablehead{
\colhead{$\alpha$~~(J2000)~~~$\delta$} &
\colhead{$Z$} &
\colhead{$Y$} &
\colhead{$J$} &
\colhead{$H$} &
\colhead{$K_s$}   \\
\colhead{($^{\rm h~~~m~~~s}$)  ~~~~($^{\rm o~~~'~~~''}$)} &
\colhead{(mag)} &
\colhead{(mag)} &
\colhead{(mag)} &
\colhead{(mag)} &
\colhead{(mag)} 
}

\startdata

05 38 47.45 $-$02 06 45.5  &  22.660  $\pm$  0.468   & $\geq$\,23.1 &   20.096 $\pm$  0.088  &  $\geq$\,19.9         &  $\geq$\,18.9     \\
05 40 14.23 $-$02 51 46.3  &  22.908 $\pm$  0.462   & $\geq$\,23.1 &   20.343 $\pm$  0.107  &  $\geq$\,19.9         &  18.769 $\pm$  0.277   \\
05 37 05.48 $-$02 25 57.9  &  23.170 $\pm$ 0.343    & $\geq$\,23.1 &   20.490  $\pm$  0.120   &  19.268 $\pm$  0.191  &  18.573 $\pm$  0.232   \\
05 40 33.21 $-$02 29 54.1\tablenotemark{a}  &   14.726  $\pm$  0.003  &   13.962  $\pm$  0.001 &   13.465  $\pm$  0.002 &   12.794  $\pm$  0.002 &   12.543  $\pm$  0.003

\enddata
\tablenotetext{a}{Extended object found in the $ZJ$ search.}
\end{deluxetable}

\setlength{\tabcolsep}{0.03in}
\addtolength{\topmargin}{1.0in}
\clearpage
\begin{landscape}
\begin{deluxetable}{cccccccccccccc}
\tablecaption{Near- and mid-infrared photometry of \so candidate members with youth features
\label{young}}
\tabletypesize{\tiny}
\tablewidth{600.0pt}
\tablehead{
\colhead{$\alpha$~~(J2000)~~~$\delta$} &
\colhead{$Z$} &
\colhead{$Y$} &
\colhead{$J$} &
\colhead{$H$} &
\colhead{$K_s$} &
\colhead{$[3.6]$} &
\colhead{$[4.5]$} &
\colhead{$[5.8]$} &
\colhead{$[8.0]$} &
\colhead{$W1\,\,3.4 \mu$m} &
\colhead{$W2\,\,4.6 \mu$m} &
\colhead{$W3\,\,12.0 \mu$m} &
\colhead{$W4\,\,22.0 \mu$m}   \\
\colhead{($^{\rm h~~~m~~~s}$)  ~~~~($^{\rm o~~~'~~~''}$)} &
\colhead{(mag)} &
\colhead{(mag)} &
\colhead{(mag)} &
\colhead{(mag)} &
\colhead{(mag)} &
\colhead{(mag)} &
\colhead{(mag)} &
\colhead{(mag)} &
\colhead{(mag)} &
\colhead{(mag)} &
\colhead{(mag)} &
\colhead{(mag)} &
\colhead{(mag)}  
}

\startdata

05 38 36.88$-$02 36 43.2  &  14.208         $\pm$      0.002     &     13.575         $\pm$      0.001    &      13.003         $\pm$      0.001   &       12.460         $\pm$      0.001    &      12.150         $\pm$      0.002    &      11.776         $\pm$      0.030     &     11.835         $\pm$      0.033      &    11.773         $\pm$      0.023      &    11.710         $\pm$      0.032      &    12.073         $\pm$      0.032      &    11.799         $\pm$      0.030     &     \nodata                               &   \nodata                            \\
05 39 43.18$-$02 32 43.3  &  14.072         $\pm$      0.002     &     13.554         $\pm$      0.001    &      13.010         $\pm$      0.001   &       12.376         $\pm$      0.001    &      11.986         $\pm$      0.002    &      11.119         $\pm$      0.029     &     10.789         $\pm$      0.029      &    10.520         $\pm$      0.029      &    9.890          $\pm$      0.029      &    11.352         $\pm$      0.025      &    10.730         $\pm$      0.023     &     8.824          $\pm$      0.032       &   6.390          $\pm$      0.070    \\
05 38 21.20$-$02 54 11.1  &  14.367         $\pm$      0.002     &     13.887         $\pm$      0.001    &      13.043         $\pm$      0.001   &       12.642         $\pm$      0.002    &      12.188         $\pm$      0.002    &      11.279         $\pm$      0.029     &     11.000         $\pm$      0.029      &    10.720         $\pm$      0.029      &    10.020         $\pm$      0.029      &    11.571         $\pm$      0.025      &    11.012         $\pm$      0.023     &     8.879          $\pm$      0.031       &   6.479          $\pm$      0.060    \\
\enddata
\tablecomments{No \textit{WISE} photometry implies quoted signal-to-noise ratio below 4.}
\end{deluxetable}
\clearpage
\end{landscape}

\clearpage

\setlength{\tabcolsep}{0.03in}
\addtolength{\topmargin}{0.1in}
\begin{landscape}
\begin{deluxetable}{cccccccccccccc}
\tablecaption{Near- and mid-infrared photometry of known \so photometric candidates
\label{known}}
\tabletypesize{\tiny}
\tablewidth{600.0pt}
\tablehead{
\colhead{$\alpha$~~(J2000)~~~$\delta$} &
\colhead{$Z$} &
\colhead{$Y$} &
\colhead{$J$} &
\colhead{$H$} &
\colhead{$K_s$} &
\colhead{$[3.6]$} &
\colhead{$[4.5]$} &
\colhead{$[5.8]$} &
\colhead{$[8.0]$} &
\colhead{$W1\,\,3.4 \mu$m} &
\colhead{$W2\,\,4.6 \mu$m} &
\colhead{$W3\,\,12.0 \mu$m} &
\colhead{$W4\,\,22.0 \mu$m}   \\
\colhead{($^{\rm h~~~m~~~s}$)  ~~~~($^{\rm o~~~'~~~''}$)} &
\colhead{(mag)} &
\colhead{(mag)} &
\colhead{(mag)} &
\colhead{(mag)} &
\colhead{(mag)} &
\colhead{(mag)} &
\colhead{(mag)} &
\colhead{(mag)} &
\colhead{(mag)} &
\colhead{(mag)} &
\colhead{(mag)} &
\colhead{(mag)} &
\colhead{(mag)}  
}

\startdata

05 39 55.91$-$02 51 21.9  &  14.157         $\pm$      0.002     &     13.575         $\pm$      0.001     &     13.012         $\pm$      0.001    &      12.541         $\pm$      0.002     &     12.235         $\pm$      0.002    &      10.732         $\pm$      0.078       &   10.645         $\pm$      0.085       &   10.770         $\pm$      0.027       &   10.702         $\pm$      0.041       &   10.768         $\pm$      0.024      &    10.569         $\pm$      0.022      &    10.338         $\pm$      0.073      &                  \nodata                     \\ 
05 40 41.45$-$02 29 07.6  &  14.324         $\pm$      0.002     &     13.607         $\pm$      0.001     &     13.017         $\pm$      0.001    &      12.492         $\pm$      0.002     &     12.188         $\pm$      0.002    &                    \nodata                  &                 \nodata                  &                 \nodata                  &                 \nodata                  &   11.951         $\pm$      0.025      &    11.686         $\pm$      0.023      &                  \nodata                 &                  \nodata                     \\ 
05 39 25.61$-$02 34 04.3  &  14.288         $\pm$      0.002     &     13.664         $\pm$      0.001     &     13.118         $\pm$      0.002    &      12.598         $\pm$      0.002     &     12.286         $\pm$      0.002    &      11.890         $\pm$      0.031       &   12.024         $\pm$      0.031       &   11.816         $\pm$      0.016       &   11.737         $\pm$      0.041       &                 \nodata                 &                  \nodata                 &                  \nodata                 &                  \nodata                     \\
\enddata
\tablecomments{All these 58 sources comply with our $ZJ$ criteria for photometric cluster membership. No \textit{WISE} photometry implies signal-to-noise ratio below 4.}
\end{deluxetable}
\end{landscape}


\setlength{\tabcolsep}{0.04in}
\addtolength{\textwidth}{0.0in}
\addtolength{\topmargin}{-1.5in}
\begin{landscape}
\begin{deluxetable}{lclr}
\tablecaption{Information about known \so photometric candidates \label{infoknown}}
\tabletypesize{\tiny}
\tablewidth{-12pt}
\tablehead{
\colhead{$\alpha$~~(J2000)~~~$\delta$} &
\colhead{Infrared excess\tablenotemark{a}} &
\colhead{Other names} &
\colhead{References} \\
\colhead{($^{\rm h~~~m~~~s}$)  ~~~~($^{\rm o~~~'~~~''}$)} &
\colhead{($\mu$m)} &
\colhead{} &
\colhead{} 
}
\startdata
05 39 55.91 $-$02 51 21.9  &   Y(4.5)  &  [SWW2004]J053956.023-025123.05  &  22,36\\       
05 40 41.45 $-$02 29 07.6  &   OL  &  Mayrit\,1797077  &  36\\      
05 39 25.61 $-$02 34 04.3  &   N  &  [SWW2004]J053925.608-023404.04,[HHM2007]1043  &  22,31\\     
05 38 28.43 $-$03 00 25.9  &   N  &  [SWW2004]J053828.433-030026.12  &  22\\       
05 37 00.30 $-$02 28 26.5  &   OL,Y(12.0)  &  [SWW2004]J053700.310-022826.34  &  22\\       
05 36 49.28 $-$02 43 54.4  &   OL,Y(12.0)  &  [SWW2004]J053649.294-024354.28  &  22\\       
05 37 35.14 $-$02 26 57.7  &   N  &  [HHM2007]107  &  31\\       
05 39 13.47 $-$02 37 39.1  &   N  &  [HHM2007]940  &  31\\
\enddata
\tablecomments{All these 58 sources comply with our $ZJ$ criteria for photometric cluster membership. }
\tablenotetext{a}{Infrared flux excesses (``Y") based on {\sl Spitzer} (4.5 and 8.0\,$\mu$m) and {\sl WISE} (12.0\,$\mu$m) data. No flux excesses are denoted with ``N". Objects off limits of the {\sl Spitzer} images are indicated with ``OL".}
\tablerefs{
(1) \citet{haromoreno53}; (2) \citet{wiramihardja91};  (3) \citet{wolk96}; (4) \citet{reipurth98};  (5) \citet{bejar99};  (6) \citet{zapatero99};  (7) \citet{zapatero00};  (8) \citet{tesisbejar};  (9) \citet{bejar01};  (10) \citet{martin01};  (11) \citet{barrado01};  (12) \citet{zapatero02};  (13) \citet{barrado02};  (14) \citet{zapatero02sori70};  (15) \citet{barrado03};  (16) \citet{jayawardhana03};  (17) \citet{mcgovern04}; (18) \citet{andrews04};  (19) \citet{scholz04};   (20) \citet{bejar04a};  (21) \citet{caballero04};  (22) \citet{sherry04};  (23) \citet{weaver04};  (24) \citet{kenyon05};  (25) \citet{burningham05};  (26) \citet{tesiscaballero};  (27) \citet{franciosini06};  (28) \citet{oliveira06};   (29) \citet{caballero06}; (30) \citet{gonzalez06};  (31) \citet{hernandez07};  (32) \citet{caballero07};  (33) \citet{zapatero07};  (34) \citet{scholz08};  (35) \citet{maxted08};  (36) \citet{caballero08b};   (37) \citet{sacco08};   (38) \citet{luhman08};  (39) \citet{lodieu09};  (40) \citet{bihain09};   (41) \citet{penaramirez11}; (42) \citet{bejar11}.
}
\end{deluxetable}
\clearpage
\end{landscape}

\setlength{\tabcolsep}{0.04in}
\addtolength{\textwidth}{0.5in}
\addtolength{\topmargin}{1.0in}
\begin{landscape}
\begin{deluxetable}{cccccccccccc}
\tablecaption{Near and mid infrared photometry of new \so photometric candidates
\label{new}}
\tabletypesize{\tiny}
\tablewidth{650.0pt}
\tablehead{
\colhead{$\alpha$~~(J2000)~~~$\delta$} &
\colhead{$Z$} &
\colhead{$Y$} &
\colhead{$J$} &
\colhead{$H$} &
\colhead{$K_s$} &
\colhead{$[3.6]$} &
\colhead{$[4.5]$} &
\colhead{$[5.8]$} &
\colhead{$[8.0]$} &
\colhead{Name} &
\colhead{Infrared excess\tablenotemark{a}}    \\
\colhead{($^{\rm h~~~m~~~s}$)  ~~~~($^{\rm o~~~'~~~''}$)} &
\colhead{(mag)} &
\colhead{(mag)} &
\colhead{(mag)} &
\colhead{(mag)} &
\colhead{(mag)} &
\colhead{(mag)} &
\colhead{(mag)} &
\colhead{(mag)} &
\colhead{(mag)} &
\colhead{} &
\colhead{} 
}

\startdata

05 38 26.50 $-$02 09 25.7  &  20.463         $\pm$      0.054    &      19.231         $\pm$      0.039     &     18.252         $\pm$      0.025     &     17.663         $\pm$      0.047     &     17.062         $\pm$      0.061     &     16.220         $\pm$      0.102       &                   \nodata                &               $\geq$\,16.1               &                   \nodata                 &           S\,Ori\,J053826.5$-$020926 &       OL  \\  
05 38 29.62 $-$03 04 38.2  &  20.585         $\pm$      0.058    &      19.416         $\pm$      0.048     &     18.404         $\pm$      0.027     &     17.731         $\pm$      0.050     &     17.049         $\pm$      0.060     &                    \nodata                &                   \nodata                &                   \nodata                &                   \nodata                 &                  S\,Ori\,J053829.6$-$030438 &       OL   \\       
05 38 29.51 $-$02 59 59.1\tablenotemark{b}     &  20.874         $\pm$      0.068    &      19.513         $\pm$      0.050     &     18.416         $\pm$      0.027     &     17.662         $\pm$      0.047     &     17.093         $\pm$      0.063     &                    \nodata                &    16.332         $\pm$      0.146       &                   \nodata                &               $\geq$\,15.2                &       S\,Ori\,J053829.5$-$025959 &       Y   \\  
05 38 24.71 $-$03 00 28.3  &  21.051         $\pm$      0.075    &      19.728         $\pm$      0.061     &     18.643         $\pm$      0.032     &     18.089         $\pm$      0.067     &     17.386         $\pm$      0.081     &                    \nodata                &    16.666         $\pm$      0.171       &                   \nodata                &               $\geq$\,15.2                &  S\,Ori\,J053824.7$-$030028 &      N   \\
\enddata
\tablecomments{No \textit{WISE} photometry with few exceptions.}
\tablenotetext{a}{Infrared flux excesses (``Y") based on {\sl Spitzer} (4.5\,$\mu$m) data. No flux excesses are denoted with ``N". Objects off limits of the {\sl Spitzer} image are indicated with ``OL".}
\tablenotetext{b}{$W1$ (3.4\,$\mu$m)\,=\,16.800         $\pm$      0.138\,mag.}
\tablenotetext{c}{T-type candidate.}
\tablenotetext{d}{Found in the extended $ZJ$ search.}
\tablenotetext{e}{$W1$ (3.4\,$\mu$m)\,=\,16.904         $\pm$      0.149\,mag; $W2$ (4.6\,$\mu$m)\,=\,16.008 $\pm$ 0.222\,mag.}
\end{deluxetable}
\clearpage
\end{landscape}



\vspace{-2cm}
\begin{deluxetable}{cccccccc}
\tablecaption{Near- and mid-infrared photometry of sources found in the search down to $J$-band completeness
\label{arrows}}
\tabletypesize{\tiny}
\tablewidth{-12pt}
\tablehead{
\colhead{$\alpha$~~(J2000)~~~$\delta$} &
\colhead{$Z$} &
\colhead{$Y$} &
\colhead{$J$} &
\colhead{$H$} &
\colhead{$K_s$}  &
\colhead{$[3.6]$}  &
\colhead{$[4.5]$}   \\
\colhead{($^{\rm h~~~m~~~s}$)  ~~~~($^{\rm o~~~'~~~''}$)} &
\colhead{(mag)} &
\colhead{(mag)} &
\colhead{(mag)} &
\colhead{(mag)} &
\colhead{(mag)} &
\colhead{(mag)} &
\colhead{(mag)} 
}

\startdata

05 39 04.51 $-$03 02 13.5  &          $\geq$\,23.1                   &          $\geq$\,21.6            &         20.320         $\pm$      0.105     &      19.542         $\pm$      0.244       &    18.642         $\pm$      0.246      & \nodata  & 17.484  $\pm$ 0.359    \\      
05 37 42.60 $-$02 58 40.3  &          $\geq$\,23.1                   &          $\geq$\,21.6            &         20.501         $\pm$      0.121     &      19.685         $\pm$      0.278       &    19.132         $\pm$      0.385       & \nodata  & 16.725 $\pm$ 0.202   \\      
05 39 43.51 $-$03 01 56.2  &          $\geq$\,23.1                   &          $\geq$\,21.6            &         20.502         $\pm$      0.121     &      19.333         $\pm$      0.202       &    18.846         $\pm$      0.297       &   \nodata & 16.410 $\pm$ 0.154   \\      
05 38 18.20 $-$02 46 36.2\tablenotemark{a}  &          $\geq$\,23.1                   &          $\geq$\,21.6            &         20.559         $\pm$      0.127     &              $\geq$\,19.9                 &    18.669         $\pm$      0.253    &  17.178 $\pm$  0.178 &  16.738 $\pm$ 0.195      \\      
05 40 38.42 $-$02 28 06.6  &          $\geq$\,23.1                   &          $\geq$\,21.6            &         20.599         $\pm$      0.131     &      19.249         $\pm$      0.188       &    18.105         $\pm$      0.152    &  \nodata  &   \nodata   \\      
05 38 27.60 $-$02 09 09.0  &          $\geq$\,23.1                   &          $\geq$\,21.6            &         20.635         $\pm$      0.135     &              $\sim$\,19.9                 &            $\geq$\,18.9      & 18.14 $\pm$  0.418 &      \nodata          \\      
05 40 24.32 $-$02 44 44.3  &          $\geq$\,23.1                   &          $\geq$\,21.6            &         20.683         $\pm$      0.140     &              $\sim$\,19.9                 &            $\geq$\,18.9    &  \nodata &        \nodata          \\
\enddata
\tablenotetext{a}{Spatially resolved by \citet{penaramirez11}.}
\tablenotetext{b}{There is only a $J$-band detection.}
\end{deluxetable}



\setlength{\tabcolsep}{0.04in}
\addtolength{\textwidth}{0.1in}
\addtolength{\topmargin}{0.8in}
\begin{landscape}
\begin{deluxetable}{cccccccccclr}
\tablecaption{Near- and mid-infrared photometry of known \so photometric candidates ($J \ge 16$\,mag) out of the $ZJ$ selection criteria
\label{known2}}
\tabletypesize{\tiny}
\tablewidth{575.0pt}
\tablehead{
\colhead{$\alpha$~~(J2000)~~~$\delta$} &
\colhead{$Z$} &
\colhead{$Y$} &
\colhead{$J$} &
\colhead{$H$} &
\colhead{$K_s$} &
\colhead{$[3.6]$} &
\colhead{$[4.5]$} &
\colhead{$[5.8]$} &
\colhead{$[8.0]$} &
\colhead{Other names} &
\colhead{References} \\
\colhead{($^{\rm h~~~m~~~s}$)  ~~~~($^{\rm o~~~'~~~''}$)} &
\colhead{(mag)} &
\colhead{(mag)} &
\colhead{(mag)} &
\colhead{(mag)} &
\colhead{(mag)} &
\colhead{(mag)} &
\colhead{(mag)} &
\colhead{(mag)} &
\colhead{(mag)} &
\colhead{} &
\colhead{}  
}

\startdata

05 39 47.05$-$02 25 24.5  &  20.298         $\pm$      0.049     &     19.679         $\pm$      0.058     &     18.937         $\pm$      0.038    &      18.363         $\pm$      0.086     &     17.988         $\pm$      0.137    &                    \nodata                  &   17.280         $\pm$      0.380       &            $\geq$\,16.1                &            $\geq$\,15.2                &         S\,Ori\,57  &  1,2\\    
05 36 53.30$-$02 24 12.9   &  21.963         $\pm$      0.134     &     20.974        $\pm$      0.188     &     19.721         $\pm$      0.066    &      19.101        $\pm$      0.164     &     18.758         $\pm$      0.274    &                     \nodata                 &                  \nodata                 &     \nodata  & \nodata & S\,Ori\,64 &  1 \\
05 38 12.61$-$02 21 37.5  &  22.130         $\pm$      0.150     &     20.908         $\pm$      0.177     &     19.904         $\pm$      0.076    &      19.505         $\pm$      0.236     &     19.049         $\pm$      0.357    &      18.214         $\pm$      0.389       &   18.677         $\pm$      0.456       &            $\geq$\,16.1                &            $\geq$\,15.2              &         S\,Ori\,67  &  1,3,4,5\\           
05 38 59.16$-$02 35 26.1  &           $\geq$\,23.1              &              $\geq$\,21.6              &     20.329         $\pm$      0.105    &      19.731         $\pm$      0.290     &     18.971         $\pm$      0.332    &      17.650         $\pm$      0.110       &   17.510         $\pm$      0.290       &            $\geq$\,16.1                &            $\geq$\,15.2                &              S\,Ori\,72  &  6\\
\enddata
\tablecomments{ None shows mid-infrared flux excesses at 8 $\mu$m nor was detected by  \textit{WISE}.}
\tablerefs{
(1) \citealt{zapatero00}; (2) \citealt{martin01}; (3) \citealt{barrado01}; (4) \citealt{scholz08}; (5) \citealt{luhman08}; (6) \citealt{bihain09}; (7) \citealt{penaramirez11}.}
\end{deluxetable}
\clearpage
\end{landscape}



\clearpage

\setlength{\tabcolsep}{0.04in}
\addtolength{\textwidth}{0.01in}
\addtolength{\topmargin}{0.8in}
\begin{landscape}
\begin{deluxetable}{ccccccccc}
\tablecaption{Near- and mid-infrared photometry of sources selected in the Western 30\arcmin--50\arcmin~rectangular region  \label{out}}
\tabletypesize{\tiny}
\tablewidth{-12pt}
\tablehead{
\colhead{$\alpha$~~(J2000)~~~$\delta$} &
\colhead{$Z$} &
\colhead{$Y$} &
\colhead{$J$} &
\colhead{$H$} &
\colhead{$K_s$} &
\colhead{$W1\,\,3.4 \mu$m} &
\colhead{$W2\,\,4.6 \mu$m} &
\colhead{$W3\,\,12.0 \mu$m} \\
\colhead{($^{\rm h~~~m~~~s}$)  ~~~~($^{\rm o~~~'~~~''}$)} &
\colhead{(mag)} &
\colhead{(mag)} &
\colhead{(mag)} &
\colhead{(mag)} &
\colhead{(mag)} &
\colhead{(mag)} &
\colhead{(mag)} &
\colhead{(mag)} 
}

\startdata

05 35 28.98$-$02 25 44.4 &14.059  $\pm$  0.002 & 13.710  $\pm$   0.001  &13.085  $\pm$   0.002 & 12.514  $\pm$   0.002 & 12.192  $\pm$   0.002 & 11.774 $\pm$ 0.025   &  11.359 $\pm$ 0.024   &  9.867  $\pm$ 0.051  \\  
05 35 28.81$-$02 28 16.4 &14.218  $\pm$  0.002 & 13.808  $\pm$   0.001  &13.227  $\pm$   0.002 & 12.666  $\pm$   0.002 & 12.418  $\pm$   0.002 & 12.202 $\pm$ 0.025   &  12.012 $\pm$ 0.025   &  \nodata  \\  
05 36 24.43$-$02 17 56.0 &14.408  $\pm$  0.003 & 13.852  $\pm$   0.001  &13.301  $\pm$   0.002 & 12.767  $\pm$   0.002 & 12.524  $\pm$   0.003 & 11.918 $\pm$ 0.023   &  11.716 $\pm$ 0.023   &  \nodata  \\  
05 36 29.27$-$02 58 02.6 &14.451  $\pm$  0.003 & 13.908  $\pm$   0.001  &13.342  $\pm$   0.002 & 12.738  $\pm$   0.002 & 12.326  $\pm$   0.002 & 11.780 $\pm$ 0.024   &  11.361 $\pm$ 0.022   &  10.221 $\pm$ 0.063  \\  
05 35 46.85$-$02 50 53.9 &14.584  $\pm$  0.003 & 14.048  $\pm$   0.001  &13.492  $\pm$   0.002 & 12.956  $\pm$   0.002 & 12.707  $\pm$   0.003 & 12.410 $\pm$ 0.026   &  12.167 $\pm$ 0.026   &  \nodata            \\
05 35 51.46$-$02 25 26.1 &15.022  $\pm$  0.004 & 14.270  $\pm$   0.001  &13.719  $\pm$   0.002 & 13.049  $\pm$   0.002 & 12.708  $\pm$   0.003 & 12.428 $\pm$ 0.027   &  12.008 $\pm$ 0.024   &  10.466 $\pm$ 0.074  \\  
05 35 38.24$-$02 50 48.0 &15.845  $\pm$  0.005 & 15.009  $\pm$   0.002  &14.303  $\pm$   0.003 & 13.827  $\pm$   0.003 & 13.483  $\pm$   0.005 & 13.225 $\pm$ 0.029   &  12.925 $\pm$ 0.032   &  \nodata            \\
05 35 56.32$-$02 17 02.7 &16.294  $\pm$  0.007 & 15.581  $\pm$   0.002  &14.937  $\pm$   0.004 & 14.476  $\pm$   0.005 & 14.133  $\pm$   0.007 & 13.897 $\pm$ 0.032   &  13.622 $\pm$ 0.040   &  \nodata            \\
\enddata
\tablenotetext{a}{Photometric candidates already identified by \citet{bejar11}}
\tablecomments{ None has \textit{WISE} 22.0 $\mu$m detections.}
\end{deluxetable}
\clearpage
\end{landscape}

\clearpage

\begin{deluxetable}{llll}
\tablecaption{Parametrization of the \so mass spectrum and mass function \label{fits}}
\tablewidth{-12pt}
\tablecolumns{4}
\tablehead{
\colhead{}    &  \multicolumn{2}{c}{Case} &   \colhead{}   \\ 
\cline{2-3} \\ 
\multicolumn{1}{l}{    Mass range    } &
\colhead{  I  } &
\colhead{    II     }&
\multicolumn{1}{l}{ Functional form}}
\startdata

19--0.35\,$M_{\odot}$      & $\alpha=$1.73$\pm$0.16   & $\alpha=$1.69$\pm$0.18 & Power law  \\
0.35--0.006\,$M_{\odot}$ & $\alpha=$0.55$\pm$0.20   & $\alpha=$0.68$\pm$0.23 & Power law  \\
\hline

19--0.006\,$M_{\odot}$ & $M_c=$0.27$\pm$0.07   & $M_c=$0.27$\pm$0.11 & Lognormal \\
                                      & $\sigma=$0.57$\pm$0.11 & $\sigma=$0.68$\pm$0.19 & \\
\hline                                      
\hline
1--0.006\,$M_{\odot}$  & $M_c=$0.25$\pm$0.07 & $M_c=$0.23$\pm$0.11 & Lognormal + Power law\tablenotemark{a} \\
                                     & $\sigma=$0.47$\pm$0.13 & $\sigma=$0.58$\pm$0.24 & \\
19--1\,$M_{\odot}$       & $\alpha=$2.20$\pm$0.19 & $\alpha=$2.20$\pm$0.19 & \\
\hline
1--0.012\,$M_{\odot}$& $M_c=$0.25$\pm$0.07 & $M_c=$0.24$\pm$0.11 & Lognormal + Power law\tablenotemark{b}\\
                                   & $\sigma=$0.46$\pm$0.13 & $\sigma=$0.54$\pm$0.23 & \\
19--1\,$M_{\odot}$     & $\alpha=$2.20$\pm$0.19 & $\alpha=$2.20$\pm$0.19 &

\enddata
\tablenotetext{a}{Resembling the functional form presented by \citet{chabrier05}.}
\tablenotetext{b}{Resembling the functional form and mass interval used in \citet{chabrier05}.}

\tablecomments{Case I uses masses derived from \citet{siess00} for the mass interval 19--1\,$M_{\odot}$, and from the Lyon models for the mass interval 1--0.006\,$M_{\odot}$. Case II uses \citet{siess00} for the interval 19--0.3\,$M_{\odot}$, and Lyon models for lower masses.}

\end{deluxetable}

\end{document}